\documentclass[a4paper,fleqn]{cas-sc}
\usepackage{amsmath}
\usepackage{amssymb}
\usepackage{graphicx}
\usepackage{mathrsfs}
\usepackage{amsfonts}
\usepackage{amsthm}
\usepackage{color}
\usepackage{titletoc}
\titlecontents{section}
              [1.1cm]
              { \normalsize}%
              {\contentslabel{2em}}%
              {}%
              {\titlerule*[0.3pc]{$\cdot$}\normalsize\contentspage\hspace*{0cm}}%
\titlecontents{subsection}
              [2cm]
              {\normalsize}%
              {\contentslabel{2em}}%
              {}%
              {\titlerule*[0.3pc]{$\cdot$}\normalsize\contentspage\hspace*{0cm}}%
\titlecontents{subsubsection}
              [2.9cm]
              {\normalsize}%
              {\contentslabel{2.5em}}%
              {}%
              {\titlerule*[0.3pc]{$\cdot$}\normalsize\contentspage\hspace*{0cm}}%

\usepackage{lipsum}
\usepackage{gensymb}
\usepackage[numbers,sort&compress]{natbib}
\usepackage[numbers]{natbib} 
\def\tsc#1{\csdef{#1}{\textsc{\lowercase{#1}}\xspace}}
\tsc{WGM}
\tsc{QE}
\tsc{EP}
\tsc{PMS}
\tsc{BEC}
\tsc{DE}
\begin{document}
\let\WriteBookmarks\relax
\def\floatpagepagefraction{1}
\def\textpagefraction{.001}
\shorttitle{Z.-X. Li, Y. Cao, and P. Yan}
\shortauthors{Z.-X. Li et~al.}
%\begin{frontmatter}

\title [mode = title]{Topological insulators and semimetals in classical magnetic systems}                      

\author[]{Z.-X. Li}[orcid=0000-0001-6673-9199]

\address[]{School of Electronic Science and Engineering and State Key Laboratory of Electronic Thin Films and Integrated Devices, University of Electronic Science and Technology of China, Chengdu 610054, China}

\author[]{Yunshan Cao}[orcid=0000-0002-3409-2578]

\author[]{Peng Yan}[orcid=0000-0001-6369-2882]
\cormark[1]
\ead{yan@uestc.edu.cn}

\cortext[cor1]{Corresponding author.}

\begin{abstract}
Pursuing topological phases in natural and artificial materials is one of the central topics in modern physical science and engineering. In classical magnetic systems, spin waves (or magnons) and magnetic solitons (such as domain wall, vortex, skyrmion, etc) represent two important excitations. Recently, the topological insulator and semimetal states in magnon- and soliton-based crystals (or metamaterials) have attracted growing attention owing to their interesting dynamics and promising applications for designing robust spintronic devices. Here, we give an overview of current progress of topological phases in structured classical magnetism. We first provide a brief introduction to spin wave, and discuss its topological properties including magnon Hall effects, topological magnon insulators, and Dirac (Weyl) magnon semimetals. Appealing proposal of topological magnonic devices is also highlighted. We then review the collective-coordinate approach for describing the dynamics of magnetic soliton lattice. Pedagogical topological models such as the Su-Schrieffer-Heeger model and the Haldane model  and their manifestation in magnetic soliton crystals are elaborated. Then we focus on the topological properties of magnetic solitons, by theoretically analyzing the first-order topological insulating phases in low dimensional systems and higher-order topological states in breathing crystals. Finally, we discuss the experimental realization and detection of the edge states in both the magnonic and solitonic crystals. We remark the challenges and future prospects before concluding this article.
\end{abstract}


\maketitle

\tableofcontents
              
\section{Introduction}\label{section1}

Since the discovery of the quantum Hall effect \cite{KlitzingRRL1980,TsuiRRL1982,ThoulessRRL1982,KlitzingRMP1986} in two-dimensional electron gas system, the topological phases of matter began to attract people's attention for their exotic physical properties. The most peculiar character of topological phase, or more precisely, the topological insulators (TIs), is that they can support chiral edge/surface states which are absent in conventional insulators. The topological edge/surface states are the modes that are confined at the boundary/surface of the system and generally have a certain chirality (clockwise or counterclockwise). These properties are topologically protected and enable them being immune from moderate disorder and/or defects, which has defined a resistance that depends only on fundamental physical constants due to the robust in-gap edge states, making possible an accurate and standardized definition of the ohm. Topological insulating phases were originally observed in electronic system \cite{HasanRMP2010,QiRMP2011,MooreN2010,KaneRRL2005_1,KaneRRL2005_2}, while the concept of TIs has been extended to a broad fields of photonics \cite{KhanikaevNM2013,LuNP2014,KhanikaevNP2017,OzawaRMP2019,BandresS2018,HarariS2018}, acoustics \cite{ZhangCP2018,YangPRL2015,HeNP2016,FleuryNC2016,ChenPRA2016,HePRL2019,MaNRP2019}, mechanics \cite{MaNRP2019,HuberNP2016,PauloseNP2015,NashPNAS2015,MitchellNP2018,PatilS2020,ChenPRL2016}, electric circuits \cite{LeeCP2018,HelbigNP2020,ZhangPRB2019,HofmannPRL2019,LiuPRA2020,HofmannPRR2020,WangNC2020,RuiNSR2019}, and very recently in spintronics \cite{ZhangPRB2013,MookPRB2014,ChisnellPRL2015,RuckriegelPRB2018,ChernyshevPRL2016,MatsumotoPRL2011,MalzNC2019}. In the past years, the research about TIs in natural and artificial materials has become one of the most active areas in physical science and engineering because of the fundamental interest and the promising application in topological devices \cite{HasanRMP2010,QiRMP2011,OzawaRMP2019,ZhangCP2018,YangPRL2015,HuberNP2016,XiaoRMP2010,FuPRL2007,HsiehN2009}. 
\begin{figure}[ptbh]
\begin{centering}
\includegraphics[width=0.6\textwidth]{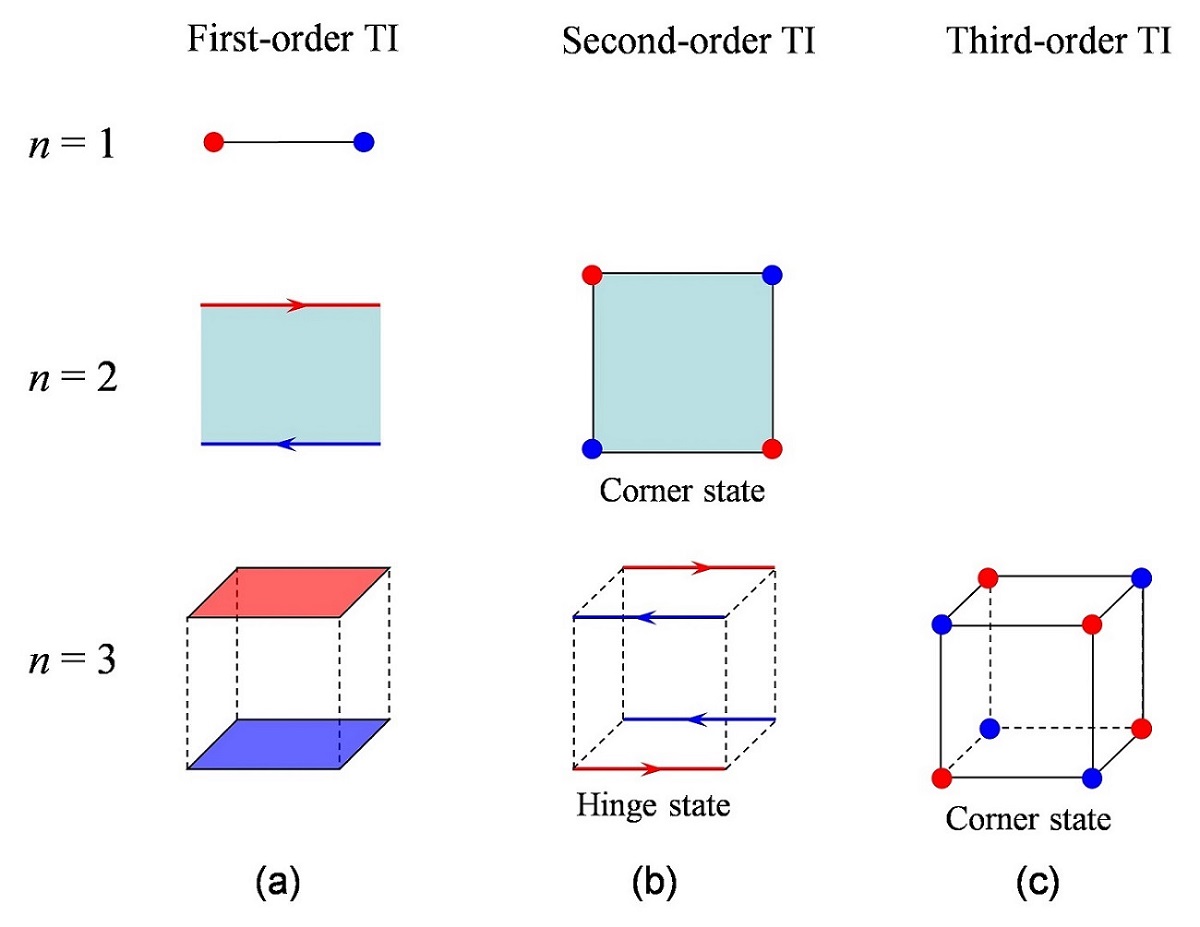}
\par\end{centering}
\caption{ The schematic plot for different types of TIs. (a) The first-order TI with in-gap modes localized at the corners for $n=1$, along the boundaries for $n=2$, and on surfaces for $n=3$. (b) The second-order TI with in-gap modes confined at the four corners for $n=2$ (corner state), and along the hinges of the system for $n=3$ (hinge state). (c) The third-order TI with in-gap modes at corners for $n=3$ (corner state).}
\label{Figure1}
\end{figure}

According to the standard bulk-boundary correspondence that the bulk property of TI dictates the character of edge or surface modes \cite{HasanRMP2010,QiRMP2011,HatsugaiPRL1993,HatsugaiPRB1993}, a conventional $n$-dimensional TI has ($n-1$)-dimensional topological edge/surface modes, called first-order TI (FOTI); see Fig. \ref{Figure1}(a). Interestingly, the concept of TIs recently is extended to higher-order cases, i.e., the so-called higher-order topological insulators (HOTIs) \cite{BenalcazarS2017,BenalcazarPRB2017,EzawaPRL2018,SongPRL2017,LangbehnPRL2017,SchindlerSA2018,QueirozPRL2019}. Different from FOTIs, a $k$th-order TI allows ($n-k$)-dimensional topological boundary modes ($2\leq k \leq n$), such as corner states and hinge states, as shown in Figs. \ref{Figure1}(b) and \ref{Figure1}(c), respectively. Due to their novel properties, the HOTIs have been investigated extensively in the broad community of photonics \cite{XiePRB2018,NohNP2018,HassanNP2019,MittalNP2019,ChenPRL2019,XiePRL2019,OtaO2019,LiNP2020,KimLSA2020,HeNC2020,ZhangAS2020,KimNp2020}, acoustics \cite{Serra-GarciaN2018,LiuNM2019,XueNM2019,NiNM2019,XuePRL2019,ZhangNP2019,ZhangNC2019,ChenPRB2019,ZhangPRL2019,ZhangAM2019,LopezCM2019,QiPRL2020,XueNC2020,NiNC2020,WeinerSA2020}, mechanics \cite{FanPRL2019,WakaoPRB2020,WuPRA2020}, electronics \cite{PetersonN2018,ImhofNP2018,LuNP2018,EzawaPRB2018_1,Serra-GarciaPRB2019,BaoPRB2019,YangPRR2020,SongAR2020,NiPRA2020}, and magnetics \cite{SilJPCM2020,Linpj2019,LiPRB2020,LiPRA2020} in the last few years. The topological description of HOTIs goes beyond the conventional bulk-boundary correspondence and is characterized by a few new topological invariants, such as the bulk polarization (Wannier center) \cite{EzawaPRL2018,XueNM2019,King-SmithPRB1993,VanderbiltPRB1993}, Green's function zeros \cite{SlagerPRB2015}, and $\mathbb{Z}_{N}$ Berry phase (quantized to $2\pi/N$) \cite{ZakPRL1989,KariyadoPRL2018,HatsugaiEPL2011,MizoguchiJPSJ2019,ArakiPRR2020,KudoPRL2019}. The HOTIs are broadening our understanding on topological insulating phases of matter. Notablely, the practical applications of higher-order topological edge states (corner states for instance) are also stimulating significant research enthusiasm in the community \cite{ZhangAM2019,LiPRB2020}. 

Topological semimetals (TSs) \cite{ArmitageRMP2018} are another exotic phase of matter with unusual gapless band structure. Typical TSs include Weyl semimetals \cite{WanPRB2011,LvPRX2015,LiuS2019} and Dirac semimetals \cite{YoungPRL2012,BorisenkoPRL2014,LiuS2014,LiuNM2014}. Weyl semimetals are characterized by the twofold degenerate points (called Weyl points or nodes) resulting from the linear crossings of two bands. The Weyl points (or nodes) are described by the momentum-space monopoles of Berry curvature and they must come in pairs with opposite chirality due to the no-go theorem. The band inversion happens between two paired Weyl nodes, leading to the generation of topologically protected Fermi-arc-like surface states. It is noted that Weyl semimetal states emerge only when at least one of the symmetries (time-reversal symmetry and inversion symmetry) is broken. In contrast, the Dirac semimetals are characterized by Dirac points with fourfold degenerate band touchings. The Dirac semimetals respect both the time-reversal and inversion symmetries, while their stability requires additional crystalline symmetries, for example, the rotation symmetry \cite{LiuS2014,LiuNM2014}. Interestingly, it is found very recently that the Weyl semimetals can support higher-order topological edge states (hinge states) \cite{WangPRL2020,WeiAR2020}, which is referred to as higher-order Weyl semimetals. At present, the topological semimetals have been studied extensively in various systems because of their exotic properties and potential applications \cite{GaoPRL2018,BurkovPRL2011,XuS2015,HuangNC2015,DengNP2016,XiongS2015,WiederPRL2016}.    

In magnetic systems, one of the most important elementary excitations is magnon (the quantized quasiparticle of spin wave) \cite{BlochZP1930,KuboPR1952,DysonPR1956,PrabhakarSP2009}, which describes the collective motion of localized spins in solids. Similar to charged currents, magnons (or spin waves) can also carry, transmit, and process information  \cite{KhitunJPD2010,ChumakNC2014,KlinglerAPL2015}. Significantly, compared with the conventional electronic devices, the magnonic devices have several outstanding advantages: (1) The propagation of spin wave (SW) does not carry electric current, which avoids the excessive energy consumption caused by Joule heating and thus improves the device performance \cite{KruglyakJPD2010,ChumakNP2015}; (2) There are more degrees of freedom for information processing by manipulating SW. For example, various logic gates can be made by controlling the phase and amplitude of SWs \cite{KlinglerAPL2015,KostylevAPL2005,LeeJAP2008}; (3) The diffusing length of SW can reach the order of millimeter or even centimeter, which is much longer than the electron spin diffusing length. This characteristic can be used to realize the long distance information transmission \cite{CornelissenNP2015,WesenbergNP2017,PirroAPL2014}; (4) The wavelength of SW can be scaled to a few nanometers or dozens of nanometers, which can improve the storage density of information and is conducive to the realization of devices miniaturization \cite{ChumakNC2014,DumasNN2014,LenkPR2011}. Because of these reasons, a new discipline---magnonics (or magnon spintronics) \cite{KruglyakJPD2010,SergaJPD2010,LenkPR2011,ChumakNP2015} emerges in recent years, which focuses on the generation, propagation, detection, and manipulation of spin waves. However, the topological properties of magnons are rarely explored until the pioneering work by Onose \emph{et al.} in 2010 \cite{OnoseS2010}---the experimental observation of the magnon Hall effect (MHE) in insulating pyrochlore ferromagnet Lu$_{2}$V$_{2}$O$_{7}$. To explain the emerging MHE that the longitudinal temperature gradient can induce the transverse thermal current, Matsumoto \emph{et al.} \cite{MatsumotoPRL2011,MatsumotoPRB2011} proposed the model of uncompensated net magnon edge currents. The concept of "topological magnon insulator" is then adopted \cite{ZhangPRB2013,MookPRB2014,RuckriegelPRB2018} to describe a large class of systems that support chiral magnon current circulating around device boundaries. The thermal MHE was subsequently observed in garnet magnet (for example, yttrium iron garnets) \cite{MadonAX2014,TanabePSS2016}, kagome and pyrochlore magnet \cite{HirschbergerPRL2015,IdeuePRB2012}, and frustrated pryocholore quantum magnet \cite{HirschbergerS2015}. The experimental and theoretical advance on topological magnons over the past decade can be summarized into the following directions: (i) The topological magnon insulator based on different lattices, such as the kagome (or pyrochlore) \cite{MookPRB2014_2,MookPRB2015_1,MookPRB2015_2,PereiroNC2014,LaurellPRL2017,ChisnellPRL2015,ChisnellPRB2016,SeshadriPRB2018,OwerreJPCM2018} and honeycomb \cite{TserkovnyakPRL2016,ChenPRX2018,McClartyPRB2018,PantaleonJPCM2019,OwerreJPCM2016,OwerreJAP2016} lattices; (ii) Different microscopic mechanisms leading to the nontrivial topology, for example, the Dzyaloshinskii-Moriya (DM) interaction \cite{KovalevPRB2018,MoulsdalePRB2019,MalkiPRB2019,IacoccaPRA2017,OwerreJPCM2017} due to the inversion symmetry broken \cite{DzyaloshinskiiJPCS1958,MoriyaPRL1960}, magnetic dipolar interaction \cite{ShindouPRB2013_2,LisenkovPRB2014,PirmoradianPRB2018}, pseudodipolar exchange interaction \cite{WangPRB2017,WangPRA2018}, and magnetic texture \cite{HoogdalemPRB2013,MochizukiNM2014,MolinaNJP2016,MookPRB2017,GarstJPD2017,DiazPRR2020}; (iii) The Dirac and Weyl magnons   \cite{FranssonPRB2016,OwerreJPCM2016,BoykoPRB2018,PershogubaPRX2018,ChenPRX2018,YuanPRX2020,MookPRL2016,SuPRB2017_1,SuPRB2017_2,ZyuzinPRB2018,OwerreSR2018,LiNC2016}; (iv) The higher-order topological magnons \cite{SilJPCM2020}; (v) Topological magnons in antiferromagnet and ferrimagnet \cite{OwerreJAP2017,NakataPRB2017,YaoNP2018,KimPRB2019,KawanoPRB2019,LiNL2018}. (vi) Novel robust magnonic devices, such as spin-wave diodes, spin-wave beam splitters, spin-wave interferometers, spin-wave logic gates, etc \cite{WangPRA2018,ShindouPRB2013_1}. 

Another important excitation in magnetic system is the magnetic soliton. Magnetic solitons \cite{KosevichPR1990} are shape-preserving and self-localized structures, with typical examples including magnetic vortex \cite{WachowiakS2002,WaeyenbergeN2006}, bubble \cite{MakhfudzPRL2012,MoonPRB2014,PetitAPL2015}, skyrmion \cite{RoblerN2006,MuhlbauerS2009,JiangS2015}, and domain wall \cite{AllwoodS2005,HayashiS2008,CatalanRMP2012}. These magnetic solitons have the characteristics of small size, easy manipulation, and high stability, and they are long-term topics in condensed matter physics for their interesting dynamics and promising applications \cite{ParkinS2008,JonietzS2010,ParkinNN2015,YuNC2012,PribiagNP2007,ZhangNJP2015,HrkacJPD2015,FertNN2013}. Similar to other (quasi-)particles, the collective dynamics of magnetic solitons exhibits the behavior of waves \cite{ShibataPRB2004,GalkinPRB2006,VogelPRL2010,JungSR2011,VogelPRB2014,HanzeAPL2014,BehnckePRB2015,HanzePRB2015,HanzeSR2016,BehnckeSR2018,GareevaJPC2018,HanSR2013,HanAPL2013,KimSR2017,MruczkiewiczPRB2016}. By mapping the massless Thiele’s equation into the Haldane model \cite{HaldanePRL1988}, Kim and Tserkovnyak \cite{KimPRL2017} predicted that the two-dimensional honeycomb lattice of magnetic vortices (or bubbles) can support chiral edge states, which has been confirmed by full micromagnetic simulations \cite{LiPRB2018}. Li \emph{et al}. \cite{LiAR2020} and Go \emph{et al}. \cite{GoPRB2020} studied the Su-Schrieffer-Heeger (SSH) \cite{SuPRL1979} states in one-dimensional magnetic soliton lattice. Li \emph{et al}. predicted theoretically the second-order topological phases (corner states) in two-dimensional breathing kagome \cite{Linpj2019}, honeycomb \cite{LiPRA2020}, and square \cite{LiPRB2020} lattice of magnetic vortices and showed that the emerging corner states are very robust against disorder and defects because of the generalized chiral symmetry. The collective motion of magnetic solitons in two dimensions is described by the generalized Thiele’s equation, which results in a wavelike equation in the artificial crystal, and this equation differs from the wave equations of its electronic, photonic, and acoustic counterparts in the following respects: (i) The nonvanishing topological charge induces a gyration term that is analogous to an effective magnetic field acting on a quasiparticle, thus breaking time-reversal symmetry. (ii) The inertial effect is taken into account by a mass term. A third-order non-Newtonian gyration term is included to capture the high-frequency behavior of the magneticsolitons and to allow one to determine the interaction parameters with high accuracy. (iii) The soliton-soliton coupling is strongly anisotropic. (iv) The conventional chiral symmetry in bipartite lattices is replaced by a more general chiral symmetry.

In this review, we give a detailed introduction to topological insulators and semimetals in magnonic and solitonic systems. The exposition is organized as follows: Section \ref{section2.1} describes the spin wave and magnon Hall effect; The summary of the studies about the topological magnon insulatrs is given in Section \ref{section2.2}; The topological magnon semimetals (including Dirac and Weyl magnons) are discussed in Section \ref{section2.3}; In Section \ref{section2.4}, the higher-order topological magnons are introduced; The concepts of topological magnonic devices are discussed in Section \ref{section2.5}; The topological structures and properties for different magnetic solitons are presented in Section \ref{section3.1}; Section \ref{section3.2} gives a brief review about the collective dynamics of magnetic solitons; Two pedagogical topological models (the SSH and Haldane models) are introduced in Section \ref{section3.3}; Sections \ref{section3.4} and \ref{section3.5} focus on the topological phases of magnetic soliton crystals. In Section \ref{section4}, we present future prospects and challenges about the topology in magnetism.
\begin{figure}[ptbh]
\begin{centering}
\includegraphics[width=0.68\textwidth]{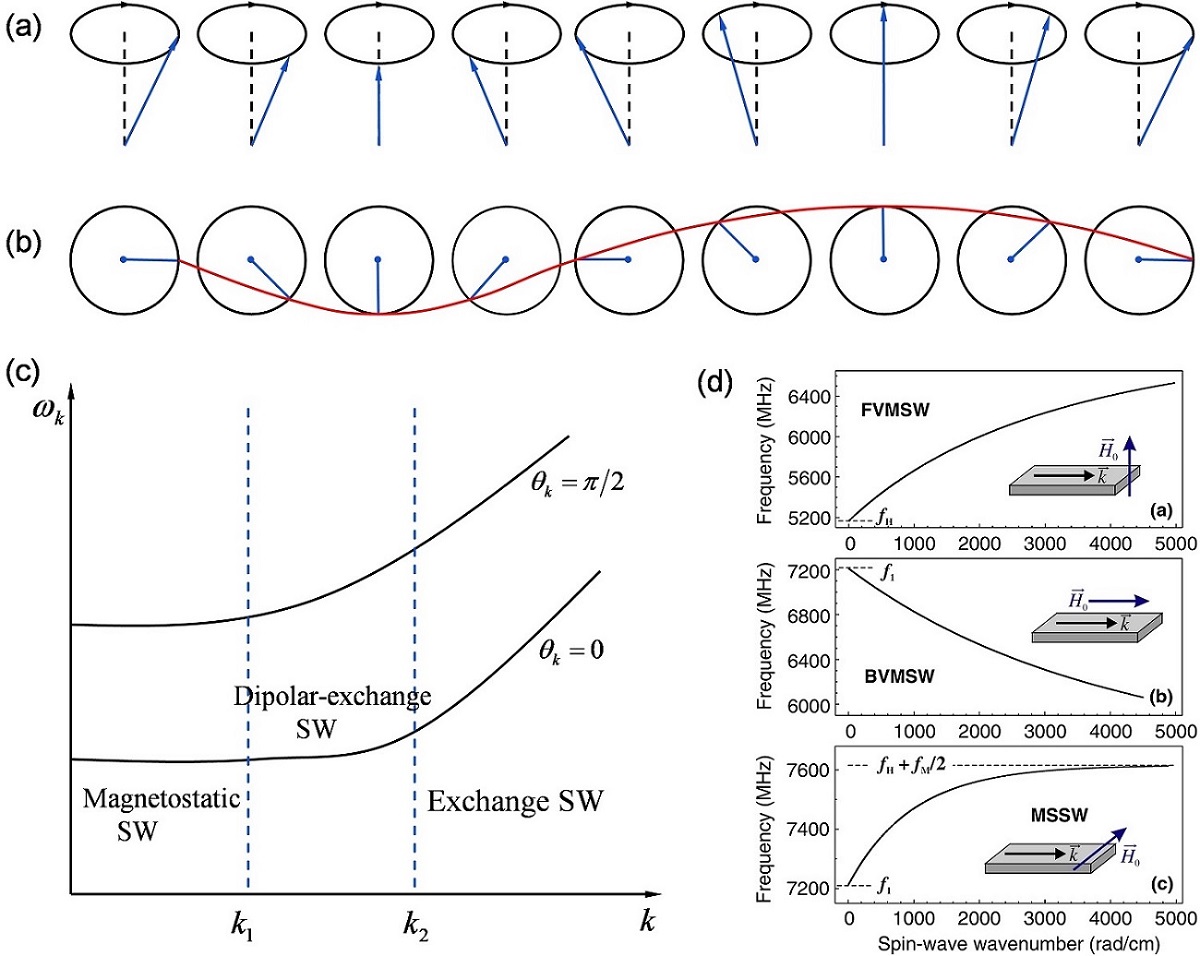}
\par\end{centering}
\caption{The side (a) and top (b) views of SW propagation in a one dimensional spin chain. (c) The dispersion relation of SW for a magnetic ellipsoid. (d) Different types of magnetostatic SWs in yttrium iron
garnet (YIG). Source: The figure (d) is taken from Ref. \cite{SergaJPD2010}.}
\label{Figure2}
\end{figure}

\section{Topological magnons}\label{section2}

In magnets, due to the short-range exchange and long-range dipolar interactions, the local oscillation of magnetic moments spreads all over the magnet in the form of SWs as shown in Figs. \ref{Figure2}(a) and \ref{Figure2}(b). Because of their different wavelengths, SWs can be divided into three types: (i) The exchange SW, with very short wavelength, where the short-range exchange interaction dominates. (ii) The dipolar (or magnetostatic) SW, with very long wavelength, where the long-range dipolar interaction dominates. (iii) The dipolar-exchange SW, where neither exchange nor dipolar interactions can be ignored. Figure \ref{Figure2}(c) plots the dispersion relation of SW in magnetic ellipsoids, with $\theta_{k}$ representing the angle between the magnetic moment and the wave vector. Due to the limitation of experimental technology, most of the researches are about magnetostatic SW. The magnetostatic SW can be further divided into the following categories: (i) The forward volume magnetostatic spin wave (FVMSW), for equilibrium magnetization perpendicular to the film surface; For in-plane magnetized films, the SW have two types with (ii) the backward volume magnetostatic spin wave (BVMSW) and (iii) magnetostatic surface spin wave (MSSW), which allows SW propagation parallel or perpendicular to the magnetic moment, respectively. The dispersion relation for different magnetostatic SWs are shown in Fig. \ref{Figure2}(d). 

\begin{figure}[ptbh]
\begin{centering}
\includegraphics[width=0.90\textwidth]{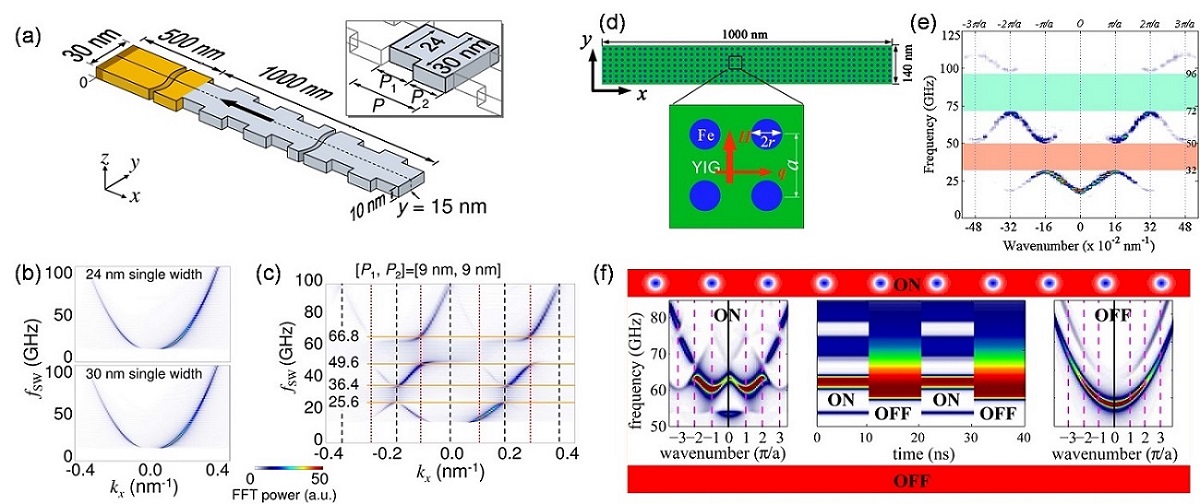}
\par\end{centering}
\caption{(a) The geometry and dimensions of nanostrip MCs with periodic modulation of widths. (b) The dispersion curves of SW in single-width nanostrips of 24 and 30 nm. (c) Dispersion curves of SW existing in the MC of $[P_{1},P_{2}]=$ [9 nm, 9 nm]. (d) Schematic of MC comprising a regular square array of circular Fe dots in a YIG matrix. (e) Dispersion relations of MC considered in (d), the dotted lines indicate the BZ boundaries. (f) The illustration and dispersion curves for uniform magnetized state and skyrmion lattice. Source: The figures are taken from Refs. \cite{LeePRL2009,MaAPL2011,MaNL2015}.}
\label{Figure3}
\end{figure}

\begin{figure}[ptbh]
\begin{centering}
\includegraphics[width=0.80\textwidth]{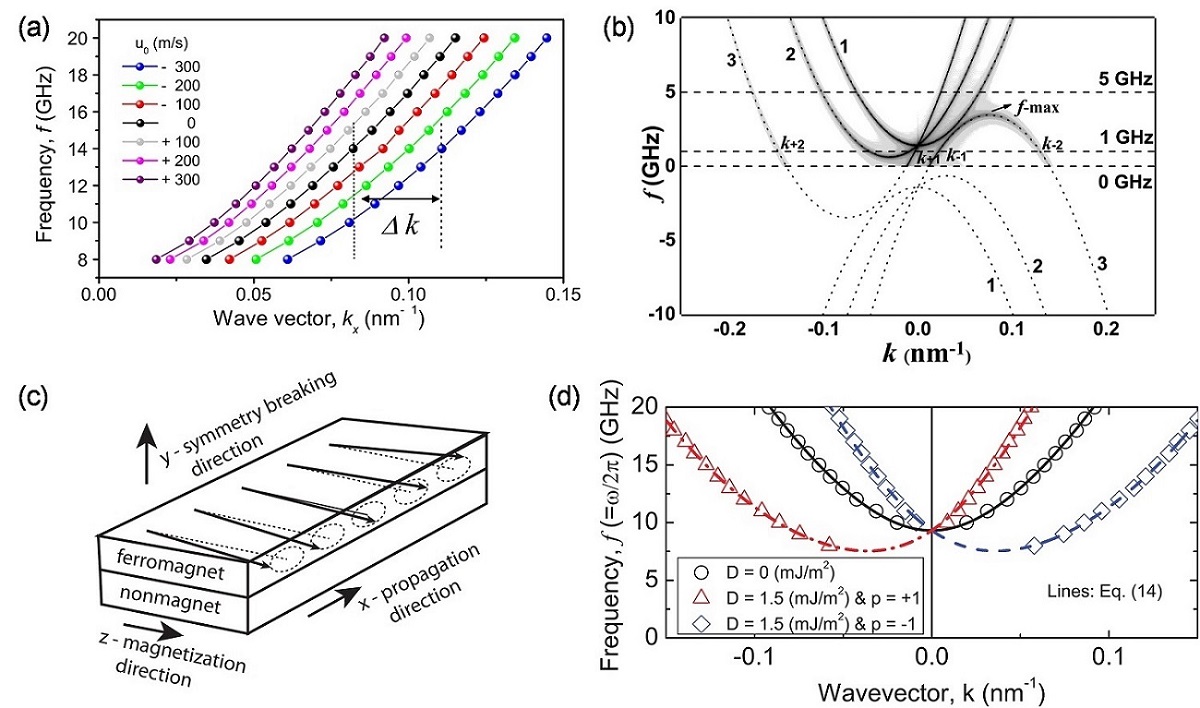}
\par\end{centering}
\caption{(a) The dispersion relations ($f$ versus $k_{x}$) at various current densities. (b) Dispersion relations of SW at different current. The curves 1, 2, and 3 correspond to the $c_{j}$ = 0, $-323$ and $-808$ m/s, respectively, $c_{j}$ is the parameter of spin transfer toque. The dotted lines represent the analytical results. (c) The illustration of the magnetic nanostrip with the symmetry breaking in the $y$-axis direction. (d) Dispersion relation of asymmetric SW induced by an interfacial DM interaction in the large-$k$ limit. Source: The figures are taken from Refs. \cite{SeoPRL2009,ZhouPRB2019,MoonPRB2013}.}
\label{Figure4}
\end{figure}

The dispersion relation of SW can be significantly modified by various approaches. One of the most effective methods is by using magnonic crystals (MCs) \cite{KrawczykJPCM2014,ChumakJPD2017}. MCs are the artificial magnetic material structures with periodic variation of magnetic or geometric parameters. Similar to the photonic or phononic crystals, the band structure of SW propagating in such structures consists of a series of allowed and forbidden frequency bands \cite{CiubotaruPRB2013,LeePRL2009,WangAPL2009}. Kim \emph{et al.} \cite{LeePRL2009} studied the one-dimensional MCs with a single nanostrip with periodic width variation, as shown in Fig. \ref{Figure3}(a). They found that there is no band gap existing when the SW propagates into the nanostrip with uniform width. However, when the width varies periodically, a series of band gaps emerge; see Figs. \ref{Figure3}(b) and \ref{Figure3}(c). Furthermore, the number, position and width of the band gaps can be tuned by modifying the device geometry. Similarly, two-dimensional MCs with periodic magnetic parameters variation can also support allowed and forbidden frequency bands \cite{MaAPL2011}, as shown in Figs. \ref{Figure3}(d) and \ref{Figure3}(e). Interestingly, Ma \emph{et al.} \cite{MaNL2015} reported a new type MC which consists of  skyrmions. They identified band daps when the SW propagates into such skyrmion lattice, as shown in Fig. \ref{Figure3}(f). In addition, the other type of magnetic soliton (such as domain wall) MCs also have been proposed \cite{LiJMMM2015,WangEPL2015}.     
Besides MCs, the spin polarized current can be used to modify the band structure of SW, too. Seo \emph{et al.} \cite{SeoPRL2009} reported that the dispersion relation of SW have a positive or negative shift depending on the direction of current; see Fig. \ref{Figure4}(a). Further, Zhou \emph{et al.} \cite{ZhouPRB2019} found that when suitable spin-polarized electrical currents are applied, the ferromagnetic system can support left-handed polarized SWs. Moreover, they confirmed that the right-handed and left-handed polarized SWs can coexist when the current density is larger than a critical value, with the dispersion relation under the different current density being plotted in Fig. \ref{Figure4}(b). What's more, Moon \emph{et al.} \cite{MoonPRB2013} showed that the interfacial DM interaction can tune the SW dispersion relation and leads to a nonreciprocal SW propagation, as showns in Figs. \ref{Figure4}(c) and \ref{Figure4}(d). Comprehensive summary on magonics can be found in early review articles \cite{KruglyakJPD2010,SergaJPD2010,LenkPR2011,ChumakNP2015}.
 
We point out that, in fermionic system (such as electron), it is easy to identify the topological phase by linear transport measurements \cite{HasanRMP2010,QiRMP2011}. While magnetic systems are bosonic, they have a simple condensate or vacuum ground state when they are in topological phases \cite{VishwanathPRX2013}. As a result, it is difficult to characterize their topological nature. However, if magnetic systems are in the excited state, the situation becomes different, because magnons can carry signatures of the topological band structure. For example, by measuring the thermal Hall conductivity, one can judge if the magnetic system is in the topological insulating phase. In what follows, we will introduce the topological properties of magnons.      

\subsection{Magnon Hall effect}\label{section2.1}
The study of the topological properties of SWs begins with the observation of MHE. It is generally known that the Hall effect occurs when the Lorentz force acts on a charge current in the presence of a perpendicular magnetic field \cite{HallAJM1879}. However, magnons are neutral quasi-particles, the realization of MHE does not resort to the Lorentz force, which is similar to the anomalous Hall effect in the metallic ferromagnets \cite{NagaosaRMP2010}.
\begin{figure}[ptbh]
\begin{centering}
\includegraphics[width=0.85\textwidth]{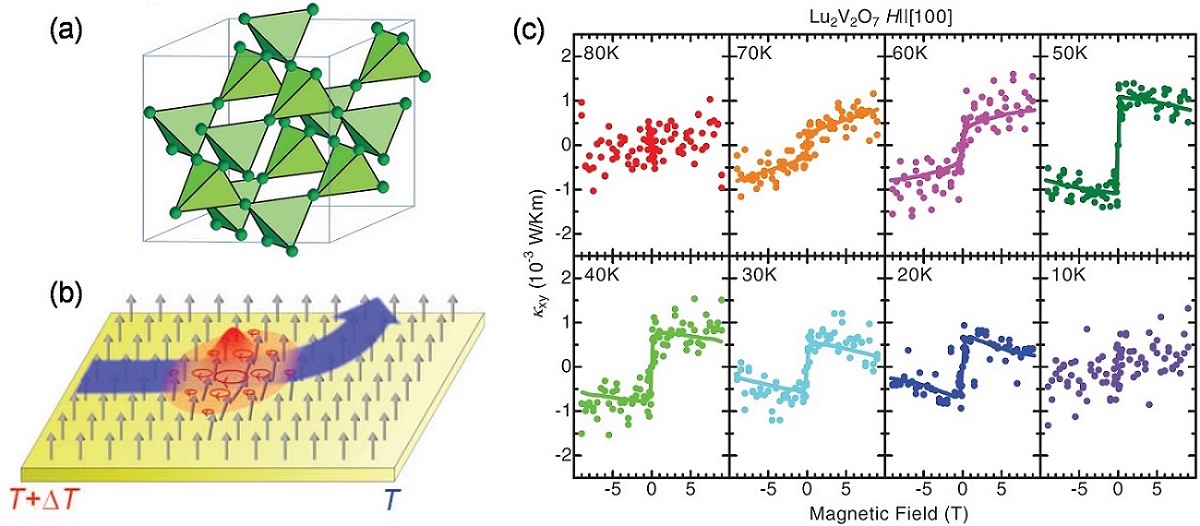}
\par\end{centering}
\caption{(a) The crystal structure of pyrochlore ferromagnet Lu$_{2}$V$_{2}$O$_{7}$, which is composed of corner-sharing tetrahedra. (b) The magnon Hall effect: the longitudinal temperature gradient leads to the transverse thermal magnon current. (c) Magnetic field variation of the thermal Hall conductivity of Lu$_{2}$V$_{2}$O$_{7}$ at various temperatures. Source: The figures are taken from Ref. \cite{OnoseS2010}.}
\label{Figure5}
\end{figure}

In 2010, Katsura \emph{et al}. \cite{KatsuraPRL2010} predicted theoretically that the intrinsic thermal Hall effect of magnons can be realized in magnets with a particular lattice structure such as kagome, which offers a promising proposal to detect the MHE via a thermal transport measurement. Later, in the same year, the MHE was observed  experimentally  by Onose \emph{et al}. \cite{OnoseS2010} in the insulating ferromagnet Lu$_{2}$V$_{2}$O$_{7}$ of pyrochlore lattice structures. The pyrochlore structure can be viewed as the stacking of alternating kagome and triangular lattices, as shown in Fig. \ref{Figure5}(a). When a temperature gradient is applied longitudinally, a transverse heat current was observed. Figure \ref{Figure5}(c) presents the measurements of transverse thermal Hall conductivity. From a broad point of view, electrons, phonons and magnons can all generate heat current. However, on the one hand, from Fig. \ref{Figure5}(c), one can see that the thermal Hall conductivity steeply increases and saturates in the low-magnetic field region, which can not be explained by either normal Hall effect (the conductivity is proportional to the magnetic field strength) or the anomalous Hall effect due to the spontaneous magnetization. On the other hand, the emergence of the decrease of the thermal Hall conductivity in the high-field region cannot be explained in terms of the phonon mechanism \cite{StrohmPRL2005,ShengPRL2006,KaganPRL2008}. Therefore, it is concluded that the transverse heat current can only be explained by the MHE. Figure \ref{Figure5}(b) plots the schematic diagram of MHE. In the model, the MHE comes from the nonzero DM interaction (induced by the spin-orbit coupling) that breaks the inversion symmetry.    

\begin{figure}[ptbh]
\begin{centering}
\includegraphics[width=0.68\textwidth]{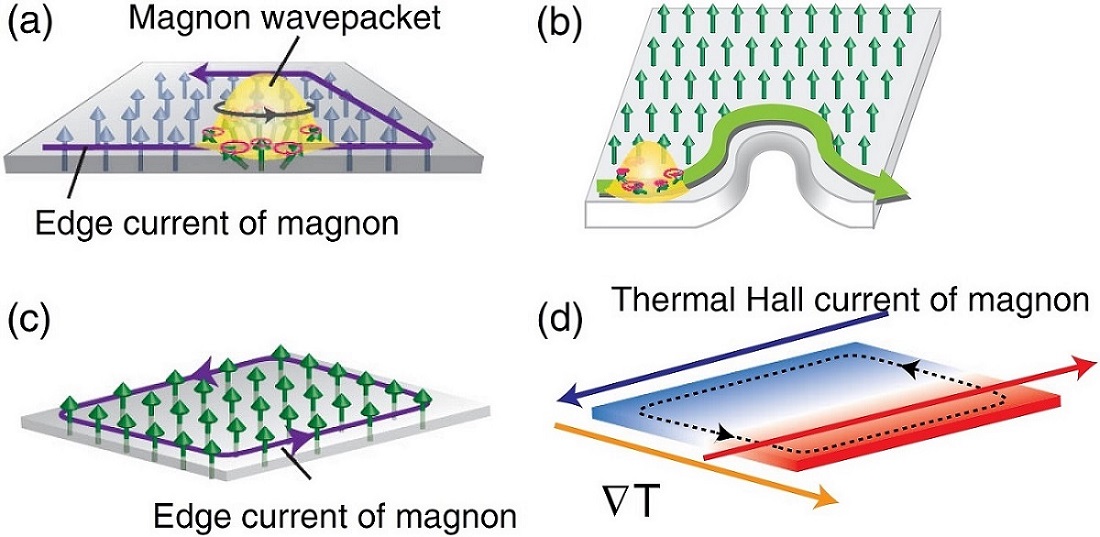}
\par\end{centering}
\caption{(a) The self-rotation of a magnon wave packet with a magnon edge current. (b) The magnon near the boundary proceeds along the boundary. (c) Magnon edge current in equilibrium. (d) Under the temperature gradient, the amount of the transverse heat current are not balanced between the two edges and a finite thermal Hall current emerges. Source: The figures are taken from Ref. \cite{MatsumotoPRL2011}.}
\label{Figure6}
\end{figure}

After the discovery of MHE, people try to understand the origin of the transverse thermal magnon current. Matsumoto \emph{et al}. \cite{MatsumotoPRL2011,MatsumotoPRB2011} demonstrated that a magnon wave packet [see Figs. \ref{Figure6}(a) and \ref{Figure6}(b)] subjected to a temperature gradient acquires an anomalous velocity perpendicular to the gradient, which is associated to the magnon edge currents. The relation between transverse thermall Hall conductivity $\kappa^{xy}$ and Berry curvature $\bf{\Omega}_{\emph{n}}(\bf{k})=-\emph{i}\langle\frac{\partial \emph{u}_{\emph{n}}}{\partial \bf{k}}|\times|\frac{\partial \emph{u}_{\emph{n}}}{\partial \bf{k}}\rangle$ is as follows:
\begin{equation}\label{Eq1}
\kappa^{xy}=-\frac{k_{B}^{2}T}{\hbar V}\sum_{n,\bf{k}}c_{2}(\rho_{n})\bf{\Omega}_{\emph{n,z}}(\bf{k}).
\end{equation}
From Eq. \eqref{Eq1}, one can clearly see that $\kappa^{xy}$ comes from the Berry curvature in momentum space. When the energy bands are close to each other (near the band crossing), the value of $\kappa^{xy}$ reaches the maximum. The MHE can be understood as follows: when the system is in equilibrium [see Fig. \ref{Figure6}(c)], the edge magnon currents exist due to the confining potential, and they circulate along the boundary. The amount of currents are equal at two edges, leading to a vanishing thermal current through the magnet. However, when the temperature gradient is applied [see Fig. \ref{Figure6}(d)], the magnons will flow from the high temperature region to the low temperature region, which breaks the balance of the heat current from the two opposite edges, leading to a finite thermal Hall current. Furthermore, Zhang \emph{et al}. \cite{ZhangPRB2013} showed that these edge magnon currents are actually SW chiral edge states resulting from the nontrivial topology of magnon bands. It is also demonstrated that the one-way chiral edge transport is topologically immune from defects and disorders. In a word, the robust MHE originates from the nontrivial band structure of magnons.   

\begin{figure}[ptbh]
\begin{centering}
\includegraphics[width=0.85\textwidth]{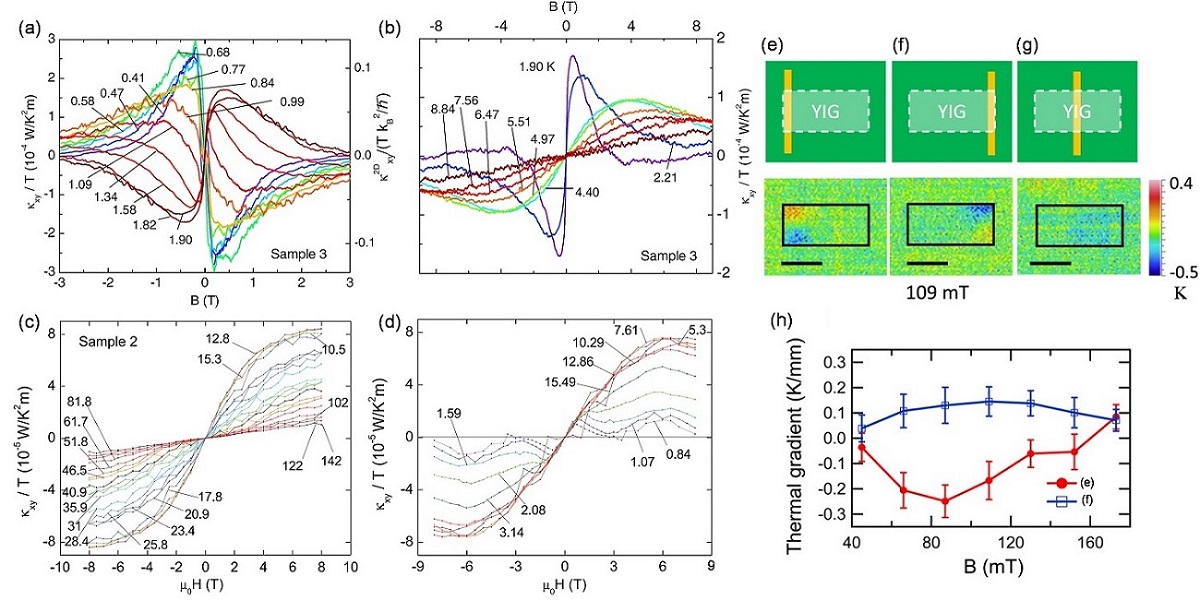}
\par\end{centering}
\caption{(a) (b) The thermal Hall conductivity measured in the kagome magnet Cu(1,3-bdc) at different temperatures. (c) (d) The thermal Hall conductivity $\kappa_{xy}/T$ versus magnetic field $H$ in a frustrated pryocholore quantum magnet Tb$_{2}$Ti$_{2}$O$_{7}$. (e)-(g) Upper figures show the configuration of YIG and the CPW, the corresponding spatial distribution of measured temperature is shown in the lower figures. (h) Thermal gradient versus magnetic field. Red points and blue squares correspond to (e) and (f). Source: The figures are taken from Refs. \cite{HirschbergerPRL2015,HirschbergerS2015,TanabePSS2016}.}
\label{Figure7}
\end{figure}

Since the discovery of MHE in pyrochlore ferromagnetic insulator \cite{OnoseS2010}, the same effect has also been observed in other magnetic materials. Hirschberger \emph{et al}. \cite{HirschbergerPRL2015} report the observation of a large thermal Hall conductivity $\kappa_{xy}$ in the kagome magnet Cu(1,3-bdc), with the main results shown in Figs. \ref{Figure7}(a) and \ref{Figure7}(b). Surprisingly, the observed $\kappa_{xy}$ undergoes a remarkable sign reversal by changing the temperature or magnetic field, which is explained by the sign change of the Chern flux between magnon bands. Besides, Hirschberger \emph{et al}. \cite{HirschbergerS2015} also report the MHE in a frustrated pryocholore quantum magnet Tb$_{2}$Ti$_{2}$O$_{7}$. The corresponding measurements of $\kappa_{xy}$ are plotted in Figs. \ref{Figure7}(c) and \ref{Figure7}(d). One can see that from 140 to 50 K, $\kappa_{xy}/T$ is $H-$linear. Below 45 K, it develops a pronounced curvature at large $H$, reaching its largest value near 12 K, which is the obvious signal for the thermal magnon current generating the transversal Hall conductivity. Furthermore, Tanabe \emph{et al}. \cite{TanabePSS2016} observe the magnon Hall-like effect for sample-edge scattering in unsaturated YIG. Figures \ref{Figure7}(e)-\ref{Figure7}(g) show the measurements of the temperature distribution with coplanar waveguide (CPW) at different positions. Thermal gradient of about 0.3 K mm$^{-1}$ was observed along the YIG [see Fig. \ref{Figure7}(e)], and the opposite sign of the thermal gradient is also detected with the CPW at the opposite side [see Fig. \ref{Figure7}(f)]. However, when the CPW is placed under the center of the YIG, no thermal gradient is observed, as shown in Fig. \ref{Figure7}(g). These results strongly indicate that the observed thermal gradient in Figs. \ref{Figure7}(e) and \ref{Figure7}(f) are attributed to the magnons at sample edges.
\begin{figure}[ptbh]
\begin{centering}
\includegraphics[width=0.90\textwidth]{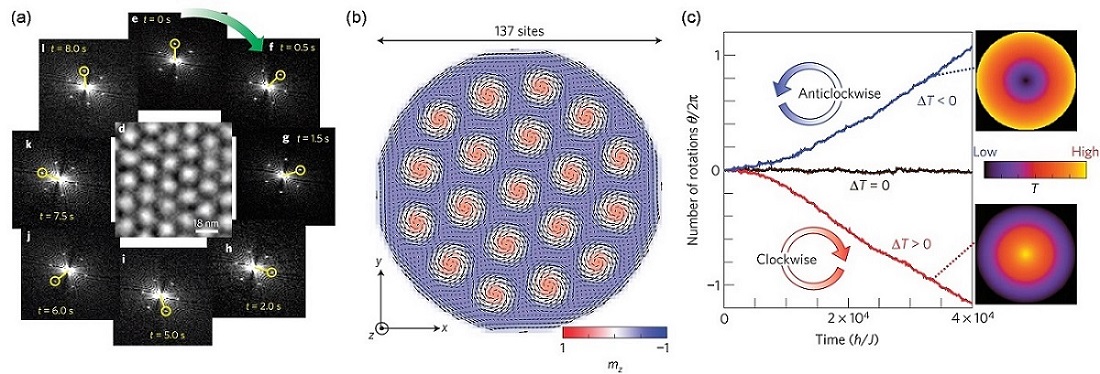}
\par\end{centering}
\caption{(a) Time profiles of Fourier transforms of temporally changing magnetic structures in the skyrmion crystal observed in MnSi by the Lorentz TEM, which exhibits a clockwise rotation. (b) The magnetic configuration of the skyrmion microcrystal confined in a circular-shaped disk at $T = 0$, with the distribution of the z-component of the magnetization being shown in the colour map. (c) Simulated number of rotations $\theta/2\pi$ as a function of time in the cases of thermal equilibrium ($\Delta T=0$), positive $T$ gradient ($\Delta T>0$), and negative $T$ gradient ($\Delta T<0$). Source: The figures are taken from Ref. \cite{MochizukiNM2014}.}
\label{Figure8}
\end{figure}

 The MHE discussed above is based on collinear ferromagnet. While Hoogdalem \emph{et al}. \cite{HoogdalemPRB2013} demonstrated theoretically that noncollinear magnetic texture (skyrmion for instance) can generate a fictitious magnetic field, which can also lead to magnon thermal Hall effect. This Hall effect is solely due to the nonzero topological charge of magnetic texture, which is therefore called topological magnon Hall effect (TMHE). Subsequently, Mochizuki \emph{et al}. \cite{MochizukiNM2014} indirectly confirmed experimentally the existence of TMHE. By using Lorentz transmission electron microscope (TEM), they observed micrometre-sized crystals of skyrmions in thin films of Cu$_{2}$OSeO$_{3}$ and MnSi exhibiting a unidirectional rotation motion, as shown in Fig. \ref{Figure8}(a). This rotational motion can be explained below: At first, the thermal gradient was generated under the electron-beam irradiation in the Lorentz TEM experiment, then the magnon current induced by the temperature gradient is deflected by the emergent magnetic field of skyrmion lattice (TMHE), which in turn gives rise to the rotation of skyrmions through the spin-transfer torque (STT). In addition, if the sign of the temperature gradient is reversed, the direction of skyrmion rotation will reverse, too, see Fig. \ref{Figure8}(c). The skyrmion-induced versions of the MHE are also studied by means of atomistic spin dynamics \cite{MookPRB2017}, in which, based on spin spiral and skyrmion lattice system, the authors predict a magnon Hall angle as large as 60$\%$.

The discovery of MHE opens the door to the study of the topological properties of magnons. The topological magnons have great potential application prospect for designing robust and flexible spintronic devices. Over the past decade, a lot of literatures have been devoted to the topological properties of magnons, including the topological magnon insulators and semimetals, which will be reviewed in next sections.     
  
\subsection{Topological magnon insulators}\label{section2.2}
Generally speaking, there are two typical systems that can support topological magnons, one is the collinear ferromagnet, while the other is noncollinear magnetic texture. Below, we will give a brief introduction about the topological magnons in these systems by several specific examples.    

\subsubsection{\text{Collinear ferromagnet}}\label{section2.2.1}
In 2014, Mook \emph{et al}. \cite{MookPRB2014} reported a detailed theoretical investigation on the nontrivial topology of magnon in kagome lattice based on a quantum-mechanical Heisenberg model \cite{HeisenbergZP1928} with the Hamiltonian $\mathcal{H}=\mathcal{H}_{\text{H}}+\mathcal{H}_{\text{DM}}$, with the Heisenberg exchange term $\mathcal{H}_{\text{H}}=-\sum_{n\neq m}J^{n}_{m}\hat{\bf{s}}_{m}\cdot \hat{\bf{s}}_{n}$, where two spin operators $\hat{\bf{s}}_{m}$ and $\hat{\bf{s}}_{n}$ at sites $n$ and $m$ are coupled by symmetric exchange parameters $J^{n}_{m}=J^{m}_{n}$, and the antisymmetric DM interaction term $\mathcal{H}_{\text{DM}}=\sum_{n\neq m}\vec{D}^{n}_{m}(\hat{\bf{s}}_{m}\times \hat{\bf{s}}_{n})$, where $\vec{D}^{n}_{m}$ is the DM vector between sites $m$ and $n$ ($\vec{D}^{n}_{m}=-\vec{D}^{m}_{n}$). For a given set of parameters $\{J^{n}_{m},D^{n}_{m}\}$, by solving the Hamiltonian $\mathcal{H}$, one can obtain the eigenvectors $|i(\bf{k})\rangle$ and eigenvalues $\varepsilon_{i}(\bf{k})$ [wave vector $\mathbf{k}=(k_{x},k_{y})$, band index $i$]. For each band $j$, the Berry curvature reads 
\begin{equation}\label{Eq2}
\mathbf{\Omega}_{j}(\mathbf{k})= i\sum_{i\neq j}\frac{\langle i(\mathbf{k})|\nabla_{\mathbf{k}}\mathcal{H}(\mathbf{k})|j(\mathbf{k})\rangle\times\langle j(\mathbf{k})|\nabla_{\mathbf{k}}\mathcal{H}(\mathbf{k})|i(\mathbf{k})\rangle}{[\varepsilon_{i}(\mathbf{k})-\varepsilon_{j}(\mathbf{k})]^{2}},
\end{equation}
and the topological invariant Chern number is given by
\begin{equation}\label{Eq3}
C_{j}=\frac{1}{2\pi}\int_{BZ}\Omega^{z}_{j}(\mathbf{k})d\mathbf{k}.
\end{equation}
Chern number is a physical quantity of particular importance for the topologically nontrivial edge modes by determining both their propagation direction and their number. There is a "bulk-boundary correspondence": the bulk property (Chern number) dictates surface/edge properties (edge magnons). The sum of Chern numbers up to the $i$th band $\nu_{i}=\sum_{j\leq i}C_{j}$ is the "winding number" of the edge states in band gap $i$. $|\nu_{i}|$ is the number of topologically nontrivial edge states in the $i$th band gap and sgn$(\nu_{i})$ determines their propagation direction.

\begin{figure}[ptbh]
\begin{centering}
\includegraphics[width=0.91\textwidth]{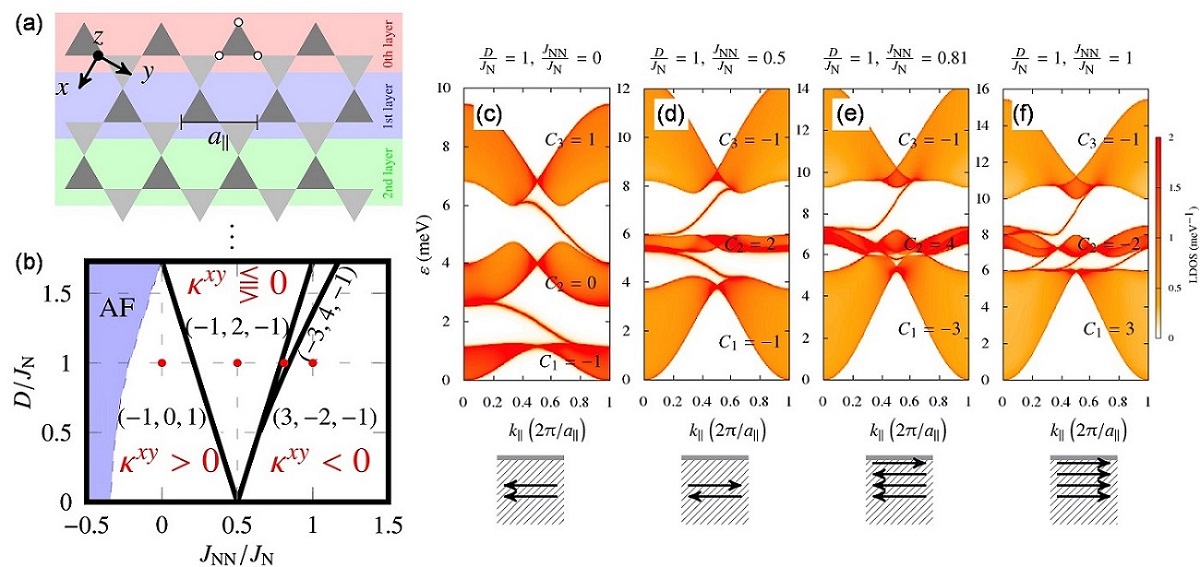}
\par\end{centering}
\caption{(a) Edge of the kagome lattice. The semi--infinite lattice is divided into the thinnest principal layers, where white dots indicate the basis consisting of three sites, and $a_{\parallel}$ is the lattice constant. (b) Topological phase diagram of the kagome lattice with regions characterized by sets $(C_{1},C_{2},C_{3})$ of Chern numbers. (c)-(f) The band structures for different topologically nontrivial phases as marked with red dots in (b). The topologically nontrivial edge magnon modes and their propagation direction are sketched at the bottom. Source: The figures are taken from Ref. \cite{MookPRB2014}.}
\label{Figure9}
\end{figure}

The ferromagnetic kagome lattice allows four topologically different phases by tuning parameters $J_{NN}/J_{N}$ and $D/J_{N}$ \cite{MookPRB2014_2}. $J_{N}$ and $J_{NN}$ represent the Heisenberg exchange constant between nearest and next-nearest sites, respectively, while the DM parameter ($D$) accounts only for the nearest-neighbor interaction. Figure \ref{Figure9}(a) shows the semi-infinite kagome lattice, and calculated topological phase diagram is presented in Fig. \ref{Figure9}(b), where the sign of the transverse thermal conductivity $\kappa^{xy}$ of the MHE is also indicated. 

Figures \ref{Figure9}(c)-\ref{Figure9}(f) plot the band structures for different topological nontrivial phases. One can illuminate the above rule (the relationship between $\nu$ and topological edge states) by considering the topological phase (3,-2,-1) [see Fig. \ref{Figure9}(d)] as an example. There are three nontrivial edge states with positive group velocity in the lowest gap because $\nu_{1}=C_{1}=3$. While in the second band gap there is only a single edge state with positive group velocity, which is in accordance with $\nu_{2}=C_{1}+C_{2}=1$. Because the sum over all Chern numbers must be zero, there are never topological nontrivial edge states above the uppermost band. For other cases [Figs. \ref{Figure9}(c), \ref{Figure9}(e), and \ref{Figure9}(f)], similar analysis can be done as well. It is worth mentioning that the phases (-1,2,1) and (-3,4,-1) can support edge modes for both propagation directions, which leads to the change of sign in $\kappa^{xy}$ when the temperature varies. At low temperatures, edge states in the first band gap are more occupied than edge states in the second band gap. Thus the heat transport is dominated by the former edge modes. However, with the increasing of temperature, the edge states in the second band gap become increasingly populated. When the temperature is high enough, the heat current is mainly mediated by these magnons, therefore, the sign of $\kappa^{xy}$ reverses.      
\begin{figure}[ptbh]
\begin{centering}
\includegraphics[width=0.80\textwidth]{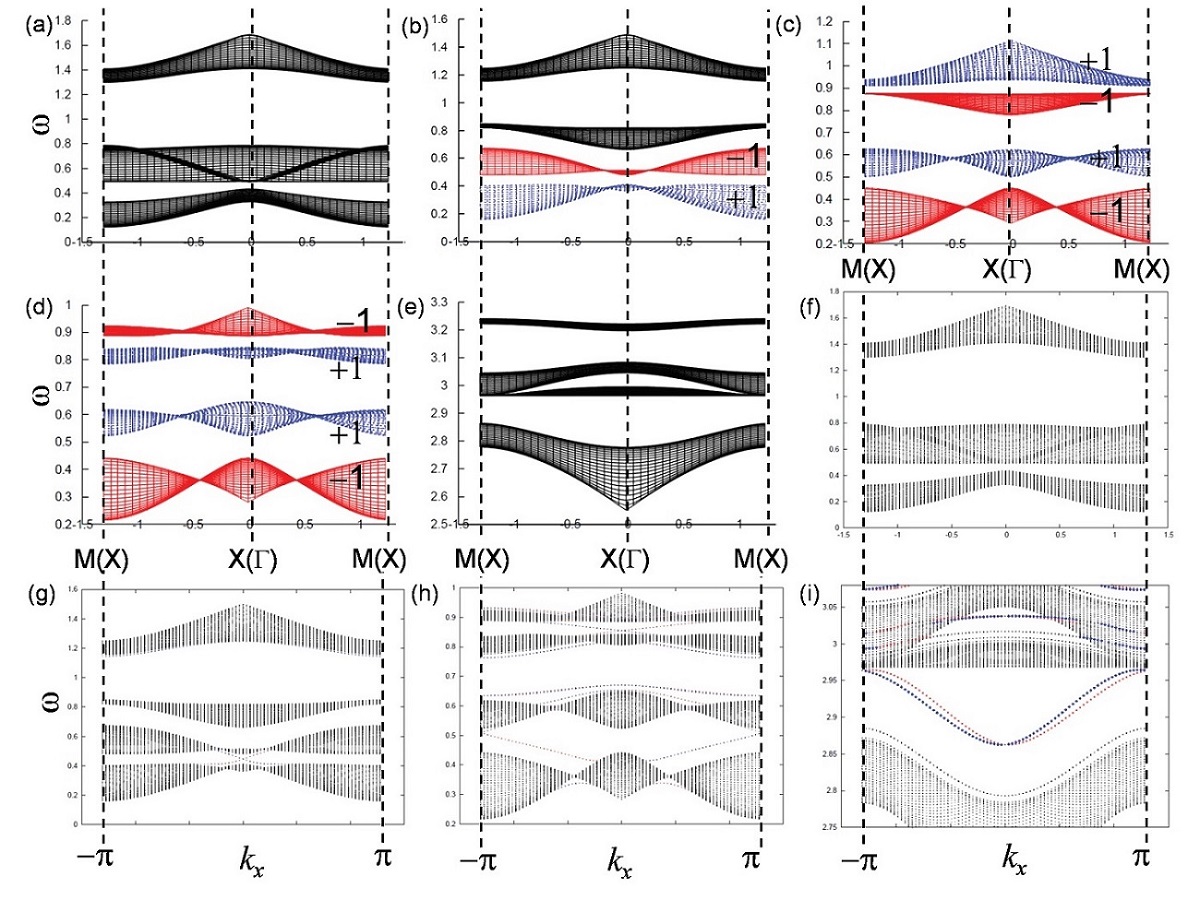}
\par\end{centering}
\caption{The band structures of infinite system for different field strength: (a) $H=0.0$, (b) $H=0.47H_{c}$, (c) $H=0.76H_{c}$, (d) $H=0.82H_{c}$, and (e) $H=2.35H_{c}$. The band structures of semi-infinite system with different field strength: (f) $H=0.0$, (g) $H=0.47H_{c}$, (h) $H=0.82H_{c}$, and (i) $H=0.35H_{c}$. Source: The figures are taken from Ref. \cite{ShindouPRB2013_2}.}
\label{Figure10}
\end{figure}

\begin{figure}[ptbh]
\begin{centering}
\includegraphics[width=0.90\textwidth]{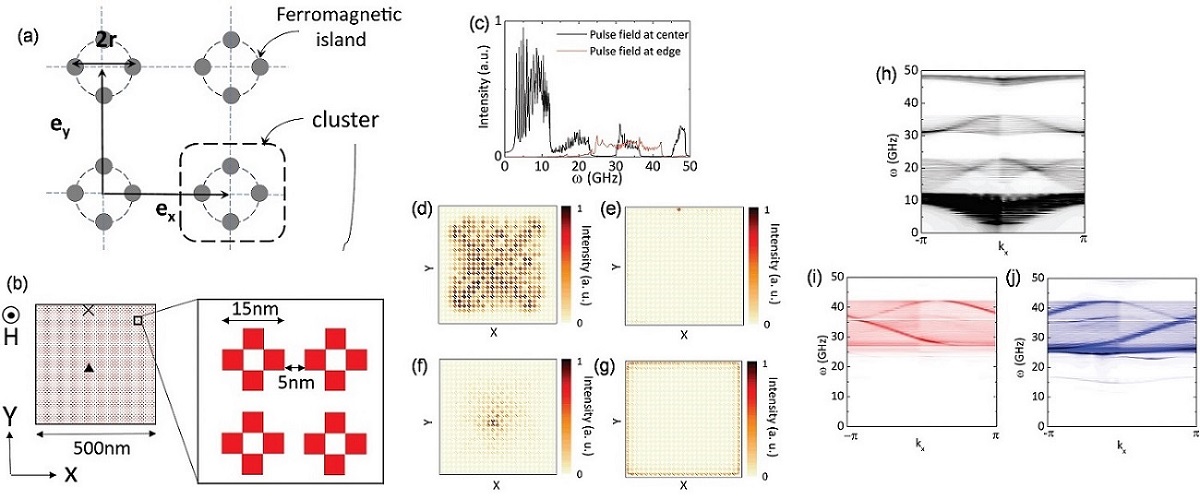}
\par\end{centering}
\caption{(a) Periodic array of ferromagnetic islands decorated square-lattice model. (b) Schematic view of a simulated system that comprises ferromagnetic nanograin. The black triangle and cross represent the positions of the pulse field. (c) Fourier power spectra for excitation at different positions. (d)-(g) Spatial distribution of the intensity, when the pulse field at center (d), (f) and the edge (e) (g); (d),(e) $\omega=10$ GHz, and (f),(g) $\omega=29$ GHz. (h) Dispersion relation with pulse field at the center. (i), (j) Dispersion relation with pulse at the edge. The Fourier transformation is taken only over the upper side (Y > L/2) for (i) and over the lower side (Y < L/2) for (j). Source: The figures are taken from Ref. \cite{ShindouPRB2013_2}.}
\label{Figure11}
\end{figure}

As mentioned above, the nontrivial topology of magnon in the kagome lattice is brought about by the strong spin-orbit coupling which manifests the DM interaction. Meanwhile, Shindou \emph{et al}. \cite{ShindouPRB2013_2} demonstrated theoretically that the magnetic dipolar interaction can also endow spin wave volume modes with nonzeros Chern number, and the propagation direction of edge states is tunable by external magnetic fields. Figure \ref{Figure11}(a) plots the model of periodic array of ferromagnetic islands decorated square-lattice. The energy of the system only includes the magnetostatic energy and the Zeeman energy. By using simple tight-binding descriptions, one can obtain the magnon Hamiltonian and the band structures under different field strengths, with the results shown in Fig. \ref{Figure10}. Here the direction of the magnetic field is perpendicular to the plane. In the absence of the magnetic field [see Fig. \ref{Figure10}(a)], the out-of-plane magnetization vanishes, so that the spin wave Hamiltonian respects both time-reversal and mirror symmetries. The Chern numbers for all bands are zero, and no chiral spin wave edge state is observed [Fig. \ref{Figure10}(f)]. With the increasing of the field strength, there appear twice band touchings of the lowest band and second-lowest band at the $\Gamma$ point when $H=0.24H_{c}$ and at $X$ points when $H=0.67H_{c}$. Here $H_{c}$ is the saturation field where all the spins become fully polarized along the field. As a result, the Chern numbers for the lowest and second lowest bands become $+1$ and $-1$, respectively for $0.24<H/H_{c}<0.67$ [Fig. \ref{Figure10}(b)], and $-1$ and $+1$, respectively for $0.67<H/H_{c}$ [Figs. \ref{Figure10}(c) and \ref{Figure10}(d)]. Correspondingly, there appears a chiral spin-wave edge mode propagating in the clockwise direction for $0.24<H/H_{c}<0.67$ [Fig. \ref{Figure10}(g)], and in the counterclockwise direction for $0.67<H/H_{c}$ [Fig. \ref{Figure10}(h)]. Besides, there appears three times band touchings of the third-lowest and highest bands at the $M$ point when $H=0.71H_{c}$, at $X$ points when $H=0.79H_{c}$, and at $\Gamma$ point when $H=0.85H_{c}$. Correspondingly, the Chern numbers for the third-lowest and the highest bands become $-1$ and $+1$ (counterclockwise edge mode) for $0.71<H/H_{c}<0.79$ [Fig. \ref{Figure10}(c)], $+1$ and $-1$ (clockwise edge mode) for $0.79<H/H_{c}<0.85$ [Figs. \ref{Figure10}(d) and \ref{Figure10}(h)].
In the limit of strong field, the system becomes effectively time-reversal symmetric, where the Chern numbers for all bands reduce to zero and no chiral spin wave edge modes is supported [Figs. \ref{Figure10}(e) and \ref{Figure10}(i)]. Yet there still exist spin wave edge modes, which have parabolic dispersions and thus support bidirectional propagations [Fig. \ref{Figure10}(i)]. 

To confirm the existence of the proposed chiral spin wave edge mode, the authors performed micromagnetic simulation for the square-lattice model, as shown in Fig. \ref{Figure11}(b). The spins are coupled via magnetic dipole-dipole interaction and no short-range exchange interaction is considered. The magnetization becomes fully polarized along $z$ direction under $H=1.02H_{c}$. The frequency power spectra for pulse at center and at edge are shown in Fig. \ref{Figure11}(c). One can clearly identify the edge state and bulk state. Spatial distributon of spin-wave excitation for different modes are presented in Figs. \ref{Figure11}(d)-\ref{Figure11}(g). When $\omega=10$ GHz, the system is in bulk state [see Figs. \ref{Figure11}(d) and \ref{Figure11}(e)], while if $\omega=29$ GHz, the system supports the edge state [see Figs. \ref{Figure11}(f) and \ref{Figure11}(g)]. The propagation of the chiral spin-wave edge mode is unidirectional, which can be clarified by the dispersion relation. Figure \ref{Figure11}(h) plots the dispersion relation when the pulse field is at the center, and no edge mode is observed. However, when the pulse field locates at the edge, by taking the Fourier transformation only over the upper (or lower) side of the sample, as shown in Fig. \ref{Figure11}(i) [Fig. \ref{Figure11}(j)], one can clearly see counterclockwise propagating chiral dispersions, which are consistent with the results in Fig. \ref{Figure10}.

\subsubsection{Noncollinear magnetic texture}\label{section2.2.2}
\begin{figure}[ptbh]
\begin{centering}
\includegraphics[width=1\textwidth]{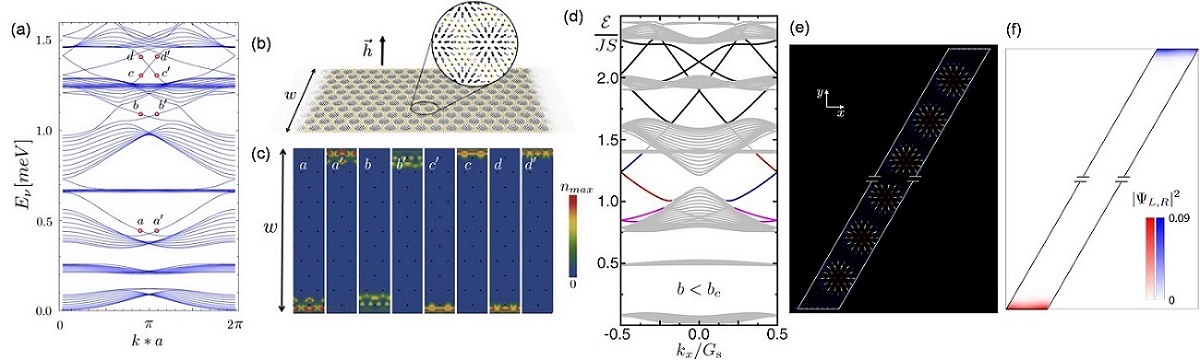}
\par\end{centering}
\caption{(a) Spin wave bands for a one-dimensional strip. (b) Strip geometry. (c) Magnon occupation for the edge modes $a, a^{'}, b, b^{'}, c, c^{'}, d$, and $d^{'}$ depicted in (a). The black dots correspond to the centers of the underlying skyrmions. (d) One-dimensional magnon spectra of an infinite strip. (e) Magnetic unit cell of the ferromagnetic skyrmion crystal in the strip geometry. (f) Probability density $|\Psi_{L,R}|^{2}$ of the left-moving (red) and right-moving (blue) magnonic edge states. Source: The figures are taken from Refs. \cite{MolinaNJP2016,DiazPRR2020}.}
\label{Figure12}
\end{figure}

As introduced in Section \ref{section2.1}, the magnetic texture can induce the topological magnon Hall effect, which indicates that there are topologically protected magnon edge states in these systems. In 2016, Roldán-Molina \emph{et al.} \cite{MolinaNJP2016} reported the topological SWs in the atomic-scale magnetic skyrmion crystal, with the schematic diagram shown in Fig. \ref{Figure12}(b). The Hamiltonian of the system contains a uniaxial anisotropy term, a nearest-neighbor ferromagnetic exchange coupling, the DM interaction, and the Zeeman energy. By solving the eigen equations numerically, one can obtain the SW band structures for a one-dimensional skyrmion crystal strip, as shown in Fig. \ref{Figure12}(a). It can be clearly seen that there are several bands allowing spin wave edge states. Figure \ref{Figure12}(c) plots the magnon occupation for the edge modes as marked in Fig. \ref{Figure12}(a), from which the localization properties can be clearly identified. Similar results are obtained by Díaz \emph{et al.} \cite{DiazPRR2020}. Figure \ref{Figure12}(d) plots the magnon band structure for one-dimensional skyrmion crystal strip, where the red and blue lines represent the bands for spin wave edge states. The magnetic unit cell of the ferromagnetic skyrmion crystal and probability density of magnonic edge states are shown in Figs. \ref{Figure12}(e) and \ref{Figure12}(f), respectively, from which the topological magnon edge states can be observed.  

\begin{figure}[ptbh]
\begin{centering}
\includegraphics[width=0.85\textwidth]{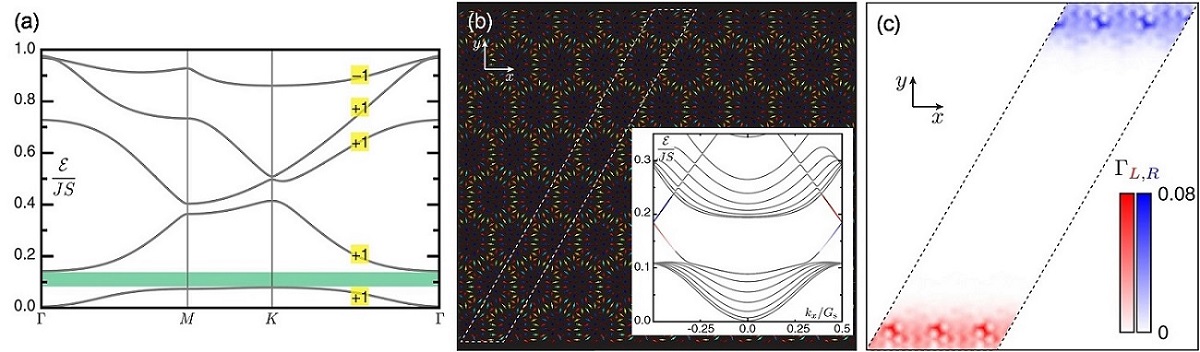}
\par\end{centering}
\caption{(a) Bulk magnon band structure of the antiferromagnetic skyrmion crystals along the loop $\Gamma-M-K-\Gamma$. Bands are labeled by their Chern numbers (yellow squares). The green rectangle highlights the first bulk magnon gap. (b) The one-dimensional antiferromagnetic skyrmion crystal strip and the corresponding spin wave band structure, where the white dashed-line loop denote the magnetic unit cell. (c) Magnonic contribution of the wave function of the edge states $\Gamma_{L,R}$ over an area spanning 3 times the magnetic unit cell. Source: The figures are taken from Ref. \cite{DiazPRL2019}.}
\label{Figure13}
\end{figure}

Furthermore, Díaz \emph{et al.} \cite{DiazPRL2019} show that the topological magnon also exists in antiferromagnetic skyrmion crystals. Figure \ref{Figure13}(a) plots the bulk band structure of antiferromagnetic skyrmion crystals along the high symmetry points of the Brillouin zone (BZ). The bulk magnon gap can be clearly identified, as marked by green rectangle. If one considers a strip of infinite length along the $x$ axis with edges located at the top and bottom of the lattice, the bands for spin wave edge states will emerge, as shown in Fig. \ref{Figure13}(b). Magnonic edge states are plotted in Fig. \ref{Figure13}(c).       

\subsection{Topological magnon semimetals}\label{section2.3}

\subsubsection{Dirac magnons}\label{section2.3.1}
In 2D systems, the Dirac materials that support Dirac-like spectra of excitations are attracting growing attention since the discovery of graphene \cite{NetoRMP2009,WehlingAP2014}. Although the Dirac dispersion was observed for fermionic quasiparticles at first, the concept has been extended to various bosonic systems, such as photonic crystals \cite{LuNP2014,OzawaRMP2019,HaldanePRL2008,MaPRL2015}, plasmonic system \cite{WeickPRL2013}, acoustic metamaterials \cite{JinNL2018}, superconducting grains \cite{BanerjeePRB2016}, and spintronics \cite{FranssonPRB2016,OwerreJPCM2016,BoykoPRB2018,PershogubaPRX2018,ChenPRX2018,YuanPRX2020}. 

\begin{figure}[ptbh]
\begin{centering}
\includegraphics[width=0.85\textwidth]{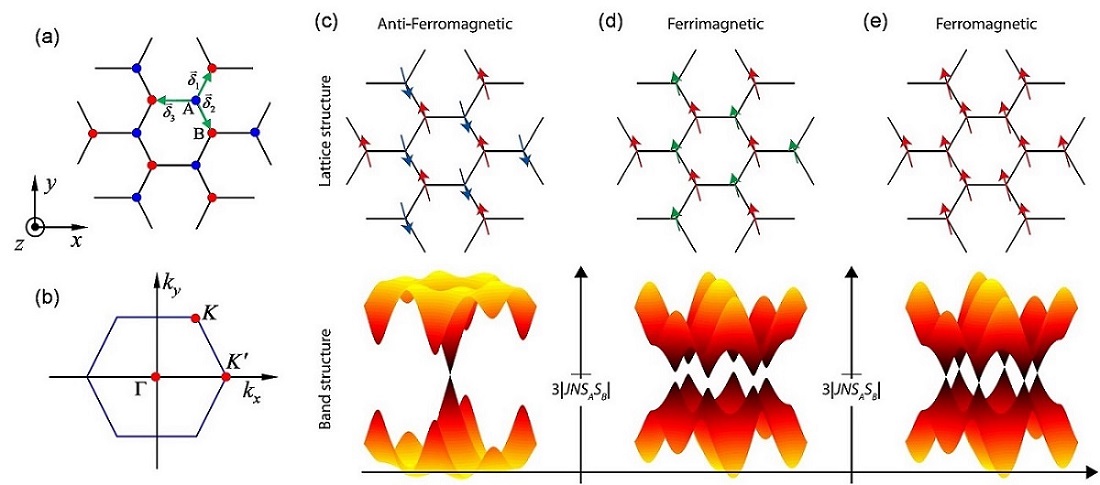}
\par\end{centering}
\caption{(a) Schematic plot of spins on a honeycomb lattice where $A$ and $B$ indicate the sublattices. (b) First BZ with high-symmetry points. The magnetic structures and calculated magnon band structures on the honeycomb lattice with (c) antiferromagnetic, (d) ferrimagnetic, and (e) ferromagnetic states. Source: The figures are taken from Ref. \cite{FranssonPRB2016}.}
\label{Figure14}
\end{figure}

For magnetic system, the simplest two-band model that exhibits Dirac points is the Heisenberg ferromagnet or antiferromagnet on the honeycomb lattice \cite{FranssonPRB2016}. The Hamiltonian can be expressed as: $\mathcal{H}=-\sum_{\langle i j \rangle}J_{ij}\mathbf{S}_{i}^{(A)}\cdot\mathbf{S}_{j}^{(B)}$, where the summation runs over nearest neighbors, $\mathbf{S}_{i}^{(A)}$ and $\mathbf{S}_{j}^{(B)}$ are the spins for two different sublattices, and $J_{ij}$ is the exchange constant. Assuming uniform ferromagnetic interaction, i.e., $J_{ij}=J>0$, by applying the Holstein-Primakoff transformation \cite{HolsteinPR1940}, the effective quadratic magnon model can be written as
\begin{equation}\label{Eq4}
\mathcal{H}_{\text{FM}}=\sum_{i}(\varepsilon_{A}a_{i}^{\dagger}a_{i}+\varepsilon_{B}b_{i}^{\dagger}b_{i})-J\sqrt{S_{A}S_{B}}\sum_{\langle ij\rangle}(a_{i}^{\dagger}b_{j}+h.c.)-3JNS_{A}S_{B}.
\end{equation}
In reciprocal space, letting $a_{i}=\sum_{\mathbf{k}}a_{\mathbf{k}}e^{i\mathbf{k}\cdot\mathbf{r}_{i}}/\sqrt{N}$ and $b_{j}=\sum_{\mathbf{k}}b_{\mathbf{k}}e^{i\mathbf{k}\cdot\mathbf{r}_{j}}/\sqrt{N}$, one can obtain 
\begin{equation}\label{Eq5}
\mathcal{H}_{\text{FM}}=\sum_{\mathbf{k}}{\varepsilon_{A}a_{\mathbf{k}}^{\dagger}a_{\mathbf{k}}+\varepsilon_{B}b_{\mathbf{k}}^{\dagger}b_{\mathbf{k}}+[\phi(\mathbf{k})a_{\mathbf{k}}^{\dagger}b_{\mathbf{k}}+h.c.]}-3JNS_{A}S_{B},
\end{equation}  
where the structure factor $\phi(\mathbf{k})=-J\sqrt{S_{A}S_{B}}\sum_{i}\text{exp} (i\mathbf{k}\cdot\vec{\delta}_{i})$ ($i=1,2,3$) is given in terms of the nearest-neighbor vector $\vec{\delta}_{i}$ [see Fig. \ref{Figure14}(a)], and $\mathbf{k}$ is wave vector. The eigenenergies can be derived
\begin{equation}\label{Eq6}
E_{\pm}(\mathbf{k})=[\varepsilon_{A}+\varepsilon_{B}\pm\Omega(\mathbf{k})]/2,
\end{equation}
where $\Omega^{2}(\mathbf{k})=\Delta^{2}+4|\phi(\mathbf{k})|^{2}$ with $\Delta=\varepsilon_{A}-\varepsilon_{B}=-3J(S_{A}-S_{B})$.

For ferromagnetic structures, $S_{A(B)}=S$ and $\Delta=0$, the configuration of ferromagnetic state and magnon band structure are shown in Fig. \ref{Figure14}(e). Around the band degeneracy points $\mathbf{K}=(2\pi/3a,2\sqrt{3}\pi/3a)$ and $\mathbf{K}^{'}=(4\pi/3a,0)$ [as marked by red dots in Fig. \ref{Figure14}(b)], the dispersion relation is linear. In the ferrimagnetic case, $S_{A}\neq S_{B}$, thus $\Delta\neq 0$, which leading to a gap opening at $\mathbf{K}$ and $\mathbf{K}^{'}$ with gap size $3J|S_{A}-S_{B}|$, as shown in Fig. \ref{Figure14}(d). However, if the system is an antiferromagnetic honeycomb lattice, the energy dispersion has degeneracy only at the $\Gamma$ point, around which the dispersion relation is also linear; see Fig. \ref{Figure14}(c). 

\begin{figure}[ptbh]
\begin{centering}
\includegraphics[width=0.90\textwidth]{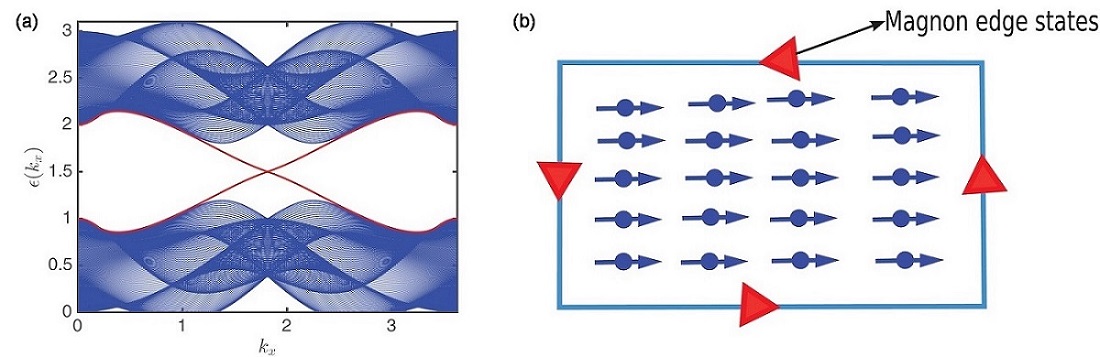}
\par\end{centering}
\caption{(a) The magnon energy band for a one-dimensional strip on the ferromagnetic honeycomb lattice with the next-nearest neighbour DM interaction. (b) Schematics of magnon edge states in topological magnon insulator material. Source: The figures are taken from Refs. \cite{OwerreJPCM2016}.}
\label{Figure15}
\end{figure}
For Dirac materials, if the inversion symmetry is broken, a gap will open at the Dirac points, leading to a TI. Topological magnon insulator can be achieved by using similar method. For the honeycomb ferromagnets, if the Hamiltonian only contains nearest neighbors (NN) exchange interaction, the magnon band structure is gapless, even if a next-nearest neighbour (NNN) interaction is considered, which only shifts the positions of the Dirac points. Owerre \cite{OwerreJPCM2016} demonstrated theoretically that, if a next-nearest neighbour DM interaction is introduced, the time reversal symmetry of the system is broken, and a gap opens at the Dirac points. The band structure for semi-infinite system is shown in Fig. \ref{Figure15}(a), from which one can clearly see the edge spin wave dispersion. Figure \ref{Figure15}(b) shows the illustration of magnon edge states.   
\begin{figure}[ptbh]
\begin{centering}
\includegraphics[width=0.90\textwidth]{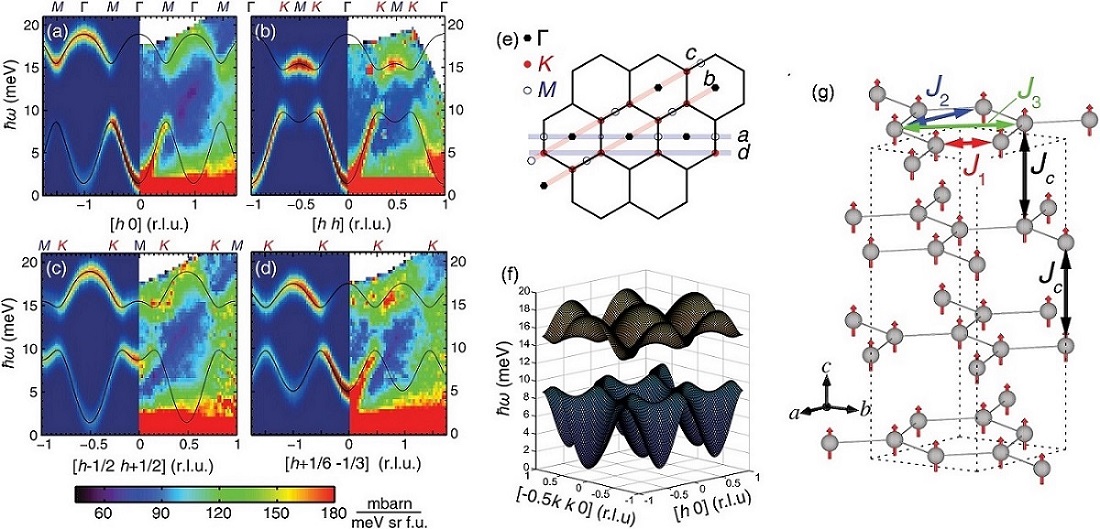}
\par\end{centering}
\caption{(a)-(d) The neutron scattering experiments at 5 K compared with theoretical calculations. The displayed cuts are described in (e) within the $(h, k)$ plane. (f) 3D view of the 2D spin-wave excitations of CrI$_{3}$. (g) Crystal and magnetic structures of CrI$_{3}$. Source: The figures are taken from Ref. \cite{ChenPRX2018}.}
\label{Figure16}
\end{figure}

From the materials point of view, chromium trihalides CrX$_{3}$ (X=F, Cl, Br and I) is a practical example of ferromagnets consisting of van der Waals-bonded stacks of honeycomb layers, which display two spin wave modes with energy dispersion similar to that for the electrons in graphene. Pershoguba \emph{et al}. \cite{PershogubaPRX2018} studied theoretically the Dirac magnons in CrX$_{3}$ (X = F, Cl, Br and I) and discussed the stability of Dirac cones affected by particle statistics and interactions. They showed that honeycomb ferromagnets can display dispersive surface and edge states. Subsequently, the gap at the Dirac points in CrI$_{3}$ was observed experimentally by Chen \emph{et al}. \cite{ChenPRX2018}. Figure \ref{Figure16}(g) plots the crystal and magnetic structures of CrI$_{3}$. By using inelastic neutron scattering, one can obtain the magnon dispersion relation. Figures \ref{Figure16}(a)-\ref{Figure16}(d) show the neutron scattering intensities experimentally observed at 5 K and the calculated dispersions. The results reveal a large gap at the Dirac points. The acoustic and optical spin wave bands are separated from each other by approximately 4 meV, which most likely arises from the next nearest-neighbor DM interaction that breaks the inversion symmetry of the lattice. These band gaps may lead to a nontrivial topological magnon insulator with magnon edge states. The observation of a large spin-wave gap indicates that the spin-orbit coupling plays an important role in the physics of topological spin excitations in honeycomb ferromagnet CrI$_{3}$.

Moreover, in CoTiO$_{3}$ with ilmenite structure, by using inelastic neutron scattering experiment, Yuan \emph{et al}. \cite{YuanPRX2020} observed Dirac magnons in this 3D quantum XY magnet. In addition, an obvious gap of order about 1 meV in the magnon dispersion is also identified. Such a gap arises from the bond-anisotropic exchange coupling, due to quantum order by disorder, which pins the order parameter to the crystal exes. The magnon spectra calculated theoretically shows that edge states connecting the bulk Dirac points can appear with zigzag edge, while vanish for armchair edge. 
\begin{figure}[ptbh]
\begin{centering}
\includegraphics[width=0.90\textwidth]{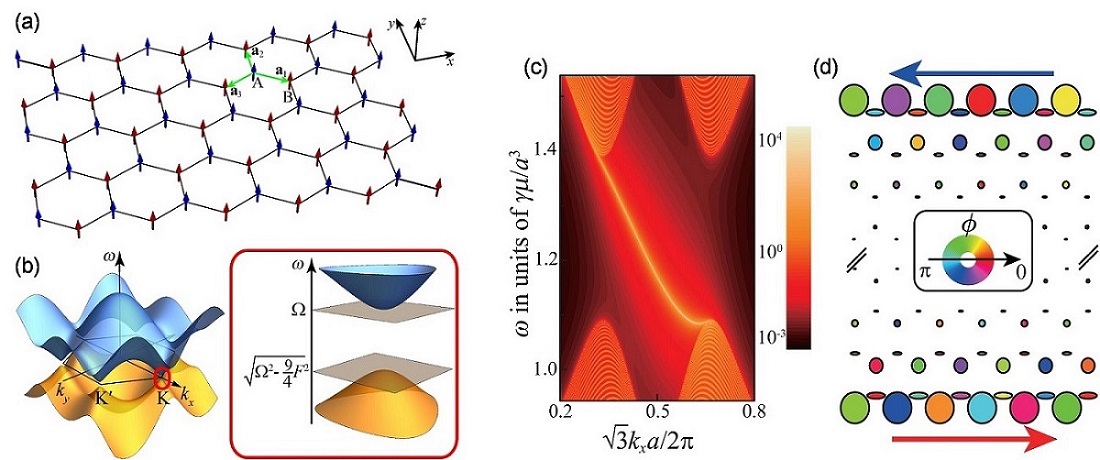}
\par\end{centering}
\caption{(a) Schematic diagram of ferromagnetic spins on a honeycomb lattice with perpendicular anisotropy (along $z$ direction). (b) The spin-wave spectrum of an infinite system in the first BZ for $J=0.1$, $F=0.5$, and $\Omega=1.3$ (in units of $\mu_{0}\mu^{2}/a^{3}$). (c) Density plot of the spectral function on the top edge. (d) Spatial distribution of the spin-wave edge state of $\omega=1.2$ ($\gamma\mu/a^{3}$). The symbol shape traces the spin precession trajectories, and the size of symbols denotes the amplitude of the spin wave at each site. Source: The figures are taken from Ref. \cite{WangPRB2017}.}
\label{Figure17}
\end{figure}

In addition to DM interaction which can open a gap at Dirac points, Wang \emph{et al}. \cite{WangPRB2017} demonstrated theoretically that the pseudodipolar exchange interaction which arises from the superexchange and atomistic spin-orbit interaction \cite{ShekhtmanPRL1992,JackeliPRL2009} can also open the gap at Dirac points and induce nontrivial topological magnon states. The 2D ferromagnetic spins on a honeycomb lattice [see Fig. \ref{Figure17}(a)] is described by a classical Hamiltonian,
\begin{equation}\label{Eq7}
\mathcal{H}=-\frac{J}{2}\sum_{\langle i,j\rangle}\mathbf{m}_{i}\cdot\mathbf{m}_{j}-\frac{1}{2}\sum_{\langle i,j\rangle}\mathcal{F}(\mathbf{m}_{i},\mathbf{m}_{j},\mathbf{e}_{ij})-\sum_{i}\frac{K_{i}}{2}m^{2}_{iz}-\mu B\sum_{i}m_{iz},
\end{equation}
where $\langle i,j\rangle$ denotes the NN sites. The first term is NN exchange interaction with exchange constant $J$. The second term is NN dipole-dipole-like pseudodipolar exchange interaction which arises from the superexchange and atomistic spin-orbit interaction \cite{ShekhtmanPRL1992,JackeliPRL2009}: $\mathcal{F}(\mathbf{m}_{i},\mathbf{m}_{j},\mathbf{e}_{ij})=F(\mathbf{m}_{i}\cdot\mathbf{e}_{ij})(\mathbf{m}_{j}\cdot\mathbf{e}_{ij})$, with $\mathbf{e}_{ij}$ the unit vector connecting sites $i$ and $j$, and $F$ being the interaction strength. The third term is the anisotropy energy with easy axis along $z$ direction, anisotropy constant $K_{i}=K_{A}$ and $K_{B}$ for sublattices A and B. The last term is the Zeeman energy from a magnetic field $B$ along $z$ direction. By neglecting damping, the LLG equation for spin $\mathbf{m}_{i}$ becomes 
\begin{equation}\label{Eq8}
\frac{\partial\mathbf{m}_{i}}{\partial t}=-\mathbf{m}_{i}\times [J\sum_{j\in\langle i\rangle}\mathbf{m}_{j}+F\sum_{j\in\langle i\rangle}(\mathbf{m}_{j}\cdot\mathbf{e}_{ij})\mathbf{e}_{ij}+Km_{iz}\mathbf{e}_{z}+B\mathbf{e}_{z}],
\end{equation}
where $K_{A}=K_{B}=K$. By solving Eq. \eqref{Eq8}, one obtains the spin-wave spectrum for an infinite system, the results are shown in Fig. \ref{Figure17} (b). The band gap at $K$ and $K'$ points is $\Delta_{g}=\Omega-\sqrt{\Omega^{2}-9F^{2}/4}$. When $F=0$, the band gap closes and Dirac cones emerge. Here, the pseudodipolar NN exchange interaction is the critical factor for band-gap opening. For a long strip with zigzag edges along $x$ direction [Fig. \ref{Figure17}(a)], the density plot of the spectral function on the top edge is shown in Fig. \ref{Figure17}(c). The negative slope of the dispersion curve indicates that the propagation direction of spin-wave edge state is counterclockwise, i.e., to the left. Similarly, the states on the bottom edge propagate unidirectionally to the right.  Figure \ref{Figure17}(d) shows spatial distribution of the edge spin-wave eigenstate.    

\subsubsection{Weyl magnons}\label{section2.3.2}
In topological magnonics, another important class of topologically nontrivial system is magnonic Weyl semimetal \cite{MookPRL2016,SuPRB2017_1,SuPRB2017_2,ZyuzinPRB2018,OwerreSR2018,LiNC2016}. Similar to the electronic Weyl semimetals \cite{WanPRB2011,XuNP2015}, the magnon bands in a magnonic Weyl semimetal are nontrivially crossing in pairs at special points (called Weyl nodes) in momentum space. The Weyl nodes are monopoles of Berry curvature and are characterized by the integer topological charge or chirality. Based on no-go theorem, the net topological charges in the entire Brillouin zone must be zero, the Weyl nodes thus must appear in pairs with opposite topological charges of $\pm1$ \cite{HosurCRP2013,NielsenPLB1983}. The magnons around Weyl nodes can be described by effective Weyl Hamiltonians and they are thus called Weyl magnons. In Weyl semimetals, the topologically protected chiral surface states between each pair of Weyl nodes exist on the system surfaces \cite{WanPRB2011,XuS2015,LvPRX2015}. The equal energy contour of these surface states form arcs, with the arc number equaling the number of paired Weyl nodes. 
\begin{figure}[ptbh]
\begin{centering}
\includegraphics[width=0.80\textwidth]{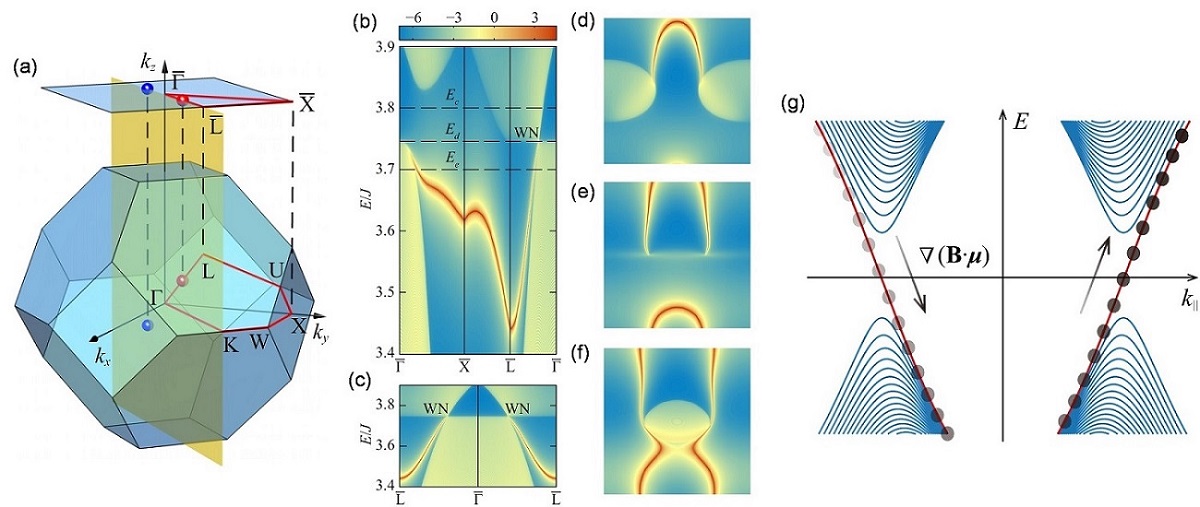}
\par\end{centering}
\caption{(a) The first bulk BZ and the first (001) surface BZ of the pyrochlore lattice. The red and blue dots represent the pair of Weyl nodes with opposite topological charges. Density plots of magnon spectral function on the top (001) surface along (b) $\overline{\Gamma XL\Gamma}$ and (c) $\overline{L\Gamma L}$. (d)–(f) Density plot of magnon spectral function on the top (001) surface in the first BZ for fixed energies of $E_{c}$, $E_{d}$, and $E_{e}$ denoted in (b). (g) Schematic diagram of the magnonic Landau levels. The red and blue curves are the zeroth and higher magnonic Landau levels, respectively. The black dots represent the magnons occupying the zeroth magnonic Landau level. The arrows indicate the magnon motion in momentum space driven by the inhomogeneous magnetic field $\mathbf{B}$. Source: The figures are taken from Ref. \cite{SuPRB2017_1}.}
\label{Figure18}
\end{figure}

Recently, Mook \emph{et al}. \cite{MookPRL2016} and Su \emph{et al}. \cite{SuPRB2017_1} show that the pyrochlore ferromagnets [see Fig. \ref{Figure5}(a)] with DM interaction are intrinsic magnonic Weyl semimetals. The effective spin Hamiltonian of the system include NN exchange interaction, NN DM interaction, and Zeeman interaction. By using the Holstein-Primakoff transformation and the Bloch theorem, one can obtain the band structures of the pyrochlore ferromagnet. Figure \ref{Figure18}(a) shows the first bulk BZ and the first (001) surface BZ of the pyrochlore lattice. The red and blue dots schematically represent the pair of Weyl nodes with opposite topological charges. Similar to the electronic Weyl semimetal, the important hallmark of magnonic Weyl semimetal is the magnon arcs on system surfaces. The density plot of magnon spectral function on the top surface along high symmetry path $\overline{\Gamma XL\Gamma}$ is shown in Fig. \ref{Figure18}(b), where one Weyl node can be identified. One can clearly see the topologically protected surface states which marked by red color with high density on the top surface. While if we consider path $\overline{L\Gamma L}$, two Weyl nodes will appear and the pair of Weyl nodes are connected by surface states, as shown in Fig. \ref{Figure18}(c). For fixed energies of $E_{c}$, $E_{d}$, and $E_{e}$ around the Weyl nodes [see Fig. \ref{Figure18}(b)], the corresponding density plot of magnon spectral function on the top surface in the first BZ are shown in Figs. \ref{Figure18}(d)-\ref{Figure18}(f), respectively. The magnon arcs due to topologically protected surface states are clearly displayed on the top surface. Besides, the authors \cite{SuPRB2017_1} showed that magnonic chiral anomaly can be realized by applying inhomogeneous electric and magnetic fields that are perpendicular to each other. The electric field is used to generate magnonic Landau level according to the Aharonov-Casher effect \cite{AharonovPRL1984}, while the magnetic field is used to drive magnon flow, as shown in Fig. \ref{Figure18}(g). The field drives magnons to move from one Weyl node to the other through the zeroth magnonic Landau level and results in the imbalance of chirality which is the signature of magnonic chiral anomaly. 
\begin{figure}[ptbh]
\begin{centering}
\includegraphics[width=0.7\textwidth]{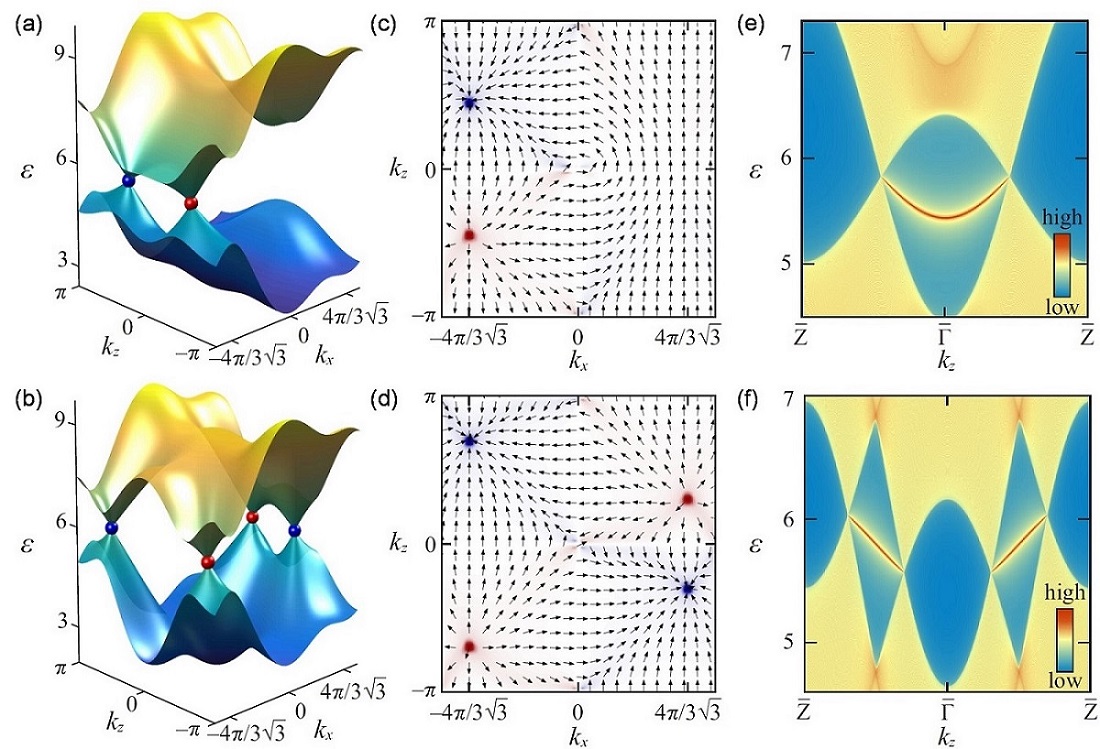}
\par\end{centering}
\caption{(a) and (b) Band structures of Weyl magnons. The Weyl nodes of chirality $\pm1$ are marked by red and blue dots, respectively. (c) and (d) Corresponding Berry curvatures of the lower magnon bands in panels (a) and (b). The arrows represent the direction of Berry curvature vectors in the $k_{x}-k_{z}$ plane with $k_{y}=0$. The background color denotes the divergence of Berry curvature, where red and blue represent positive and negative values, respectively. (e) and (f) Density plots of the front (100) surface spectral functions along $\overline{Z\Gamma Z}$ for the energy bands in panels (a) and (b). Source: The figures are taken from Ref. \cite{SuPRB2017_2}.}
\label{Figure19}
\end{figure}
\begin{figure}[ptbh]
\begin{centering}
\includegraphics[width=0.7\textwidth]{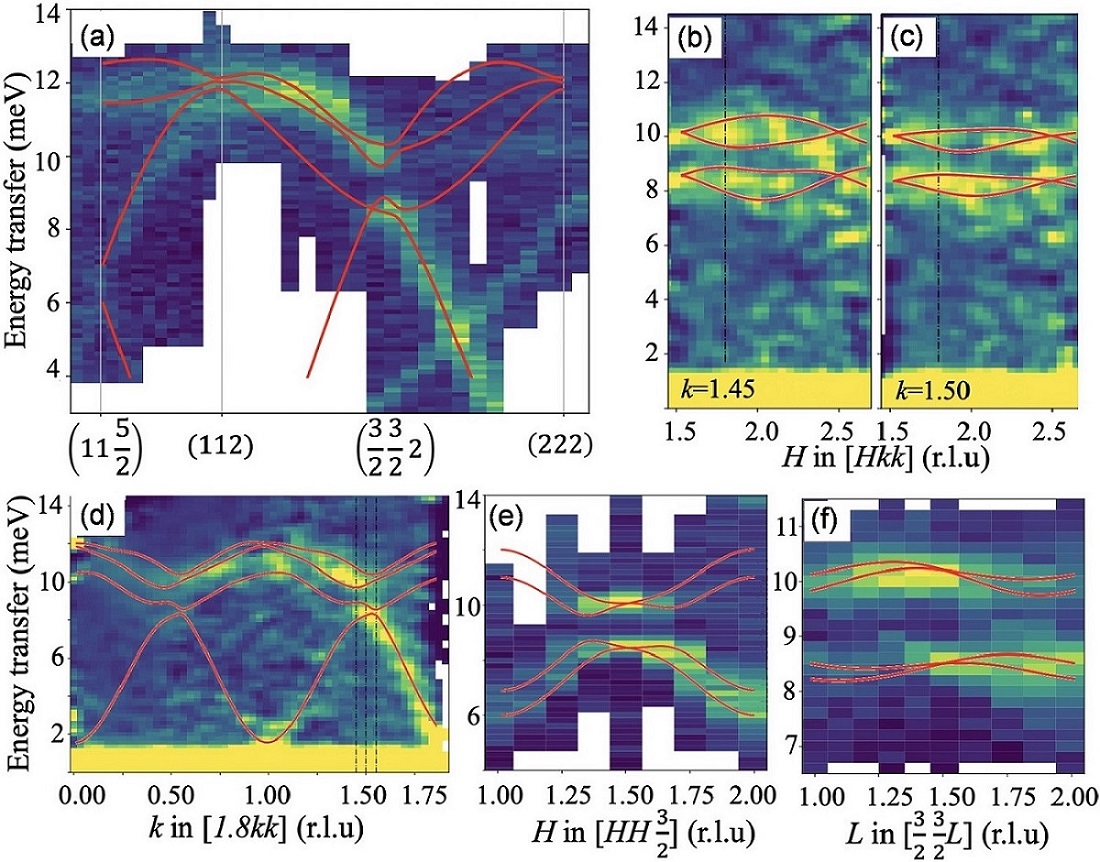}
\par\end{centering}
\caption{Momentum-energy cuts along different directions in reciprocal space. All measurements were done at 2 K without applying magnetic field. Data in (a), (e), and (f) were measured at the triple-axis spectrometer, while (b)–(d) are cuts from the time-of-flight data set. The thin red lines represent the magnon dispersion calculated with the fitted value of the DM interaction. Source: The figures are taken from Ref. \cite{ZhangPRR2020}.}
\label{Figure20}
\end{figure}

Furthermore, Su \emph{et al.} \cite{SuPRB2017_2} show theoretically that the stacked honeycomb ferromagnets can also support Weyl magnons. The spin Hamiltonian of the system include NN intralayer ferromagnetic exchange interaction, anisotropy energy, NN interlayer exchange interaction, DM interaction, and Zeeman interaction. By analyzing the magnon Hamiltonian, one can obtain the phase diagram and identify five distinct phases: topological nontrivial phase which can support topologically protected in-gap surface states, trivial phase, and three different magnonic Weyl semimetal phases. Noticeably, two Weyl semimetal phases have one pair of Weyl nodes at different positions in momentum space, while one Weyl semimetal phase can support two pairs of Weyl nodes. To visualize the Weyl nodes, two different parameters are adopted in the magnonic Weyl semimetal phases (one or two pairs of Weyl nodes). The energy bands in the $k_{x}-k_{z}$ plane for fixed $k_{y}=0$ are shown in Figs. \ref{Figure19}(a) and \ref{Figure19}(b), where one and two pairs of Weyl nodes emerge. The red and blue dots denote chirality $\pm1$ of Weyl nodes. The Berry curvatures are shown in in Figs. \ref{Figure19}(c) and \ref{Figure19}(d), in which the black arrows represent the direction of Berry curvatures projected onto the $k_{x}-k_{z}$ plane and the background color represents the divergence of Berry curvature with red for positive and blue for negative values, respectively. Apparently, the Weyl nodes demonstrated in Figs. \ref{Figure19}(a) and \ref{Figure19}(b) correspond to the monopoles of Berry curvature. The spectral functions on the front (100) surface along $\overline{Z\Gamma Z}$ of the first (100) surface BZ are shown in Figs. \ref{Figure19}(e) and \ref{Figure19}(f), respectively. The surface states with high density (red color) on the front surface between Weyl nodes can be clearly seen. Near the energy of Weyl nodes, these surface states form magnon arcs on sample surfaces. 

In spite of the theoretical progress, the experimental evidence of Weyl magnons is rather rare. The main reason is that Weyl points often locate far away from the center of Brillouin zone and their frequency are thus very high (hundreds of gigahertz). It is difficult to generate such high-frequency magnons by the mature microwave technology. Very recently, by using inelastic neutron scattering technology, Zhang \emph{et al.} \cite{ZhangPRR2020} observed the magnonic Weyl states in multiferroic ferrimagnet Cu$_{2}$OSeO$_{3}$. They show that, in the absence of DM interaction, two pairs of degenerate Weyl nodes with the topological charge $+2$ and $-2$ are located at the Brillouin zone center and boundary. When considering the NN DM interaction, these Weyl nodes are shifted away from the high-symmetry points into a position that sensitively depends on the direction and magnitude of the DM interaction vector. Figure \ref{Figure20} shows the comparison of the experimental and calculated magnon spectra.    

\subsection{Higher-order topological magnons}\label{section2.4}
So far, in magnetic system, most of the studies on topological phases are the first-order, while there are only few references discussing the higher-order topological phases. In 2019, Li \emph{et al.} \cite{Linpj2019} first proposed the higher-order topological insulator in magnetic system, the discussion of which will be present in Section \ref{section3.5}. Later, Sil \emph{et al.} \cite{SilJPCM2020} reported the second-order topological magnonic phases in the ferromagnetic breathing kagome lattice. In this subsection, we focus on this model.   

\begin{figure}[ptbh]
\begin{centering}
\includegraphics[width=0.8\textwidth]{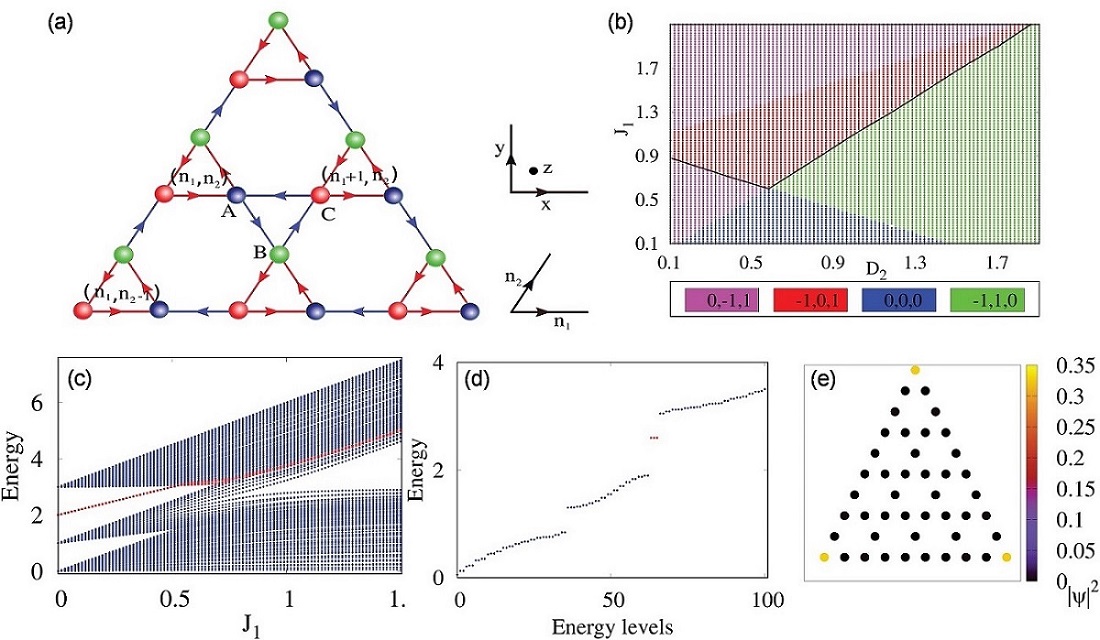}
\par\end{centering}
\caption{(a) A triangular replica of ferromagnetic breathing kagome lattice. (b) Topological phase diagram in $J_{1}-D_{2}$ parameter space with $J_{2}=1.0$ and $D_{1}=0.1$. Four different phases are separated by
different colors as shown in the lower panel by considering the Chern number $C_{1}, C_{2}, C_{3}$. The solid
black line separates topologically nontrivial and trivial phases. (c) Energy spectrum of the breathing kagome lattice with varying $J_{1}$ for $L=15$ with $J_{2}=1.0$ and $D_{1}=0$. (d) Energy of the same system is plotted with respect to energy levels for $J_{1}=0.3$. (e) Probability distribution of a particular eigenstate corresponding to a corner state energy for $J_{1}=0.2$ and $L=5$. Source: The figures are taken from Ref. \cite{SilJPCM2020}.}
\label{Figure21}
\end{figure}

As introduced in Section \ref{section2.2}, the first-order topological magnon can be observed in kagome lattice. Sil \emph{et al.} \cite{SilJPCM2020} found that the ferromagnetic breathing kagome lattice can support second-order topological phases under suitable conditions. Here the Hamiltonian contains Heisenberg exchange interaction, Zeeman term, and DM interaction. Figure \ref{Figure21}(a) shows the lattice structure, with the unit cell comprising of three sites A, B, and C, where the Heisenberg exchange coupling strength is $J_{1}$ for the red lines and $J_{2}$ for the blue lines, the DM interaction strength $D_{1}$ ($D_{2}$) points towards $z$ ($-z$) direction between upward (downward) triangles, and the external magnetic field $H$ is applied to force the magnetic moments magnetized along $z$ direction. In the absence of DM interaction, it is found that the kagome ferromagnet ($J_{1}=J_{2}$) is topologically trivial \cite{MookPRB2014}, while the breathing configuration ($J_{1}\neq J_{2}$) can support the second-order topological phase. The energy spectrum of finite lattice is shown in Fig. \ref{Figure21}(c), from which one can see that when $0<J_{1}<0.5$, there exist degenerate states, with $J_{2}$ being fixed to 1. The energy of the same system is plotted in Fig. \ref{Figure21}(d) with respect to energy levels for $J_{1}=0.3$, which clearly shows the existence of three degenerate states. Furthermore, Fig. \ref{Figure21}(e) shows the distribution of probability density for these degenerate states, one can identify that these states are corner states localized at each corner. Intersetingly, the breathing kagome lattice with non-zero DM interaction ($J_{1}\neq J_{2}$ and $D_{1}\neq D_{2}$) exhibits a rich topological phase diagram which includes distinct first- and second-order topological magnon insulating phases as well as coexistence of them. The phase diagram is presented in Fig. \ref{Figure21}(b), with $J_{2}$ and $D_{1}$ being fixed to 1 and 0.1, respectively. On the one hand, by calculating the topological invariant Chern number, four different phases are identified and separated by different colors; see Fig. \ref{Figure21}(b). On the other hand, the solid black line separates topologically nontrivial (lower portion) and trivial (upper portion) phases by considering the topological invariant bulk polarization, which was defined in Section \ref{section3.5.1}. Therefore, one can obtain the first-order topological phase in red and magenta portions above the solid black line, while the green and magenta portions beneath the solid black line host both first- and second-order topological nontrivial phases, and the blue portion only supports the second-order topological phase.         

\subsection{Topological magnonic device}\label{section2.5}
In magnon spintronics, one key topic is how to control spin wave propagation in a designed way. However, conventional spin waves are very sensitive to the device geometry, internal and external perturbations, which makes spin wave devices inflexible and fragile. Besides, it is difficult for conventional spin wave to realize unidirectional propagation. To design tunable and stable spin wave devices, the topologically robust spin waves are indispensable. Wang \emph{et al}. \cite{WangPRB2017} have predicted the topological chiral spin wave edge state on a ferromagnetic 2D honeycomb lattice, as shown in Fig. \ref{Figure17}. Based on these results, they further proposed \cite{WangPRA2018} the concept of topological magnonic device, including spin-wave diode, spin-wave beam splitters, and spin-wave interferometers.
\begin{figure}[ptbh]
\begin{centering}
\includegraphics[width=1\textwidth]{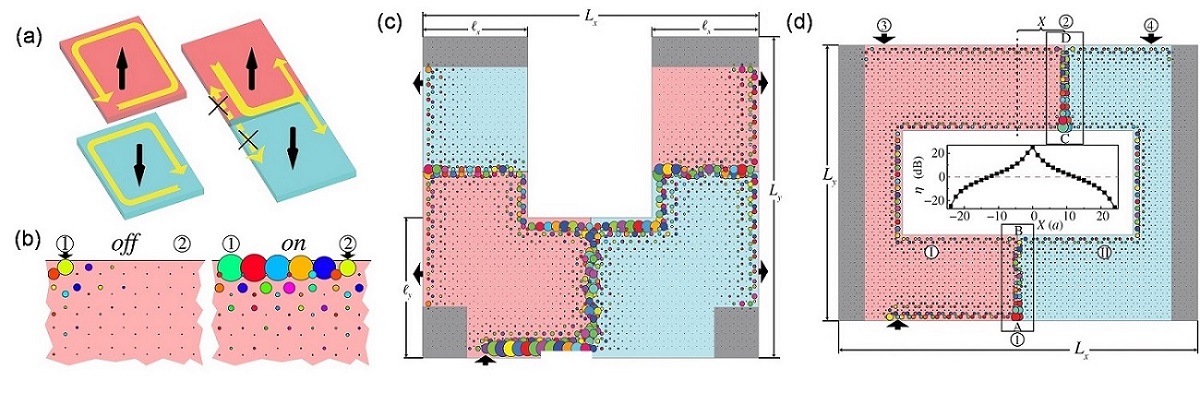}
\par\end{centering}
\caption{(a) Schematic illustration of SW propagation. The red and cyan regions denote domains in which spins point to the $+z$ and $-z$ directions, respectively. The yellow arrows denote the spin wave propagation direction. The illustrations of SW diode (b), SW beam splitters (c) and SW interferometers (d). Source: The figures are taken from Ref. \cite{WangPRA2018}.}
\label{Figure22}
\end{figure}

As demonstrated in Ref. \cite{WangPRB2017}, the topologically protected chiral edge spin waves exist in the band gap and propagate in a certain direction with respect to magnetization direction, i.e., counterclockwise (clockwise) for magnetization along the +$z$ (-$z$) direction, as shown in the left panel of Fig. \ref{Figure22}(a). Owing to the unidirectional property of topological magnons, a segment of a sample edge can be used as a spin-wave diode. The right panel of Fig. \ref{Figure22}(b) shows a snapshot of spin waves when the excitation field is applied at position \textcircled{\scriptsize{2}}. When a spin wave beam is excited and propagates to position \textcircled{\scriptsize{1}}, an \emph{on} state is presented. Conversely, if the excitation field is applied at position \textcircled{\scriptsize{1}}, no spin wave can be detected at position \textcircled{\scriptsize{2}}, which means an \emph{off} state, as shown in the left panel of Fig. \ref{Figure22}(b).   

Since the topological magnons propagate in opposite directions in different domains, as shown in the left panel of Fig. \ref{Figure22}(a), their propagation towards the domain wall can neither penetrate into it nor be reflected by it. It must move along the domain wall. When the spin-wave beam reaches the other edge, it will split into two beams propagating in opposite directions, as shown in the right panel of Fig. \ref{Figure22}(a). Thus, a domain wall is essential for a 1:2 beam splitters. Figure \ref{Figure22}(c) illustrates an example of a 1:4 spin wave beam splitters with three domain walls that separate the $m_{z}=+1$ domains (the pink areas) from the $m_{z}=-1$ domains (the cyan areas). The figure shows a snapshot of the spin-wave pattern when a excitation field of frequency $\omega=12$ is continuously applied at the site marked by the inward arrow in the bottom edge. It is clearly shown that a spin-wave beam splits into four beams eventually.    
\begin{figure}[ptbh]
\begin{centering}
\includegraphics[width=0.80\textwidth]{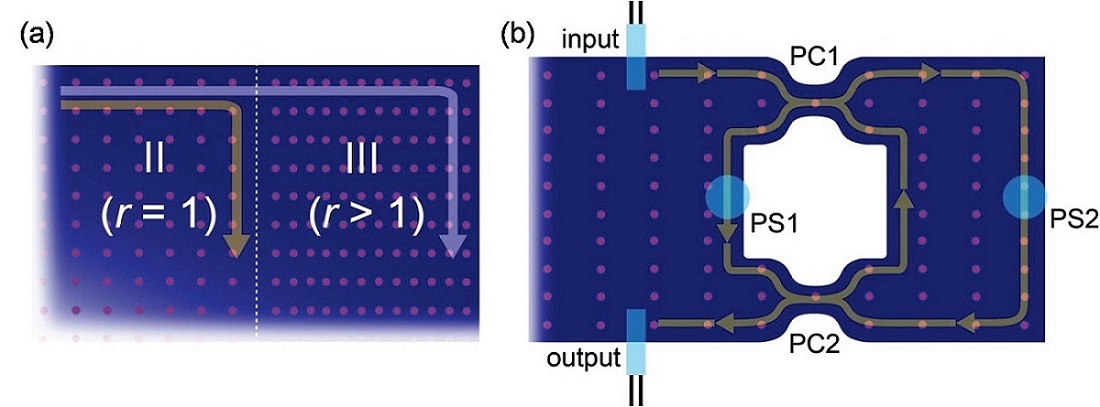}
\par\end{centering}
\caption{Examples of magnonic ciruit made by chiral magnonic edge modes. (a) Schematic picture of spin-current splitter. (b) Magnonic analog of the Fabry-Perot interferometer. Source: The figures are taken from Ref. \cite{ShindouPRB2013_1}.}
\label{Figure23}
\end{figure}

Moreover, spin-wave interferometer is an important element in magnonics. By utilizing topological magnons, one can design a robust, reconfigurable spin-wave interferometer. Figure \ref{Figure22}(d) is a proposal of  a Mach-Zehnder–type spin-wave interferometer with two different domains separated by domain walls. A topological magnon generated at the site marked by the inward arrow enters the first domain wall of lengh $AB$. Then the spin wave beam splits evenly to beam \textcircled{\scriptsize{\uppercase\expandafter{\romannumeral1}}} and \textcircled{\scriptsize{\uppercase\expandafter{\romannumeral2}}}. After traveling a certain distance, the two beams recombine and enter the second domain wall of lengh $CD$, then spin wave can go to either \textcircled{\scriptsize{3}} and \textcircled{\scriptsize{4}}. Their intensities should depend on the interference of the two beams inside the second domain wall. Remarkablely, by placing the second domain wall at different positions or by changing the length of the second domain wall, the relative phase of the two interfered spin waves can be tuned. 

Interestingly, Shindou \emph{et al}. \cite{ShindouPRB2013_1} showed that the magnonic crystal (MC) can induce topological chiral magnonic edge mode as well. The magnonic crystal is composed of YIG and Fe, and the periodic array of holes is introduced into YIG, where Fe is filled inside every hole, as shown in Fig. \ref{Figure23}. By using the chiral magnonic edge mode, the spin wave splitter and interferometer also can be realized. In Fig. \ref{Figure23}(a), the MC in phase \uppercase\expandafter{\romannumeral2} is connected with the other MC in phase \uppercase\expandafter{\romannumeral3}, where the phase \uppercase\expandafter{\romannumeral2} and phase \uppercase\expandafter{\romannumeral3} are different topological nontrivial phases with Chern number $\mathcal{C}_{2}=2$ and $\mathcal{C}_{3}=1$, respectively, and $r$ is the geometric parameter. The two chiral edge modes propagating along the boundary of the theMC in phase II are spatially divided into two, where one mode goes along the boundary of the MC in the phase III, while the other goes along the boundary between these two MCs. This configuration realizes a spin-wave current splitter. The Fabry-Perot interferometer is made up of a couple of chiral spin wave edge modes encompassing a single topological MC [see Fig. \ref{Figure23}(b)]. A unidirectional spin-wave is induced in a chiral mode [“input” in Fig. \ref{Figure23}(b)]. Then the spin wave is divided into two chiral edge modes at a point contact (PC1). Two chiral propagations merge into a single chiral propagation at the other point contact (PC2). Depending on a phase difference between these two, the superposed wave exhibits either a destructive or a constructive interference, which is detected as an electric signal from the other antenna (“output”). Here the application of magnetic fields (PS1 and PS2) can change the velocities of the two chiral edge modes locally.  

In this section, we have reviewed the topological properties of magnons, including topological magnon insulators and semimetals. The topological magnons based spintronic devices have the obvious advantages over the conventional magnonic devices. Firstly, the topological magnons are confined at the boundary of the system, while conventional magnons spread all over the system. From the energy point of view, the topological magnons based devices have lower energy consumption. By using the topological magnons, one can miniaturize the device to a greater extent. Secondly, the conventional magnons are inflexible and fragile, while topological magnons are very robust against defects and disorder, which enables topological magnons to propagate further.      

From the point of view of practical application, there is a demand for low frequency mode, while the frequency of topological magnons often ranges from a few dozen to a few hundred gigahertz. Fortunately, in magnetic system, there exists another important excitation---magnetic soliton---and its collective oscillation frequency is much lower than magnons. In the next section, we will discuss the topological phases of magnetic soliton in artifical lattices. 
     
\section{Topological solitonic insulators}\label{section3}
The magnetic soliton represents an important nonlinear excitation in magnetic system, which can exhibit the behavior of waves and topological phases eventually, due to the soliton-soliton interaction. Remarkably, the spintronic devices based on magnetic solitons have a lot of advantages over their electronic counterpart. For example, the nano-oscillators based on magnetic vortices or skyrmions are very robust and flexible \cite{PribiagNP2007,HrkacJPD2015,ZhangNJP2015}; By using the skyrmion as the carrier of information, the data storage density can be greatly improved, and the current density required for encoding information can be significantly reduced \cite{NagaosaNN2013,FinocchioJPD2016,FertNRM2017,ZhangJPCM2020,JonietzS2010,YuNC2012}; It is very convenient to realize various logic operations by using skyrmion \cite{ChauwinPRA2019,ZhangPRA2020,LuoNL2018,ZhangSR2015,YangPRB2018}; There are many ways to manipulate magnetic solitons \cite{IwasakiNN2013,PsaroudakiPRL2018,WangPRB2015,KongPRL2013,YangOE2018,JiangPRL2020}, which makes the spintronic devices reconfigurable and tunable. In this section, we focus on the collective dynamics and the topological insulator state in artificial magnetic-soliton lattice.    
 
\subsection{Structures and properties of magnetic solitons}\label{section3.1}
\begin{figure}[ptbh]
\begin{centering}
\includegraphics[width=0.98\textwidth]{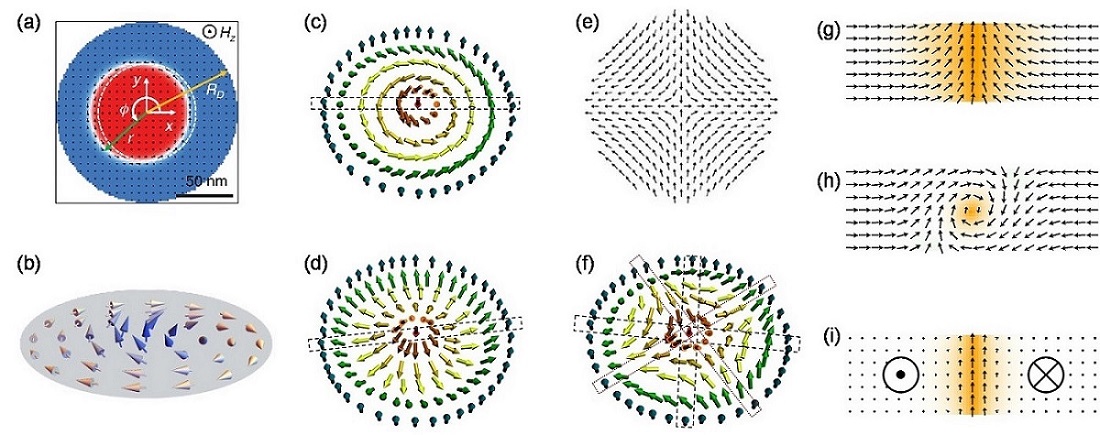}
\par\end{centering}
\caption{The micromagnetic structures of (a) magnetic bubble, (b) vortex, (c) Bloch-type skyrmion, (d) Néel-type skyrmion, (e) antivortex, (f) antiskyrmion, (g) Néel-type, (h) vortex-type and (i) Bloch-type domain walls. Source: The figures are taken from Refs. \cite{MoonPRB2014, KimPRL2017,NayakN2017,MironovPRB2010}.}
\label{Figure24}
\end{figure}

Topology is a study of geometry or space that can keep some properties invariant under a continuous variation of the order parameter. The continuous variation means that the variations do not need to be the same for every position in physical space, but change continuously as a function of position. Magnetic soliton is a manifestation of topology in condensed matter physics. Generally, the magnetic solitons in two dimensions can be characterized by their topological charges
\begin{equation}\label{Eq9}
Q\equiv\frac{1}{4\pi}\int\int \mathbf{m}\cdot(\frac{\partial\mathbf{m}}{\partial x}\times\frac{\partial\mathbf{m}}{\partial x}) dxdy,
\end{equation}
which counts how many times the local normalized magnetization $\mathbf{m}$ wraps the unit sphere. The typical magnetic solitons include the magnetic bubble, vortex, skyrmion, and domain wall, with the micromagnetic structures shown in Fig. \ref{Figure24}. The topological charges for magnetic bubble and skyrmion are $\pm1$, while it turns to $\pm1/2$ for vortex configurations. Topological charge is an invariant indicating that the trivial structure (for example, ferromagnetic state) can not continuously deform into a topological spin texture because of the topological protection. Magnetic solitons with the same topological charge are homotopic. 

The low-energy dynamics of the magnetic vortex (or skyrmion) can be described by the massless Thiele's equation \cite{KimPRL2017,ThielePRL1973} whithin the rigid approximation:
\begin{equation}\label{Eq10}
\mathcal{G}\hat{z}\times \frac{d\textbf{U}_{j}}{dt}-\alpha \mathcal{D} \frac{d\textbf{U}_{j}}{dt}+\textbf{F}_{j}=0,
\end{equation}
\begin{figure}[ptbh]
\begin{centering}
\includegraphics[width=0.75\textwidth]{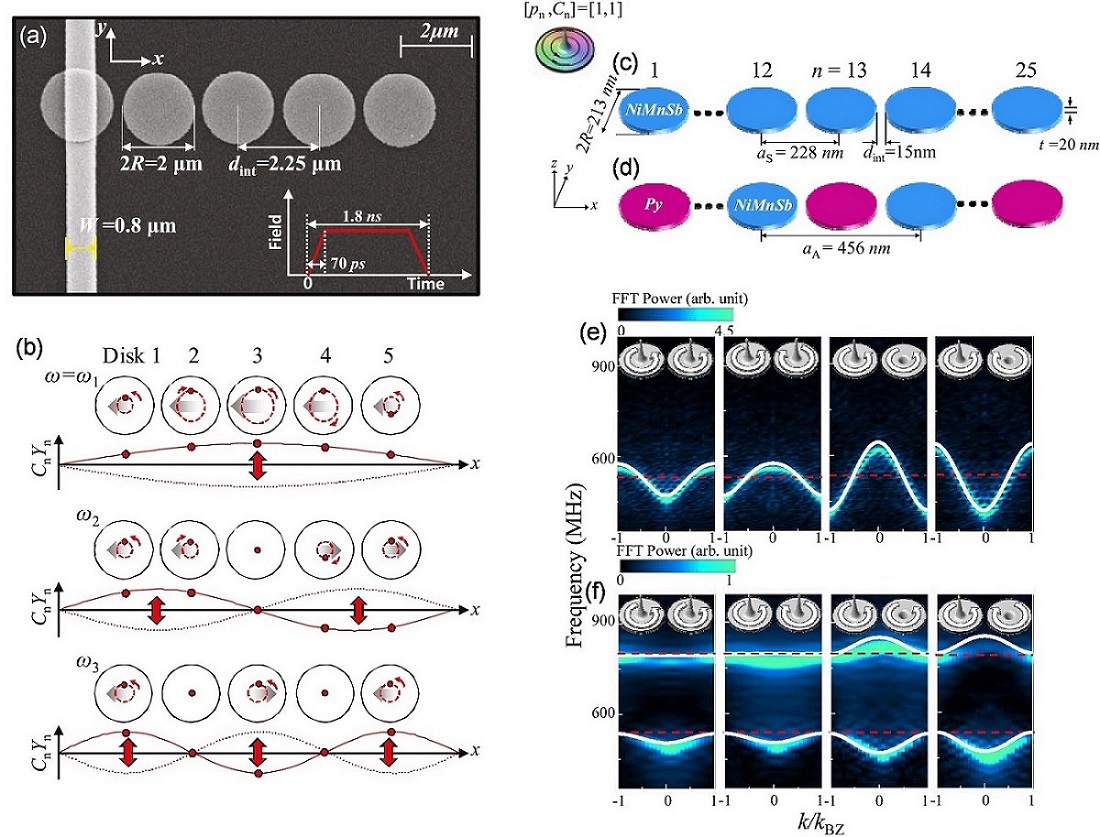}
\par\end{centering}
\caption{ (a) SEM image of an array of five Py disks of identical dimensions and center-to-center distance and with a stripline for application of local magnetic field pulses to the left-end disk. The inset shows a schematic drawing of the field pulse used in the experiment. (b) Spatial distributions of the individual disks’ core positions for the discrete modes. The core trajectories are noted by the dashed lines inside the individual disks. Each dot on each trajectory represents the core position in the given disk. Schematic illustration of 1D chains comprising 25 indentical disks of (c) pure NiMnSb and (d) alternating NiMnSb and Py. Dispersion relations of collective vortex-gyration excitations in chains of (e) pure NiMnSb and (f) alternating NiMnSb and Py, for indicating four different $[p_{n}, C_{n}]$ orderings. Source: The figures are taken from Refs. \cite{HanSR2013,HanAPL2013}.}
\label{Figure25}
\end{figure}where $\mathbf{U}_{j}$ is the position vector of the vortex core, $\mathcal{G} = -4\pi$$Qd M_{s}$/$\gamma$ is the gyroscopic constant with $Q$ the topological charge, $d$ is the thickness of ferromagnetic layer, $M_{s}$ is the saturation magnetization, and $\gamma$ is the gyromagnetic ratio. $\alpha D$ is the viscous coefficient with $\alpha$ being the Gilbert damping constant. The conservative force $\textbf{F}_{j}=-\partial \mathcal{W} / \partial \mathbf U_{j}$ where $\mathcal{W}$ is the potential energy of the system. For a single vortex, the potential energy have the parabolic type: $\mathcal{W}=\mathcal{W}_{0}+\mathcal{K}\mathbf U_{j}^{2}/2$, where $\mathcal{W}_{0}$ is the energy of system when vortex core locates at the center of the nanodisk and $\mathcal{K}$ is the spring constant. By neglecting the damping term, we can derive the gyration frequency of an isolated vortex with $\omega_{0}=\mathcal{K}/|\mathcal{G}|$. 

However, it is well known that magnetic vortices and skyrmions in particular manifest an inertia in their gyration motion \cite{MakhfudzPRL2012,ButtnerNP2015}. The mass effect thus should be taken into account for describing the vortex (or skyrmion) oscillation. Therefore, the Thiele's equation can be generalized as:  
\begin{equation}\label{Eq11}
  -\mathcal{M}\frac{d^{2}\textbf{U}_{j}}{dt^{2}}+\mathcal{G}\hat{z}\times \frac{d\textbf{U}_{j}}{dt}-\alpha \mathcal{D} \frac{d\textbf{U}_{j}}{dt}+\textbf{F}_{j}=0,
\end{equation}  
with $\mathcal{M}$ being the inertial mass of magnetic soliton. Similarly, we can calculate the gyration frequency of magnetic solitons with:
\begin{equation}\label{Eq12}
  \omega_{\pm}=-\mathcal{G}/2\mathcal{M}\pm\sqrt{(\mathcal{G}/2\mathcal{M})^{2}+\mathcal{K}/\mathcal{M}}.
\end{equation}  
The positive and negative values of the $\omega$ in Eq. \eqref{Eq12} indicate that there are two kinds of gyration modes with clockwise and counterclockwise direction, respectively. It is noted that the Thiele's equation containing higher-order terms can be derived from Landau-Lifshitz-Gilbert (LLG) equation (see Section \ref{section3.4.2} for details).       

\subsection{Collective dynamics of magnetic soliton crystals}\label{section3.2}
As discussed above, the oscillation of magnetic soliton lattice [including one-dimensional (1D), two-dimensional (2D), and three-dimensional (3D) structures] have the properties of waves. For 1D case, Fig. \ref{Figure25}(a) shows the SEM image of a sample with an array of five Py disks, which have identical dimensions and initial configuration (vortex states). Here, the polarization and chirality of vortices array are $P=[+1,-1,+1,-1,+1]$ and $C=[-1,-1,-1,-1,+1]$, respectively. A current pulse of 1.8 ns duration is applied into the electrode stripline to trigger an excitation of vortex gyration in the first disk. The gyration motion of the first vortex can propagate to other vortices because of the dipolar interaction between disks. Then, the fast Fourier transformation (FFT) of the core position for all votices are calculated and the spectra show that the system have five discrete wave modes. Figure \ref{Figure25}(b) shows the trajectories of the vortices cores in the individual disks for different frequencies (we choose three modes as examples). One can clearly see that the collective gyration of vortices is similar to a standing wave. 

\begin{figure}[ptbh]
\begin{centering}
\includegraphics[width=0.75\textwidth]{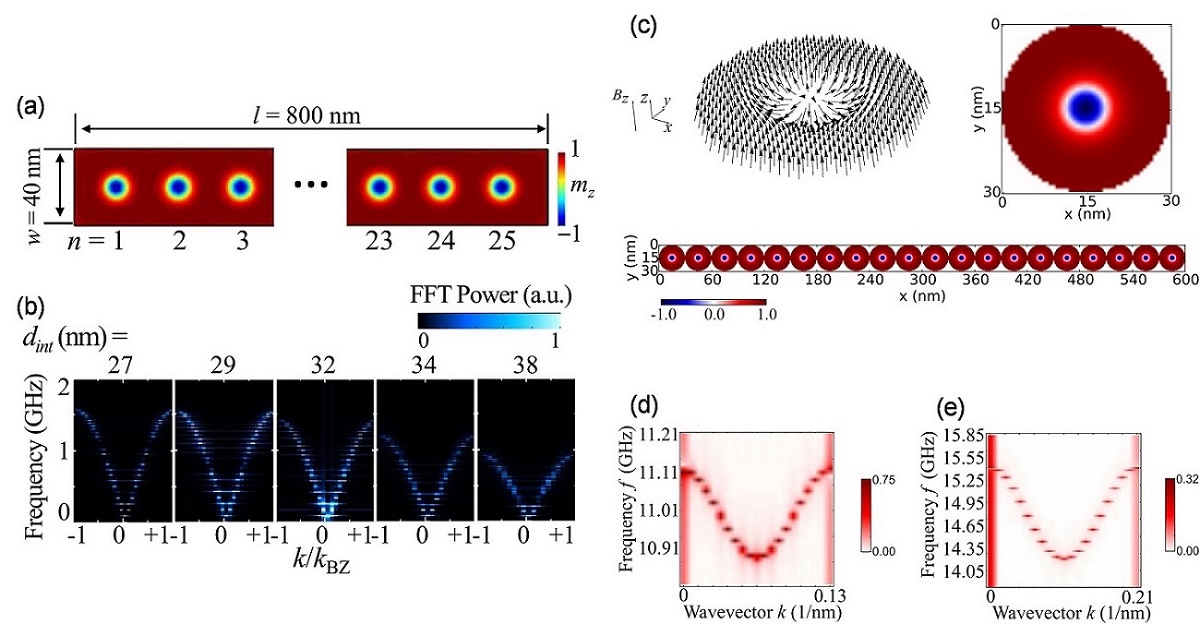}
\par\end{centering}
\caption{ (a) 1D skyrmion array in nanostrip of indicated dimensions comprising 25 skyrmions. (b) Dispersion curves of 1D skyrmion chains for different interdistances. (c) The schematic illustrations of single skyrmion nanodisk and the array of 20 nanodisks, with diameter 30 nm, the magnetization z-component are plotted. Dispersion relation of the breathing excitation in an array of (d) Diameter 50 nm disks under external magnetic field $B_{z}=0.1$ T (e) Diameter 30 nm disks under external magnetic field $B_{z}=0.2$ T. Source: The figures are taken from Refs. \cite{KimSR2017,MruczkiewiczPRB2016}.}
\label{Figure26}
\end{figure}

Furthermore, if we consider the 1D lattice containing more magnetic vortices, the nature of waves will be more feasible. Figures \ref{Figure25}(c) and \ref{Figure25}(d) show the 1D chains comprising 25 disks of pure NiMnSb and alternating NiMnSb and Py, respectively. All disks are in vortex states and have the same dimension. The spectra of collective vortex-gyration excitations for pure NiMnSb and alternating NiMnSb and Py are shown in Figs. \ref{Figure25}(e) and \ref{Figure25}(f), respectively, which look like the dispersion relations of waves (spin wave for instance). Besides, the influence of different $[P_{n},C_{n}]$ orderings on the band structure are also given. Similarly, the 1D lattice containing many skyrmions can also present the band structure. Figure \ref{Figure26}(a) plots the schematic diagram of 1D skyrmion array in nanostrip comprising 25 skyrmions, and the dependence of the dispersion relations on $d_{int}$ is shown in Fig. \ref{Figure26}(b), respectively, where $d_{int}$ is the distance between nearest neighbor skyrmions. It can be seen that as $d_{int}$ increases, the band width $\Delta\omega$ and the angular frequency $\omega_{\text{BZ}}$ at wavevector $k=k_{\text{BZ}}$ decrease. The decrease of the total energy density with increasing $d_{int}$ would result in the decrease of $\Delta\omega$ and $\omega_{\text{BZ}}$. The 1D skyrmions lattice can be composed of nanodisks array, as shown in Fig. \ref{Figure26}(c). The collective breathing excitation of skyrmion nanodisks exhibits the dispersive band structure; see Figs. \ref{Figure26}(d) and \ref{Figure26}(e). 

\begin{figure}[ptbh]
\begin{centering}
\includegraphics[width=0.8\textwidth]{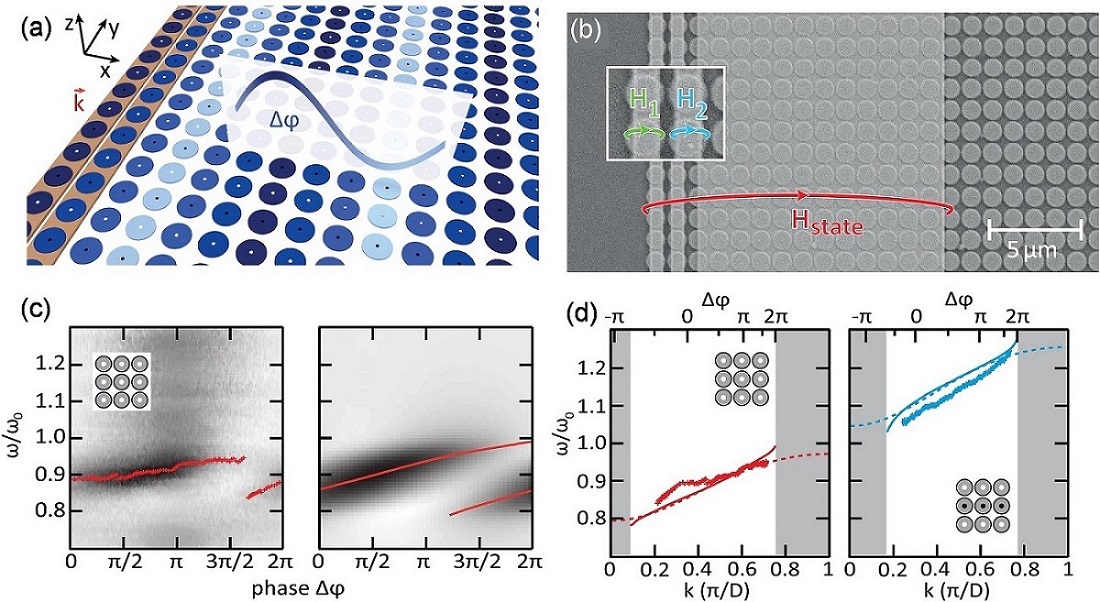}
\par\end{centering}
\caption{ (a) Schematic representation of gyration waves propagating through a two-dimensional vortex crystal imprinted by a phase-shifted excitation of the first two columns of vortices. (b) Scanning electron micrograph of a vortex crystal covered by copper striplines. The two thin striplines are used for band structure measurements. (c) Measured and calculated absorption intensity as a function of excitation frequency and phase difference for homogeneous polarization. (d) Band structure measurements (markers), calculations based on the extended Thiele model including damping (lines) and without damping (dashed lines) for homogeneous polarization pattern and horizontally striped polarization pattern. Source: The figures are taken from Ref. \cite{BehnckePRB2015}.}
\label{Figure27}
\end{figure}

Considering 2D magnetic soliton systems, without loss of generality, we choose the 2D vortex lattice \cite{BehnckePRB2015} as an example. Figures \ref{Figure27}(a) and \ref{Figure27}(b) show the schematic diagram and scanning electron micrograph of the vortex crystal, respectively. Figure \ref{Figure27}(c) depicts absorption spectra depending on the phase difference between the exciting magnetic fields for homogeneous polarization patterns. The red lines represent the maxima of the absorption determined by Lorentzian fits. It can be seen that by increasing the phase difference between the exciting magnetic fields the resonance frequencies increase. Figure \ref{Figure27}(d) plots the dispersion relation combining the measured absorption (markers) with the calculations using the extended Thiele's model (lines) and calculations for an infinite crystal without damping (dashed lines) obtained from Ref. \cite{ShibataPRB2004} with different polarization patterns. The lower $x$ axis is the wave number $k$ while the upper $x$ axis shows the phase difference $\Delta\varphi$. These results indicate that the band structure of 2D vortices lattice can be reprogrammed by the polarization pattern.

By using ferromagnetic resonance spectroscopy and scanning transmission X-ray microscopy, the collective dynamics of 3D vortex crystals have been studied by Hänze \emph{et al}. \cite{HanzeSR2016}. They find that the spectra of the vortex arrangements are directly linked to the chirality and polarity of the vortices.  

\subsection{Two pedagogical models}\label{section3.3}
To have a better understanding about the topological properties of soliton systems, one can map the Hamiltonian into well-known topological models. Pedagogical topological models include the Su-Schrieffer-Heeger (SSH) model and Haldane model. In this section, we give a brief introduction about these two models to facilitate our readers interpreting the following results.
\subsubsection{The Su-Schrieffer-Heeger model}\label{section3.3.1}
The SSH model is a simple tight-binding model with spontaneous dimerization proposed by Su, Schrieffer, and Heeger to describe the one-dimensional polyacetylene \cite{SuPRL1979}. Over the past decades, the SSH model has been generalized to various different systems \cite{HasanRMP2010,DelplacePRB2011,LiNC2013,GaneshanPRL2013} and attracted growing interest for demonstrating the fundamental topological physics. 
\begin{figure}[ptbh]
\begin{centering}
\includegraphics[width=0.70\textwidth]{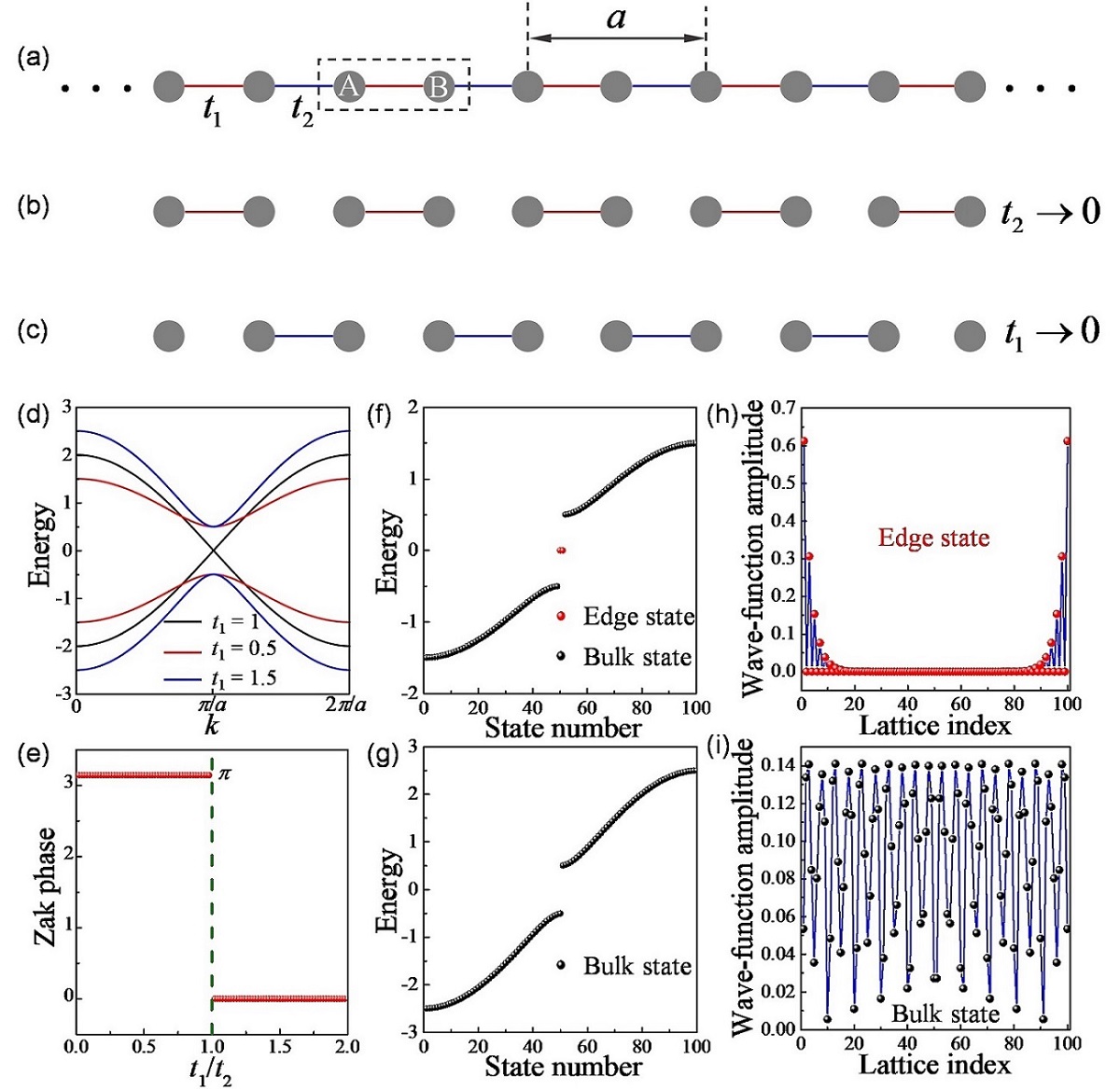}
\par\end{centering}
\caption{(a) Illustration of the SSH model. The schematic plot of the finite system for two limit cases, $t_{2}\rightarrow 0$ (b) and $t_{1}\rightarrow 0$ (c). (d) Band structure of infinite system for different $t_{1}$, with $t_{2}$ being fixed to 1. (e) Dependence of the Zak phase on the different values $t_{1}/t_{2}$. Eigenvalue of the finite system which contains 100 lattice with $t_{1}/t_{2}=0.5$ (f) and $t_{1}/t_{2}=1.5$ (g). The spatial distribution of the wave-function amplitude for the edge (h) and bulk (i) states.} 
\label{Figure28}
\end{figure}Figure \ref{Figure28} (a) plots the illustration of the SSH model, where A and B represent two sublattices, and $t_{1}$ and $t_{2}$ are the alternating intracellular and intercellular hopping parameters, respectively. For simplicity, here we only consider the positive values of hopping parameters. The Hamiltonian reads
\begin{equation}\label{Eq13}
\mathcal{H}=\sum_{n}(t_{1}c^{\dag}_{A,n}c_{B,n}+t_{2}c^{\dag}_{A,n+1}c_{B,n})+h.c.,
\end{equation}
with $c_{A,n}$ (or $c_{B,n}$) the annihilation operators localized on site $A$ (or $B$) of the $n$-th cell. By using a Fourier transformation, the Hamiltonian can be written in the form of
\begin{equation}\label{Eq14}
 \mathcal {H}=\left(
 \begin{matrix}
   0 & t_{1}+t_{2}e^{-ika} \\
   t_{1}+t_{2}e^{ika} & 0
  \end{matrix}
  \right),
\end{equation}
where $k$ is the wave vector, and $a$ is the lattice constant, as shown in Fig. \ref{Figure28}(a). The spectrum of the SSH model thus can be obtained, $E=\pm\sqrt{t_{1}^2+t_{2}^2+2t_{1}t_{2}\text{cos}ka}$. For $t_{1}=t_{2}$, the band structure of the system is gapless, while a gap opens at $k=\pi/a$ when $t_{1}\neq t_{2}$, leading to an insulating phase, as shown in Fig. \ref{Figure28}(d). The topological invariant Zak phase can be used to judge if these insulating phases are topological:
\begin{equation}\label{Eq15}
\mathbb{Z}=i\int_{0}^{2\pi/a}\Psi^{\dag}(k)\nabla_{k}\Psi(k)dk\ \  (\text{mod}\ 2\pi),
\end{equation}  
where $\Psi(k)$ is the Bloch wave function of the energy band. Figure \ref{Figure28}(e) plots the dependence of the Zak phase $\mathbb{Z}$ on the value of $t_{1}/t_{2}$. It can be clearly seen that $\mathbb{Z}$ is quantized to $\pi$ when $t_{1}/t_{2}<1$ and to 0 otherwise, which indicates that the system allows two topologically distinct phases for $t_{1}/t_{2}<1$ and $t_{1}/t_{2}>1$. 

Bulk-boundary correspondence indicates the existence of robust edge states when the system is in topological phase. Figure \ref{Figure28}(f) plots the spectrum of a finite system which contains 100 lattices when $t_{1}/t_{2}=0.5$. One can clearly see that the system can support degenerate "zero mode" marked by red dots. Further, it is found that its wave function spatial distribution is highly localized in both end of the system, as shown in Fig. \ref{Figure28}(h). Noticeably, these edge states are topologically protected and are immune from moderate disorder and defects. Moreover, the bulk modes (marked by black dots) are identified, with the wave function spreading all over the system, as shown in Fig. \ref{Figure28}(i). The spectrum of the trivial system is also plotted in Fig. \ref{Figure28}(g), where $t_{1}/t_{2}=1.5$. In this case, one can only observe the bulk modes. Remarkably, these results can be intuitively understood as follow: (i) For the case of $t_{2}\rightarrow 0$, the system is in trivial phase [see Fig. \ref{Figure28}(e)], one can clearly see that there are no uncoupled lattice [see Fig. \ref{Figure28}(b)]. The system thus can only support bulk states; (ii) When $t_{1}\rightarrow 0$, the system is in topological phase, and we can identify isolated lattice emerging at both ends of the system, which corresponds to the edge state [see Fig. \ref{Figure28}(c)].      
         
\subsubsection{The Haldane model}\label{section3.3.2}
Another important topological model is the Haldane model \cite{HaldanePRL1988}, which can realize the quantum Hall effect in the honeycomb lattice (graphene). Graphene is a two-dimensional form of carbon, with the conduction band and valence band touching each other at high-symmetry points (Dirac points) in the BZ \cite{NetoRMP2009}. Near those points, the system has a linear dispersion. The tight-binding model with the NN coupling reads:
\begin{equation}\label{Eq16}
\mathcal{H}=\sum_{\langle ij\rangle} t_{1}c_{i}^{\dag}c_{j}+h.c..
\end{equation}
After the Fourier transformation, the tight-binding Hamiltonian takes the following form
\begin{equation}\label{Eq17}
\mathcal{H}=\sum_{\mathbf{k}} t_{1}[e^{i(\frac{\sqrt{3}}{2}k_{x}a+\frac{1}{2}k_{y}a)}+e^{i(-\frac{\sqrt{3}}{2}k_{x}a+\frac{1}{2}k_{y}a)}+e^{-ik_{y}a}]a_{A,\mathbf{k}}^{\dag}a_{B,\mathbf{k}}+h.c..
\end{equation} 
Near the high-symmetry points $\pm K=(\pm\frac{4\pi}{3\sqrt{3}a},0)$, we can expand the wave vector as $\mathbf{k}=\mathbf{K}+\mathbf{p}$, with $|\mathbf{p}|\ll|\mathbf{K}|$. Then the effective Hamiltonian in terms of $\mathbf{p}$ becomes:
\begin{equation}\label{Eq18}
\mathcal{H}=\sum_{\mathbf{k}}-\frac{3t_{1}}{2} (p_{x}a+ip_{y}a)a_{A,\mathbf{k}}^{\dag}a_{B,\mathbf{k}}+h.c..
\end{equation}
In terms of the basis $\{\phi^{A}(\mathbf{p}),\phi^{B}(\mathbf{p})\}$, the Hamiltonian reads: $h(\mathbf{p})=-\frac{3t_{1}}{2}(\pm p_{x}\sigma_{x}-p_{y}\sigma_{y})$, where $\sigma_{x}$ and $\sigma_{y}$ are Pauli matrixes. Then we can clearly see that the band structure of graphene is gapless at high-symmetry points $\pm K$, and has a linear dispersion near these points.   
\begin{figure}[ptbh]
\begin{centering}
\includegraphics[width=0.80\textwidth]{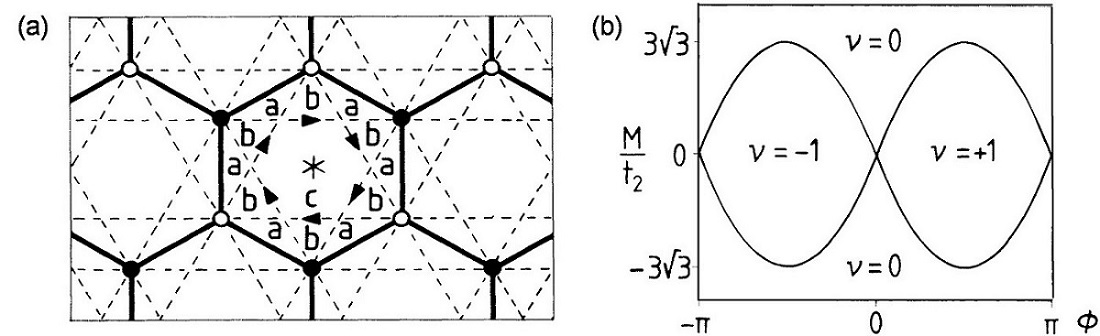}
\par\end{centering}
\caption{(a) The honeycomb-net model, where the solid and dashed lines represent the neareat-neighbor and second-neighbor bonds, respectively. Open and solid points mark the $A$ and $B$ sublattices. Arrows on second-neighbor bonds mark the directions of positive phase hoppong in the state with broken time-reversal invariance. (b) Phase diagram of the system. Zero-field quantum Hall effect ($\sigma^{xy}=\nu e^{2}/h$, where $\sigma^{xy}$ is the transverse conductivity and $\nu=\pm1$) emerges if $|M/t_{2}|<3\sqrt{3}|\text{sin}\phi|$. Source: The figures are taken from Ref. \cite{HaldanePRL1988}.} 
\label{Figure29}
\end{figure}

The degeneracy at the Dirac points is protected by inversion ($\mathbb{P}$) and time-reversal ($\mathbb{T}$) symmetry. By breaking these symmetries the degeneracy can be lifted and gap will open at Dirac points, leading to a topologically non-trivial phase. Conventionally, the realization of quantum Hall state requires a strong magnetic field \cite{KlitzingRRL1980,TsuiRRL1982}. In 1988, Haldane \cite{HaldanePRL1988} proposed that the $\mathbb{T}$ symmetry of the graphene can be broken with a magnetic field that is zero on average in the unit cell, which brings a periodic local magnetic-flux density $B(\mathbf{r})$ in the $\hat{z}$ direction normal to the 2D plane. The Haldane model was formulated by introducing the NNN hopping:
\begin{equation}\label{Eq19}
\mathcal{H}=\sum_{\langle ij\rangle} t_{1}c_{i}^{\dag}c_{j}+\sum_{\langle\langle ij\rangle\rangle} t_{2}c_{i}^{\dag}c_{j}.
\end{equation}
The displacement of the nearest $A-A$ and $B-B$ hopping can be written as: $\mathbf{b}_{1}=\mathbf{a}_{2}-\mathbf{a}_{3}$, $\mathbf{b}_{2}=\mathbf{a}_{3}-\mathbf{a}_{1}$, and $\mathbf{b}_{3}=\mathbf{a}_{1}-\mathbf{a}_{2}$, with basis vectors $\mathbf{a}_{1}=(\sqrt{3}a/2,a/2)$, $\mathbf{a}_{2}=(-\sqrt{3}a/2,a/2)$, and $\mathbf{a}_{3}=(0,-a)$. The closed path of the NNN hopping contains non-trivial phase $\phi=2\pi(2\Phi_{a}+\Phi_{b})/\Phi_{0}$, where $\Phi_{a}$ and $\Phi_{b}$ are the fluxes through the regions of the unit cell marked by $a$ and $b$ in Fig. \ref{Figure29}(a), and $\Phi_{0}=|h/e|$ is the flux quantum. The NN hopping parameter $t_{1}$ is unaffected by the magntic flux, while the NNN hopping $t_{2}$ suffers from a shift $t_{2}\rightarrow t_{2}\text{exp}(i\phi)$. 

In the momentum space, by using the basis $\{\phi^{A}(\mathbf{k}),\phi^{B}(\mathbf{k})\}$, the Hamiltonian can be expressed as:
\begin{equation}\label{Eq20}
\mathcal{H}(\mathbf{k})=2t_{2}\text{cos}(\phi)\sum_{i}\text{cos}(\mathbf{k}\cdot\mathbf{b}_{i})\sigma_{0}+\sum_{i}t_{1}\text{cos}(\mathbf{k}\cdot\mathbf{a}_{i})\sigma_{x}+\sum_{i}t_{1}\text{sin}(\mathbf{k}\cdot\mathbf{a}_{i})\sigma_{y}+[M-2t_{2}\text{sin}\phi\sum_{i}\text{sin}(\mathbf{k}\cdot\mathbf{b}_{i})]\sigma_{z},
\end{equation}
where the masses ($M$) with opposite signs at high-symmetry points $\pm K$ are introduced because of the $\mathbb{P}$ symmetry. Since the $\mathbb{T}$ symmetry is broken by applying local magnetic-flux, a gap emerges at the high-symmetry points.

To achieve the topological non-trivial phase, we require $M^{2}<(3\sqrt{3}t_{2}\text{sin}\phi)^{2}$ [one can refer to Eqs. \eqref{Eq46}-\eqref{Eq49} in Section \ref{section3.4.2} for calculation details]. Figure \ref{Figure29}(b) shows the phase diagram of the spinless electron model, where we have assumed $|t_{2}/t_{1}|<1/3$, to guarantee that the two bands are separated by a finite gap.   

\subsection{First-order topological phases}\label{section3.4}
The collective dynamics of magnetic solitons has received significant recent attention, as introduced in Section \ref{section3.2}, while the possible topological phase is rarely discussed. In 2017, a pioneering work about the topological phase in magnetic soliton lattice was made by Kim \emph{et al.} \cite{KimPRL2017}. By solving the massless Thiele's equation and mapping it into the Haldane model, it is found that the solitons (vortex and bubble) arranged as a honeycomb lattice can support a chiral edge mode with the propagation direction being associated with the topological charge of the constituent solitons. Soon after that, Li \emph{et al.} \cite{LiPRB2018} generalized the approach by including both a second-order inertial term and a third-order non-Newtonian gyroscopic term to interpret the emerging multiband nature of chiral edge states observed in honeycomb lattice of magnetic skyrmions. Interestingly, the realization of SSH states in one-dimensional magnetic soliton lattice was also reported recently \cite{LiAR2020,GoPRB2020}. In this section, we aim to review the theory of the first-order topological insulating phase emerging in low dimensional magnetic soliton systems.    
 
\subsubsection{One-dimensional lattice}\label{section3.4.1}
\begin{figure}[ptbh]
\begin{centering}
\includegraphics[width=0.95\textwidth]{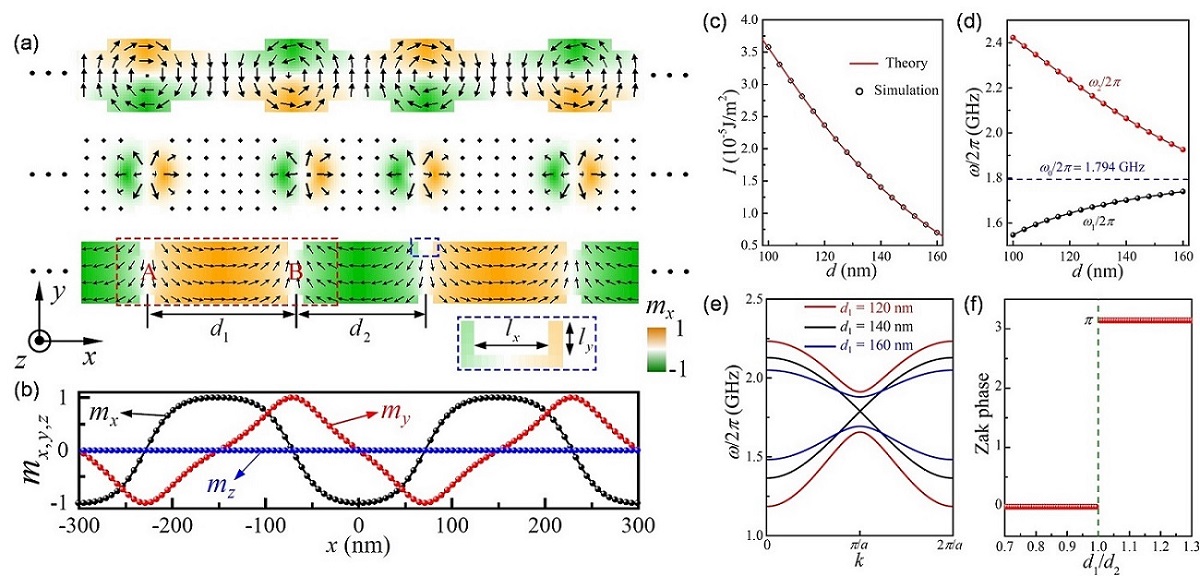}
\par\end{centering}
\caption{(a) Illustration of the vortex, skyrmion, and DW racetrack with periodic pinnings. The micromagnetic structure of N$\acute{\text{e}}$el-type DWs pinned by cuboid notches with $l_{x}=20$ nm, $l_{y}=9$ nm, and $l_{z}=5$ nm is plotted, with the unit cell containing two DWs at sites A and B. $d_{1}$ and $d_{2}$ are the intracellular and intercellular distances between notches, respectively. (b) The components of normalized magnetization along the center of DW racetrack with $d_{1}=160$ nm and $d_{2}=140$ nm. (c) Dependence of the coupling strength $\mathcal{I}$ on $d$. Black circles denote simulation results and red solid line represents the analytical formula. (d) The two eigen frequencies of a DW-DW pair varying with $d$. (e) Band structure of an infinite DW racetrack for different intracellular lengths: $d_{1}=$120, 140, and 160 nm, with $d_{2}$ being fixed to 140 nm. (f) Dependence of the Zak phase on the ratio $d_{1}/d_{2}$. Source: The figures are taken from Ref. \cite{LiAR2020}.}
\label{Figure30}
\end{figure}

To discuss the topological insulating phases in one-dimensional magnetic soliton lattice, without loss of generality, we choose the DW as the representative example. Figure \ref{Figure30}(a) plots the illustration of various one-dimensional magnetic soliton lattice. The Landau-Lifshitz-Gilbert (LLG) equation can be used to describe the magnetization dynamics \cite{ZhangPRL2004,ThiavilleEPL2005}:
\begin{equation}\label{Eq21}
\frac{\partial\mathbf{m}}{\partial t}=-\gamma\mathbf{m}\times\mathbf{H}_{\text{eff}}+\alpha\mathbf{m}\times\frac{\partial\mathbf{m}}{\partial t}+\mathbf{\Gamma}_{\text{st}},
\end{equation}
where $\mathbf{m}=\mathbf{M}/M_{s}$ is the unit magnetization vector with the saturated magnetization $M_{s}$, $\gamma$ is the gyromagnetic ratio, and $\alpha$ is the Gilbert damping constant.
The effective field $\mathbf{H}_{\mathrm{eff}}$ comprises the external field, the exchange field, the magnetic anisotropic field, and the dipolar field. $\mathbf{\Gamma}_{\text{st}}$ is the spin-transfer or spin-orbit torque. For the case of spin-transfer torque, $\mathbf{\Gamma}_{\text{st}}=b_{J}(\hat{J}\cdot\nabla)\mathbf{m}-\beta b_{J}\mathbf{m}\times(\hat{J}\cdot\nabla)\mathbf{m}$ with $b_{J}=JPg\mu_{B}/2|e|M_{s}$ and $\hat{J}$ being the flow direction of the spin-polarized current. Here $J$ is the charge current density, $P$ is the spin polarization, $g$ is the $g$-factor, $\mu_{B}$ is the Bohr magneton, and $e$ is the (negative) electron charge.

The collective-coordinate or $\{q,\phi\}$ method provides a simple, yet accurate description of the motion of complex DWs [see Fig. \ref{Figure30}(b)] \cite{WalkerJAP1974,LiPRB2004}:
\begin{equation}\label{Eq22}
\begin{aligned}
 (1+\alpha^{2})\frac{dq_{j}}{dt} =&\gamma \alpha H_{\text{pin},j}\Delta_{j}+\frac{1}{2}\gamma(N_{z}-N_{y})\Delta_{j} M_{s}\text{sin}2\phi_{j},\\
  (1+\alpha^{2})\frac{d\phi_{j}}{dt} =&\gamma H_{\text{pin},j}-\frac{1}{2}\gamma\alpha(N_{z}-N_{y})M_{s}\text{sin}2\phi_{j},
\end{aligned}
\end{equation}
%\begin{equation}\label{Eq2}
%\begin{aligned}
% (1+\alpha^{2})\frac{dq_{j}}{dt} =&\gamma \alpha H_{\text{pin},j}\Delta+\frac{1}{2}\gamma(N_{z}-N_{y})\Delta M_{s}\text{sin}2\phi_{j}-(1+\alpha \beta)b_{J},\\
%  (1+\alpha^{2})\frac{d\phi_{j}}{dt} =&\gamma H_{\text{pin},j}-\frac{1}{2}\gamma\alpha(N_{z}-N_{y})M_{s}\text{sin}2\phi_{j}+\frac{\alpha}{\Delta}b_{J}-\frac{c_{J}}{\Delta},
%\end{aligned}
%\end{equation}
where the collective coordinates $q_{j}$ and $\phi_{j}$ are the position and tilt angle of the $j$-th DW, respectively, $H_{\text{pin},j}$ includes the pinning field from both the notch and the DW-DW interaction, $N_{y}$ and $N_{z}$ are the demagnetizing factors along the $y$- and $z$-axis of the nanostrip, respectively, and $\Delta_{j}=\sqrt{2A/\left\{2K_{u}+\mu_{0}M_{s}^{2}\left[(N_{y}-N_{x})+(N_{z}-N_{y})\text{sin}^{2}\phi_{j}\right]\right\}}$ represents the DW width with $A$ the exchange stiffness, $K_{u}$ the magnetocrystalline anisotropy constant, and $\mu_{0}$ being the vacuum permeability. Here we discuss the collective genuine oscillation of DW lattice near the pinning notch, the spin torque in Eq. \eqref{Eq21} thus can be dropped tentatively.

From the energy point of view, $H_{\text{pin},j}$ can be expressed as the spatial derivative of the total potential:
\begin{equation}\label{Eq23}
H_{\text{pin},j}=-\frac{1}{2\mu_{0}M_{s}L_{y}L_{z}}\frac{\partial U}{\partial q_{j}},
\end{equation}
where $L_{y}$ and $L_{z}$ are the width and thickness of the nanostrip, respectively, and $U$ is the total energy of the system: $U=\sum_{j}\mathcal {K}q_{j}^{2}/2+\sum_{j\neq k}\mathcal {I}(d_{jk})q_{j}q_{k}/2$. Here $\mathcal {K}$ is the spring constant determined by the shape of the notch and $\mathcal {I}(d_{jk})$ is the coupling constant depending on the distance $d_{jk}$ between DWs. Generally, the DW-DW interaction can be divided into three parts: the monopole-monopole ($\propto1/d_{jk}$), the exchange ($\propto1/d_{jk}^{2}$), and the dipole-dipole ($\propto1/d_{jk}^{3}$) \cite{PivanoPRB2020}. The explicit form of $\mathcal{I}(d)$ can be obtained from micromagnetic simulations in a self-consistent manner. Considering a small $\phi$ and neglecting the dissipation terms, we can arrive at the linear form of Eq. \eqref{Eq22}:
\begin{equation}\label{Eq24}
\mathcal{M}\frac{d^{2}q_{j}}{dt^{2}}+\mathcal {K}q_{j}+\sum_{k\in\langle j\rangle}\mathcal{I}(d_{jk})q_{k}=0,
\end{equation}
where the $\mathcal{M}=2\mu_{0}L_{y}L_{z}/\gamma^{2}(N_{z}-N_{y})\Delta$ is the effective mass of a single DW with $\Delta=\sqrt{2A/\left[2K_{u}+\mu_{0}M_{s}^{2}(N_{y}-N_{x})\right]}$ and $\langle j\rangle$ is the set of the nearest neighbors of $j$. Here, $\mathcal{I}(d_{jk})=\mathcal {I}_{1}$ ($\mathcal {I}_{2}$) when $j$ and $k$ share an intracellular (intercellular) connection with $\mathcal {I}_{1,2}=\mathcal {I}(d_{1,2})$ ($d_{1}$ and $d_{2}$ are the alternating intersite lengths). We thus have mapped the governing equation to a Su-Schrieffer-Heeger problem \cite{SuPRL1979}. The analytical formula of $\mathcal{I}(d)$ can be obtained by the micromagnetic simulation of a DW-DW pair separated by an arbitrary distance. Symbols in Fig. \ref{Figure30}(c) are numerical results and the solid curve is theoretical formula $\mathcal{I}(d)=c_{1}/d+c_{2}/d^{2}+c_{3}/d^{3}$, with $c_{1}=-9.2635\times10^{-12}$ J m$^{-1}$, $c_{2}=2.294\times10^{-18}$ J, and $c_{3}=-1.0111\times10^{-25}$ J m. Figure \ref{Figure30}(d) plots the $d$-dependence of the out-of-phase and in-phase DW-oscillation frequencies, that is, $\omega_{1}$ and $\omega_{2}$ respectively, in the simple two-DW system. It shows that $\omega_{1}$ increases while $\omega_{2}$ decreases for an increasing $d$. One naturally expects that $\omega_{1}=\omega_{2}=\omega_{0}$ when $d\rightarrow\infty$, with $\omega_0=\sqrt{\mathcal{K}/\mathcal{M}}$ corresponding to the oscillation frequency of an isolated DW. By measuring $\omega_{0}$ in experiments, one can determine the pinning-potential stiffness $\mathcal{K}$. This approach, however, suffers from an issue that the dynamics of a single DW can be easily modified by structure defects and material randomness, and it thus cannot precisely determine the genuine profile of the pinning potential. Below, a topological method is introduced to overcome this issue.

Considering a one-dimensional DW lattice, as plotted in Fig. \ref{Figure30}(a), where the dashed red rectangle represents the unit cell and the basis vector is $\textbf{a}=a\hat{x}$ with $a=d_{1}+d_{2}$. The band structure of the collective DW oscillations can be computed by a plane wave expansion $q_{j}=q_{j}\exp\big[i(\omega t+nka)\big]$, where $j=A, B$ for different sublattices, $n$ is an integer, and $k$ is the wave vector. The Hamiltonian then can be expressed in momentum space as:
\begin{equation}\label{Eq25}
 \mathcal {H}=\left(
 \begin{matrix}
   \mathcal{K} & \mathcal{I}_{1}+\mathcal{I}_{2}e^{-ika} \\
   \mathcal{I}_{1}+\mathcal{I}_{2}e^{ika} & \mathcal{K}
  \end{matrix}
  \right).
\end{equation}
Solving \eqref{Eq25} gives the dispersion relation:
\begin{equation}\label{Eq26}
\omega_{\pm}(k)=\sqrt{\frac{\mathcal {K}\pm\sqrt{\mathcal {I}_{1}^{2}+\mathcal {I}_{2}^{2}+2\mathcal {I}_{1}\mathcal{I}_{2}\text{cos}ka}}{\mathcal {M}}},
\end{equation}
where $+(-)$ represents the optical (acoustic) branch. The bulk band structures for different geometric parameters are plotted in Fig. \ref{Figure30}(e), where $d_{2}$ is fixed to $140$ nm and magnetic parameters of Ni \cite{PivanoPRB2020} are adopted. For $d_{1}=d_{2}$, the two bands merge together [black curve in Fig. \ref{Figure30}(e)], while a gap opens at $k=\pi/a$ when $d_{1}\neq d_{2}$ [red and blue curves in Fig. \ref{Figure30}(e)], leading to an insulating phase. Moreover, Figure \ref{Figure30}(f) shows the dependence of the Zak phase \cite{SuPRL1979} $\mathbb{Z}$ on the ratio $d_{1}/d_{2}$. It is observed that $\mathbb{Z}$ is quantized to 0 when $d_{1}/d_{2}<1$ and to $\pi$ otherwise, indicating two topologically distinct phases in the two regions. 

\begin{figure}[ptbh]
\begin{centering}
\includegraphics[width=0.95\textwidth]{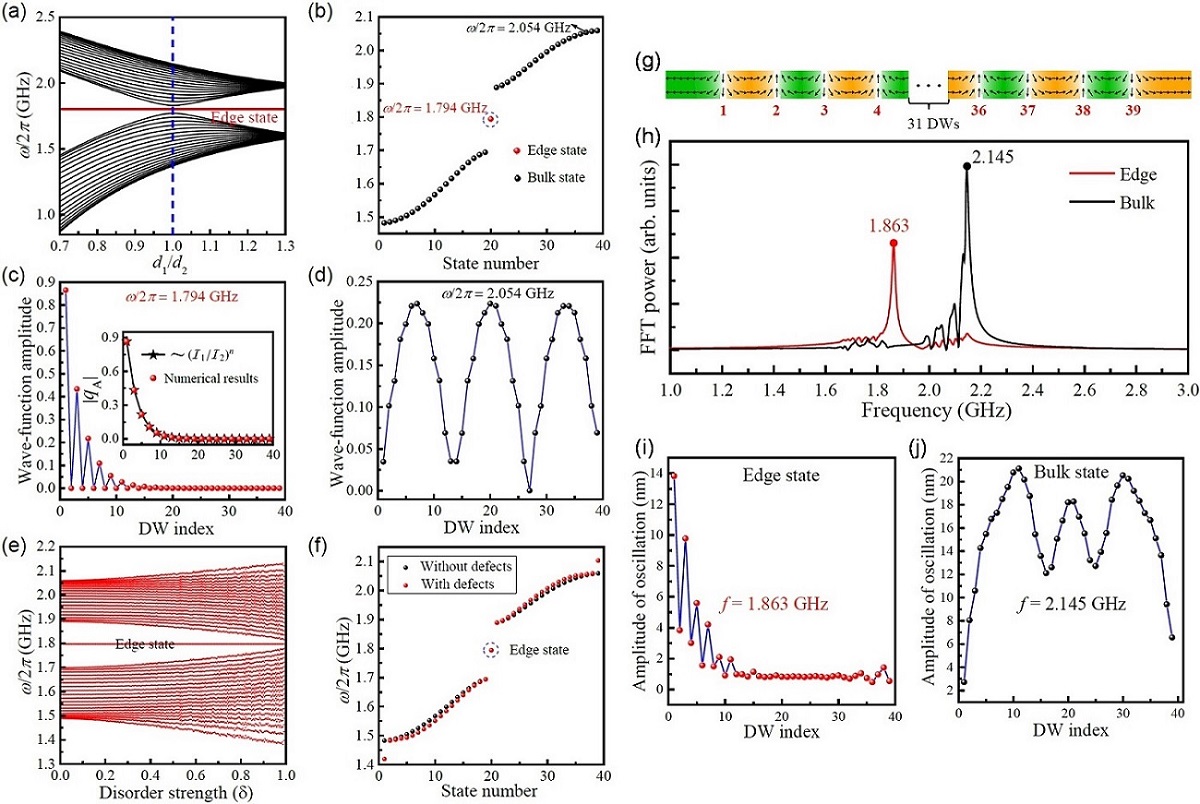}
\par\end{centering}
\caption{ (a) Spectrum of a finite DW racetrack for different $d_{1}/d_{2}$. The dotted blue line denotes the boundary separating two topologically distinct phases; the red segment represents the in-gap mode. (b) Eigenfrequencies of the DW lattice with $d_{1}/d_{2}=8/7$. The spatial distribution of DW-oscillation amplitude for the edge (c) and bulk (d) states. (Inset) Comparison between analytical and numerical results. (e) Spectrum of disordered DW racetracks. (f) Spectrum with (red dots) and without (black dots) defects. (g) Schematic plot of a finite racetrack containing $39$ DWs, with $d_{1}=160$ nm and $d_{2}=140$ nm. (h) The temporal Fourier specra of the DWs oscillation at edge (1st DW) and bulk (20th DW) positions. The spatial distribution of amplitude of DW oscillations for edge (i) and bulk (j) states. Source: The figures are taken from Ref. \cite{LiAR2020}.}
\label{Figure31}
\end{figure}

To verify the bulk-boundary correspondence, we consider the finite system containing an odd number (e.g., 39) of DWs. Numerical results of spectrum are shown in Fig. \ref{Figure31}(a), where the in-gap state (red line) emerges for all ratios $d_{1}/d_{2}\neq1$. We first consider the case $d_{1}/d_{2}=8/7\ (>1)$. Figure \ref{Figure31}(b) plots the eigenfrequencies of the system, showing that there is one in-gap mode marked by red dot. Further, it is found that its spatial distribution is highly localized at the left end of the racetrack [see Fig. \ref{Figure31}(c)], in contrast to its bulk counterpart shown in Fig. \ref{Figure31}(d). We adopt the \emph{Ans\"{a}tze} for the localized mode as $q_{j}=q_{j}\text{exp}(i\omega_{0} t)z^{n}$ with $|z|<1$. The edge state then can be solved by the equations:
\begin{equation}\label{Eq27}
\begin{aligned}
(\mathcal{I}_{1}+\mathcal{I}_{2}z)q_{A}(n)&=0,\ \text{for}\  n=2,3,...,\\
(\mathcal{I}_{1}+\mathcal{I}_{2}z^{-1})q_{B}(n)&=0,\ \text{for}\  n=1,2,3,...,
\end{aligned}
\end{equation}
with the boundary condition $\mathcal{I}_{1}q_{B}(1)=0$ (A-site DW is in the outmost left boundary).
Because $\mathcal{I}_{1,2}\neq 0$, we obtain $q_{B}(n)=0$ $\forall n$ and $z=-\mathcal{I}_{1}/\mathcal{I}_{2}$. The wave function of A-site DWs therefore follows an exponentially decaying formula $|q_{A}|=|q_{0}|(\mathcal{I}_{1}/\mathcal{I}_{2})^{n}$ for $n=1,2,3,...$. Analytical result agrees excellently with numerical calculations, as plotted in the inset of Fig. \ref{Figure31}(c). Furthermore, we expect that the edge state becomes localized in the right end instead if $d_{1}/d_{2}<1$ and the localized modes emerge in both ends as the magnetic racetrack contains an even number of DWs.

The topological robustness of the edge states can be verified by analysing the spectrum of the system including disorder and defects, with results presented in Figs. \ref{Figure31}(e) and \ref{Figure31}(f), respectively. Here the disorder is introduced by assuming that the coupling parameters $\mathcal{I}_{1}$ and $\mathcal{I}_{2}$ have a random variation, i.e., $\mathcal{I}_{1}\rightarrow \mathcal{I}_{1}(1+\delta N)$, $\mathcal{I}_{2}\rightarrow \mathcal{I}_{2}(1+\delta N)$, with $\delta$ the disorder strength and $N$ a uniformly distributed random number between $-1$ and 1. As to the defects, we assume $\mathcal{I}_{1}$ and $\mathcal{I}_{2}$ suffering from a shift ($\mathcal{I}_{1}\rightarrow 10\mathcal{I}_{1}$, $\mathcal{I}_{2}\rightarrow 0.1\mathcal{I}_{2}$) on the second and fourth DWs. From Figs. \ref{Figure31}(e) and \ref{Figure31}(f), we observe that the edge state is very robust against these disorder and defects, while the bulk states are sensitive to them.

The micromagnetic simulations are utilized to confirm the theoretical predictions above. A system containing $39$ interacting DWs in Ni nanostrip of length $7000$ nm is considered, as shown in Fig. \ref{Figure31}(g). To obtain the spectra of DW oscillations, a sinc-function magnetic field is applied along the $x$-axis. To find the frequency range of the edge and bulk states, we analyze the temporal Fourier spectra of the DW racetrack at two different positions (DW 1 and DW 20, for example). Figure \ref{Figure31}(h) shows the results, with peaks of the red and black curves denoting the positions of edge and bulk bands, respectively. We then apply a sinusoidal magnetic field $\textbf{h}(t)=h_0\sin(2\pi ft)\hat{x}$ over the whole system to excite the edge and bulk modes by choosing two frequencies $f=1.863$ and $2.145$ GHz, respectively, as marked in Fig. \ref{Figure31}(h). The spatial distribution of DW-oscillation amplitude for these two modes are plotted in Figs. \ref{Figure31}(i) and \ref{Figure31}(j), respectively, from which one can clearly identify the localized and extended nature of the edge and bulk states, respectively. Full micromagnetic simulations are well consistent with the analytical results. 

By including the STT term in Eq. \eqref{Eq22}, we obtain the generalized Landau-Lifshitz-Gilbert equation:
\begin{equation}\label{Eq28}
\begin{aligned}
 (1+\alpha^{2})\frac{dq}{dt} =&\gamma \alpha H_{\text{pin}}\Delta+\frac{1}{2}\gamma(N_{z}-N_{y})\Delta M_{s}\text{sin}2\phi-(1+\alpha\beta)b_{J},\\
  (1+\alpha^{2})\frac{d\phi}{dt} =&\gamma H_{\text{pin}}-\frac{1}{2}\gamma\alpha(N_{z}-N_{y})M_{s}\text{sin}2\phi+\frac{b_{J}}{\Delta}(\alpha-\beta).
\end{aligned}
\end{equation}
By linearizing Eq. \eqref{Eq28} and neglecting the dissipation terms, we have:
\begin{equation}\label{Eq29}
\frac{d^{2}q}{dt^{2}}+\frac{\mathcal{K}}{\mathcal{M}}q=-\beta\gamma(N_{z}-N_{y})M_{s}b_{J}.
\end{equation}
The solution of \eqref{Eq29} can be written as:
\begin{equation}\label{Eq30}
q(t)=q_0\text{exp}(i\omega_{0}t)-\frac{\beta\gamma(N_{z}-N_{y})M_{s}b_{J}}{\omega^{2}_{0}}.
\end{equation}
From Eq. \eqref{Eq30}, we find that the STT does not modify the DW-oscillation frequency but causes a shift to its equilibrium position
\begin{equation}\label{Eq31}
  X=|\langle q(t)\rangle|=\frac{\beta\gamma(N_{z}-N_{y})M_{s}b_{J}}{\omega^{2}_{0}}.
\end{equation}
The non-adiabaticity parameter $\beta$ can therefore be accurately quantified by experimentally measuring the slope of $X-b_{J}$ curve, i.e., $\lambda=\beta\gamma(N_{z}-N_{y})M_{s}/\omega^{2}_{0}$. 

The rubust topological edge of DW lattice can be utilized as the DW frequency standard, which can be used to accurately measure the pinning profile and to finally resolve the controversy about the $\beta$ parameter. Noticeably, these general results are applicable to other types of soliton (e.g., magnetic vortex, skyrmion, etc). 

\begin{figure}[ptbh]
\begin{centering}
\includegraphics[width=0.55\textwidth]{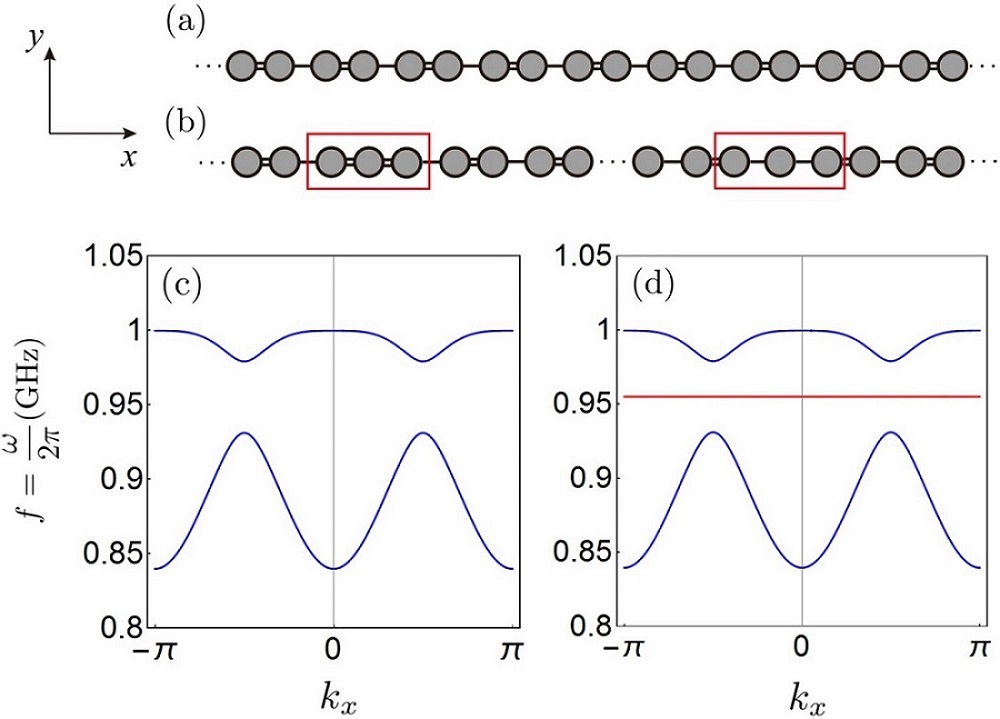}
\par\end{centering}
\caption{ A schematic illustration of the staggered 1D chain of magnetic nanodisks (a) without and (b) with the domain-wall defect. Band structure of the system (c) without and (d) with the domain-wall defect. Source: The figures are taken from Ref. \cite{GoPRB2020}}
\label{Figure32}
\end{figure}
Recently, Go \emph{et al.} \cite{GoPRB2020} studied a metamaterial composed of the magnetic soliton disks structured in a one-dimensional bipartite chain, as shown in Fig. \ref{Figure32}(a). The system supports two bands and no bound states appear in the absence of defects [see Fig. \ref{Figure32}(c)]. However, when a pair of domain-wall and anti-domain-wall defects are introduced [see Fig. \ref{Figure32}(b)], the authors show the existence of a midgap state bounded at a domain wall connecting topologically distinct two configurations [see Fig. \ref{Figure32}(d)], which mimics the electronic SSH model.

\subsubsection{Two-dimensional lattice}\label{section3.4.2}
To analysis the collective dynamics of magnetic vortices on honeycomb lattice, Kim \emph{et al.} \cite{KimPRL2017} begin with the massless Thiele's equation [Eq. \eqref{Eq10}]. 
Different from the single isolated vortex, the potential energy $\mathcal{W}$ should include the contributions both from the confinement of a single disk and the interaction between  disks: $\mathcal {W}=\sum_{j}\mathcal {K}\textbf{U}_{j}^{2}/2+\sum_{j\neq k}U_{jk}/2$ with $U_{jk}=\mathcal {I}_{\parallel}U_{j}^{\parallel}U_{k}^{\parallel}-\mathcal {I}_{\perp}U_{j}^{\perp}U_{k}^{\perp}$ \cite{ShibataPRB2004,KimPRL2017,ShibataPRB2003}. Here, $\mathcal {I}_{\parallel}$ and $\mathcal {I}_{\perp}$ are the longitudinal and transverse coupling constants due to the anisotropic nature of dipole-dipole interactions, respectively.

Impose $\mathbf{U}_{j}=(u_{j},v_{j})$ and defining $\psi_{j}=u_{j}+i v_{j}$, Eq. \eqref{Eq10} can be simplied as follow (here the topological charge $Q$ is chosen to $-1/2$): 
\begin{equation}\label{Eq32}
   i\frac{d\psi_{j}}{dt}=\omega_{K}\psi_{j}+\sum_{k\in\langle j\rangle}(\zeta\psi_{k}+\xi e^{i2\theta_{jk}}\psi^{*}_{k}),
\end{equation}
where $\omega_{K}=\mathcal{K}/\mathcal{|G|}$, $\zeta=(\mathcal {I}_{\parallel}-\mathcal {I}_{\perp})/2\mathcal {|G|} $, $\xi=(\mathcal {I}_{\parallel}+\mathcal {I}_{\perp})/2\mathcal{|G|}$, $\theta_{jk}$ is the angle of the direction $\hat{e}_{jk}$ from the $x$-axis, $\hat{e}_{jk}=(\mathbf{R}_{k}^{0}-\mathbf{R}_{j}^{0})/|\mathbf{R}_{k}^{0}-\mathbf{R}_{j}^{0}|$, and $\langle j\rangle$ is the set of nearest neighbors of $j$. Here we have neglected the dissipation. We then expand the complex variable as:
\begin{equation}\label{Eq33}
  \psi_{j}=\chi_{j}(t)\exp(-i\omega_{0}t)+\eta_{j}(t)\exp(i\omega_{0}t),
\end{equation}
For vortex gyrations with $Q=-1/2$, one can justify $|\chi_{j}|\gg|\eta_{j}|$. By substituting Eq. \eqref{Eq33} into Eq. \eqref{Eq32}, one can obtain:
\begin{equation}\label{Eq34}
i\frac{d\psi_{j}}{dt}=(\omega_{K}-\frac{3\xi^{2}}{2\omega_{K}})\psi_{j}+\zeta\sum_{k\in\langle j\rangle}\psi_{k}-\frac{\xi^{2}}{2\omega_{K}}\sum_{l\in\langle\langle j\rangle\rangle}\text{cos}(2\bar{\theta}_{jl})\psi_{l}-i\frac{\xi^{2}}{2\omega_{K}}\sum_{l\in\langle\langle j\rangle\rangle}\text{sin}(2\bar{\theta}_{jl})\psi_{l},
\end{equation}
where $\bar{\theta}_{jl}=\theta_{jk}-\theta_{kl}$ is the relative angle from the bond $k\rightarrow l$ to the bond $j\rightarrow k$ with $k$ between $j$ and $l$, $\langle\langle j\rangle\rangle$ is the set of the second-nearest neighbors of $j$.
\begin{figure}[ptbh]
\begin{centering}
\includegraphics[width=0.95\textwidth]{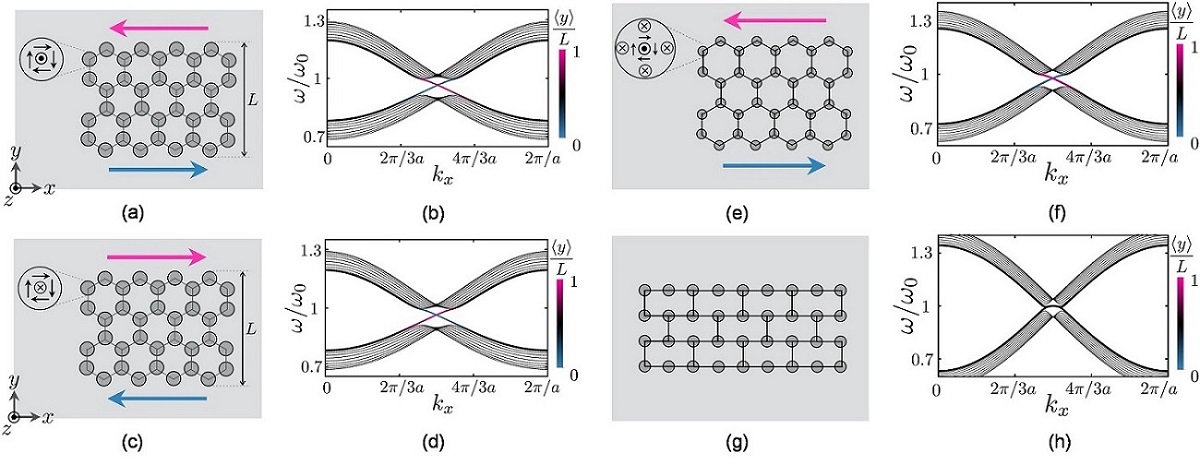}
\par\end{centering}
\caption{ Schematic illustrations of physically separated ferromagnetic disks in a honeycomb lattice with zigzag edges; vortices have the topological charge (a) $Q=1/2$ and (c) $Q=-1/2$. The arrows along the edges represent the directions of the chiral mode. (b) and (d) The one dimensional dispersions for the system shown in (a) and (c), respectively. Schematic illustrations of magnetic bubbles with the topological charge $Q=1$ in a (e) honeycomb lattice and (g) deformed hexagon lattice (the angles between nearest bonds are multiples of $\pi/2$). (f) and (h) The corresponding dispersions of the system shown in (e) and (g), respectively. Source: The figures are taken from Ref. \cite{KimPRL2017}.}
\label{Figure33}
\end{figure}

Equation \eqref{Eq34} is similar to the Haldane model for electrons in a honeycomb lattice \cite{HaldanePRL1988,KanePRL2005}, where the last term in the right-hand side represents the next-to-nearest hopping that breaks the time reversal symmetry, leading to the existence of a chiral edge state. Figures \ref{Figure33}(a) and \ref{Figure33}(c) plot the schematic illustrations of vortices in a honeycomb lattice with zigzag edges for $Q=1/2$ and $Q=-1/2$, respectively. The corresponding band structures for gyration modes are shown in Figs. \ref{Figure33}(b) and \ref{Figure33}(d). One can clearly identify the chiral edge states. Further, it can be seen that the chirality of the edge modes reverses when the topological charge of vortices switches the sign. In addition, if the vortices are replaced by magnetic bubbles [see Fig. \ref{Figure33}(e)], the chiral edge states still exist, as shown in Fig. \ref{Figure33}(f). However, if the last term in the right-hand side of Eq. \eqref{Eq34} vanishes, e.g. we set $\bar{\theta}_{jl}=\pm\pi/2$ [see Fig. \ref{Figure33}(g)], the time reversal symmetry of the system maintains and the dispersion relation is gapless, leading the disappearance of chiral edge states, see Fig. \ref{Figure33}(h).            

It is well known that the magnetic bubbles and skyrmions manifest an inertia in their gyration motion \cite{MakhfudzPRL2012,SchuttePRB2014,ButtnerNP2015,YangOE2018}. Therefore, if we want to develop an accurate theory about the coupled magnetic soliton (including vortex, bubble, and skyrmion) oscillations, a second-order inertial term and higher-order corrections should be taken into account. Next, we derive the generalized form of the Thiele’s equation from the original LLG equation. In terms of the tensor notation, the LLG equation can be written as
\begin{equation}\label{Eq35}
\frac{1}{\gamma}\dot{m}_{i}+\epsilon_{jki}m_{j}H_{k}^{\text{eff}}-\frac{\alpha}{\gamma}\epsilon_{jki}m_{j}\dot{m}_{k}=0,
\end{equation}
We can write an alternative form of the LLG equation
\begin{equation}\label{Eq36}
\mathbf{m}\times\mathbf{H}^{\text{t}}=0,
\end{equation}
with $\mathbf{H}^{\text{t}}$ the total effective magnetic field and $H_{i}^{\text{t}}=H_{i}^{1}+H_{i}^{2}+H_{i}^{3}$. Here $H_{i}^{1}=-\frac{1}{\gamma}\epsilon_{jki}m_{j}\dot{m}_{k}$, $H_{i}^{2}=-\frac{\alpha}{\gamma}\dot{m}_{i}$, and $H_{i}^{3}=H_{i}^{\text{eff}}$ are the gyroscopic equivalent field, the dissipative equivalent field, and the effective field, respectively. Besides, we can define the local force density $f_{i}^{\nu}=-M_{s}H_{j}^{\nu}\frac{\partial m_{j}}{\partial r_{i}}$, with $\nu=1,2,3$. For all $r_{i}$, we require the balance of forces $f_{i}^{1}+f_{i}^{2}+f_{i}^{3}=0$. Next, we assume that the steady-state magnetization depends on not only the position of the guiding center but also its velocity and acceleration, and we thus have $m_{j}=m_{j}[\mathbf{r}-\mathbf{R}(t),\dot{\mathbf{R}}(t),\ddot{\mathbf{R}}(t)]$, where $\mathbf{R}=(R_{x},R_{y})$ is the position of the magnetic soliton guiding center. We have 
\begin{equation}\label{Eq37}
\dot{m}_{j}=\frac{\partial m_{j}}{\partial R_{i}}\dot{R}_{i}+\frac{\partial m_{j}}{\partial \dot{R}_{i}}\ddot{R}_{i}+\frac{\partial m_{j}}{\partial \ddot{R}_{i}}\dddot{R}_{i}.
\end{equation}  
Then the different forces can be calculated. First of all, $f_{i}^{1}=g_{ij}^{1}\dot{R}_{j}+g_{ij}^{2}\ddot{R}_{j}+g_{ij}^{3}\dddot{R}_{j}$, with $g_{ij}^{1}=-\frac{M_{s}}{\gamma}\epsilon_{kpq}m_{k}\frac{\partial m_{p}}{\partial r_{i}}\frac{\partial m_{q}}{\partial R_{j}}$, $g_{ij}^{2}=-\frac{M_{s}}{\gamma}\epsilon_{kpq}m_{k}\frac{\partial m_{p}}{\partial r_{i}}\frac{\partial m_{q}}{\partial \dot{R}_{j}}$, and $g_{ij}^{3}=-\frac{M_{s}}{\gamma}\epsilon_{kpq}m_{k}\frac{\partial m_{p}}{\partial r_{i}}\frac{\partial m_{q}}{\partial \ddot{R}_{j}}$. Secondly, the dissipation term $f_{i}^{2}$ can be ignored due to the small damping. Finally, since the spins propagate in a steady manner, only externally applied fields contribute to the reversible energy force. Thus $f_{i}^{3}=f_{i}^{\text{ex}}=-M_{s}H_{j}^{\text{ex}}\frac{\partial m_{j}}{\partial r_{i}}$.

For the gyroscopic term, we can define vector $g_{t}^{\nu}=-\frac{1}{2}\epsilon_{tij}g_{ij}$, with $\nu=1,3$ such that $g_{ij}^{1}\dot{R}_{j}=\epsilon_{ijk}g_{i}^{1}\dot{R}_{j}$ and $g_{ij}^{3}\dddot{R}_{j}=\epsilon_{ijk}g_{i}^{3}\dddot{R}_{j}$. Then a new vector can be defined
\begin{equation}\label{Eq38}
\mathbf{G}=\int dV \mathbf{g}^{1}.
\end{equation}
It is obvious that $\mathbf{G}=(0,0,\mathcal{G})$ when a two-dimensional system is considered. Then we have
\begin{equation}\label{Eq39}
\mathcal{G}=-\frac{M_{s}}{\gamma}\int dV[\mathbf{m}\cdot(\partial_{r_{x}}\mathbf{m}\times\partial_{r_{y}}\mathbf{m} )]=-\frac{4\pi dQM_{s}}{\gamma}.
\end{equation}
Likewise, there exists a third-order gyroscopic term of the magnetic soliton as  
\begin{equation}\label{Eq40}
\mathbf{G}_{3}=(0,0,\mathcal{G}_{3})=\int dV \mathbf{g}^{3}.
\end{equation}
where
\begin{equation}\label{Eq41}
\mathcal{G}_{3}=\frac{dM_{s}}{2\gamma}\int dS[\mathbf{m}\cdot(\partial_{r_{x}}\mathbf{m}\times\partial_{\ddot{R}_{y}}\mathbf{m})-\mathbf{m}\cdot(\partial_{r_{y}}\mathbf{m}\times\partial_{\ddot{R}_{x}}\mathbf{m})].
\end{equation}
We can also define a mass tensor
\begin{equation}\label{Eq42}
\mathcal{M}_{ij}=\int dV g_{ij}^{2}=\frac{dM_{s}}{\gamma}\int dS[\mathbf{m}\cdot(\partial_{r_{i}}\mathbf{m}\times\partial_{\dot{R}_{j}}\mathbf{m})].
\end{equation}
Here we assume that $\mathcal{M}_{xx}=\mathcal{M}_{yy}=-\mathcal{M}$ and $\mathcal{M}_{xy}=\mathcal{M}_{yx}=0$, where
\begin{equation}\label{Eq43}
\mathcal{M}=\frac{dM_{s}}{\gamma}\int dS[\mathbf{m}\cdot(\partial_{r_{x}}\mathbf{m}\times\partial_{\dot{R}_{x}}\mathbf{m})].
\end{equation}
Eventually, the LLG equation is simplified to the generalized Thiele's form:
\begin{equation}\label{Eq44}
  \mathcal {G}_{3}\hat{z}\times\frac{d^{3}\textbf{U}_{j}}{dt^{3}}-\mathcal {M}\frac{d^{2}\textbf{U}_{j}}{dt^{2}}+\mathcal {G}\hat{z}\times \frac{d\textbf{U}_{j}}{dt}+\textbf{F}_{j}=0,
\end{equation}
where $\mathbf{U}_{j}= \mathbf R_{j} - \mathbf R_{j}^{0}$ is the displacement of the magnetic soliton center from its equilibrium position $\mathbf R_{j}^{0}$, $\mathcal{G}$ is gyroscopic parameter, $\mathcal {M}$ is the effective mass of the magnetic soliton \cite{MakhfudzPRL2012,SchuttePRB2014,ButtnerNP2015,YangOE2018}, $\mathcal {G}_{3}$ is the third-order non-Newtonian gyroscopic coefficient \cite{MertensPRB1997,IvanovJETPL2010,CherepovPRL2012}, and $\mathbf{F}_{j}=\int dV \mathbf{f}_{j}^{\text{ex}}$ is the external force. It should be noted that the above derivation is purely phenomenological and detailed microscopic mechanisms are still needed to clarify the origin of the solition mass and its non-Newtonian behavior. 
\begin{figure}[ptbh]
\begin{centering}
\includegraphics[width=0.85\textwidth]{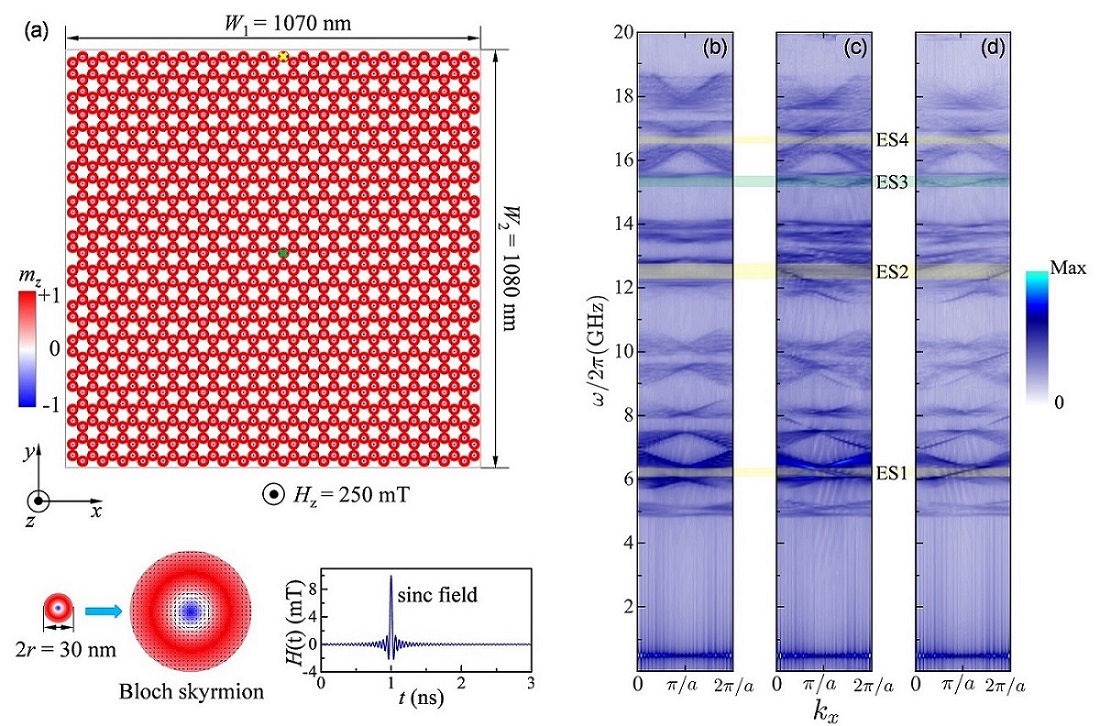}
\par\end{centering}
\caption{ (a) Illustration of the honeycomb lattice with size 1070 × 1080 nm$^{2}$, including 984 Bloch skyrmions with the topological charge $Q=+1$. A uniform magnetic field is applied along the $z$ axis to stabilize the skyrmions. Green and yellow crosses denote the positions of the driving fields. The band structure of skyrmion gyrations when the exciting field is in the film center (b) and at the film edge by evaluating the Fourier spectrum over the upper (c) and the lower (d) parts of the honeycomb lattice. Source: The figures are taken from Ref. \cite{LiPRB2018}.}
\label{Figure34}
\end{figure}

From the aspect of micromagnetic simulations, we consider a two-dimensional honeycomb lattice with 984 identical magnetic nanodisks to demonstrate the chiral edge states. Figure \ref{Figure34}(a) shows the sketch. Each nanodisk contains a Bloch-type skyrmion made of MnSi \cite{TomaselloSR2015} which supports the bulk Dzyakoshinskii-Moriya interaction \cite{YuN2010,SekiS2012}. Here, the distance between nearest neighbor disks is equal to the disk diameter, indicating that skyrmions can strongly interact with each other mediated by exchange coupling. Figure \ref{Figure34}(b) plots the band structure of the collective skyrmion oscillations when the exciting field (sinc-function magnetic field) locates in the lattice center [marked by green cross in Fig. \ref{Figure34}(a)]. It can be clearly seen that there are no bulk states in the gaps (the shaded areas). Interestingly, when the exciting field is located at the edge of the lattice [marked by yellow cross in Fig. \ref{Figure34}(a)], the band structures are obviously different. Figures \ref{Figure34}(c) and \ref{Figure34}(d) show the dispersion relations of the system by performing the FFT over the upper ($W_{2}/2<y<W_{2}$) and the lower ($0<y<W_{2}/2$) parts of the lattice, respectively. One can easily find four edge states appear in the gaps, labeled as ES1-ES4. Furthermore, by analysing the group velocity $d\omega/dk_{x}$ of these edge states, the chirality can be identified: ES1 and ES2 counterclockwise propagate, while ES4 behaves oppositely, ES3 shows a bidirectional propagation and it is thus non-chiral.     

\begin{figure}[ptbh]
\begin{centering}
\includegraphics[width=0.85\textwidth]{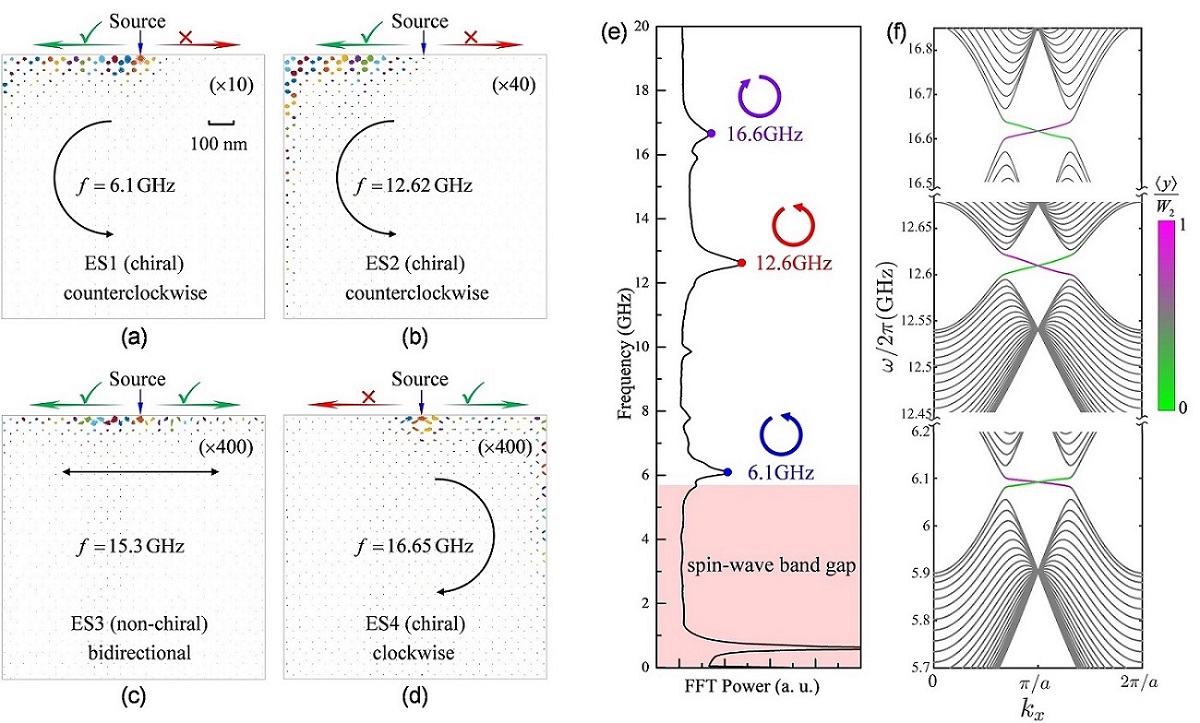}
\par\end{centering}
\caption{ Snapshot of the propagation of edge states with frequency (a) f = 6.1, (b) 12.62, (c) 15.3, and (d) 16.65 GHz at t = 40 ns. The oscillation amplitudes of the skyrmion guiding centers have been magnified in suitable multiples for better observation, as labeled in each figures. (a) Resonant spectrum of skyrmion gyrations when the exciting field is applied over the whole system. (b) Band structure of the system by solving Eq. (5) numerically. Source: The figures are taken from Ref. \cite{LiPRB2018}}
\label{Figure35}
\end{figure}

By plotting the propagation of gyration motion of skyrmions in real space for different modes, one can further confirm the chirality of these edge states. The excitation of edge modes can be realized by applying a sinusoidal field $\textbf{h}(t)=h_0\sin(2\pi ft)\hat{x}$ on one nanodisk at the top edge, indicated by the blue arrows in Figs. \ref{Figure35}(a)-\ref{Figure35}(d). Here, four representative frequencies are chosen to visualize the propagation for different edge states. One can clearly observe the unidirectional propagation of these modes with either a counterclockwise manner [ES1 and ES2 shown in Figs. \ref{Figure35}(a) and \ref{Figure35}(b), respectively] or a clockwise one [ES4 shown in Fig. \ref{Figure35}(d)]. In contrast, the propagation of ES3 is bidirectional, as shown in Fig. \ref{Figure35}(c). This non-chiral mode can be simply explained in terms of the Tamm-Shockley mechanism \cite{TammPZS1932,ShockleyPR1939} which predicts that the periodicity breaking of the crystal potential at the boundary can lead to the formation of a conducting surface/edge state. Furthermore, the propagation of the edge states is shown to be immune from the defects, while the Tamm-Shockley mode is not.  

The generalized Thiele’s equation [Eq. \eqref{Eq44}] can be used to theoretically explain the multiband chiral skyrmionic edge states observed above. After a straightforward derivation, Eq. \eqref{Eq44} can be simplified to the following form:
\begin{equation}\label{Eq45}
 \hat{\mathcal {D}}\psi_{j}=(\omega_{K}-3\xi^{2}/2\bar{\omega}_{K})\psi_{j}+\zeta\sum_{k\in\langle j\rangle}\psi_{k}
  -(\xi^{2}/2\bar{\omega}_{K})\sum_{l\in\langle\langle j\rangle\rangle}\text{cos}(2\bar{\theta}_{jl})\psi_{l} -i(\xi^{2}/2\bar{\omega}_{K})\sum_{l\in\langle\langle j\rangle\rangle}\text{sin}(2\bar{\theta}_{jl})\psi_{l},
\end{equation}
where the differential operator $\hat{\mathcal {D}}=i\omega_{3}\frac{d^{3}}{dt^{3}}-\omega_{M}\frac{d^{2}}{dt^{2}}+i\frac{d}{dt}$, $\omega_{3}=G_{3}/|G|$, $\omega_{M}=M/|G| $, $\omega_{K}=K/|G|$, and $\bar{\omega}_{K}=\omega_{K}-\omega^{2}_{0}\omega_{M}$ with $\omega_{0}$ satisfying the condition: $\omega_{K}=\omega_{M}\omega_{0}^{2}\pm(\omega_{3}\omega_{0}^{3}-\omega_{0})$, for clockwise and counterclockwise skyrmion gyrations, respectively. The key parameters $G_{3}$, $M$, and $K$ can be determined from micromagnetic simulations in a self-consistent manner \cite{LiPRB2018}. Figure \ref{Figure35}(e) plots the spectrum for collective skyrmion oscillations with three strong resonance peaks above the spin-wave band gap. By solving Eq. \eqref{Eq45} with the periodic boundary condition along $x$-axis and the zigzag termination at $y=0$ and $y=W_{2}$, the band structure of the skyrmion gyrations near the resonance frequencies $\omega_{0,1}/2\pi=6.1$ GHz, $\omega_{0,2}/2\pi=12.6$ GHz, and $\omega_{0,3}/2\pi=16.6$ GHz can be obtained, as shown in Fig. \ref{Figure35}(f). The average vertical position of the modes $\langle{y}\rangle\equiv\sum_{j}R^{0}_{j,y}|\textbf{U}_{j}|^2/\sum_{j}|\textbf{U}_{j}|^2$ are also shown in Fig. \ref{Figure35}(f), where $R^{0}_{j,y}$ is the equilibrium position of the skyrmion projected onto the $y$ axis, represented by different colors: closer to magenta indicating more localized at the upper edge. 

It is interesting to note that the chirality of ES4 is opposite to those of ES1 and ES2. This result can be understood by the sign change of Chern number. First of all, Eq. \eqref{Eq45} can be mapped into the Haldane model with the following Hamiltonian
\begin{equation}\label{Eq46}
 \mathcal{H}=\sum_{\langle jk\rangle}t_{1}\psi_{j}^{*}\psi_{k}+\sum_{\langle\langle jk\rangle\rangle}t_{2}\text{exp}(i2\pi\phi)\psi_{j}^{*}\psi_{k}+c.c.,
\end{equation}
with $t_{1}=\zeta$, $t_{2}=-\xi^{2}/2\bar{\omega}_{K}$, and $\phi=1/3$. According to Bloch's theorem, Eq. \eqref{Eq46} then can be block diagonalized in the $k$-space as
\begin{equation}\label{Eq47}
 \mathcal{H}=\sum_{i=0,x,y,z}h_{i}(\mathbf{k})\sigma_{i},
\end{equation}
with $h_{0}(\mathbf{k})=-2t_{2}\text{cos}(2\pi\phi)(\text{cos}2\sqrt{3}k_{x}r+2\text{cos}\sqrt{3}k_{x}r\text{cos}k_{y}r)$, $h_{x}(\mathbf{k})=-t_{1}(\text{cos}2k_{y}r+2\text{cos}k_{x}r\text{cos}k_{y}r)$, $h_{y}(\mathbf{k})=-t_{1}(\text{sin}2k_{y}r-2\text{cos}\sqrt{3}k_{x}r\text{sin}k_{y}r)$, and $h_{z}(\mathbf{k})=-2t_{2}\text{sin}(2\pi\phi)(\text{sin}2\sqrt{3}k_{x}r-2\text{sin}\sqrt{3}k_{x}r\text{cos}k_{y}r)$.
The two-band model Eq. \eqref{Eq47} is similar to the problem of a single spin under a $\mathbf{k}$-dependent magnetic field $h(\mathbf{k})$. The variation of the parameter $\mathbf{k}$ in the first Brillouin zone forms a closed surface $\mathbf{S}$. The Berry curvature then reads
\begin{equation}\label{Eq48}
 \mathbf{\Omega}_{\eta}=\eta\frac{\mathbf{e}_{h}}{2h^{2}},
\end{equation}
where $h=\sqrt{h_{x}^{2}+h_{y}^{2}+h_{z}^{2}}$ is the amplitude of the "external magnetic field" $\mathbf{h}=(h_{x},h_{y},h_{z})$ with the unit direction along $\mathbf{e}_{h}=\mathbf{h}/h$. Here $\eta=\pm$ labels the band index. The Chern number for the two-band system is given by
\begin{equation}\label{Eq49}
C_{\eta}=\eta\frac{1}{2\pi}\oint_{\mathbf{S}}\mathbf{\Omega}_{\eta}(h)\cdot d\mathbf{S}.
\end{equation}
It is therefore straightforward to calculate the Chern number $C_{\eta}=-\eta$ if the degeneracy point $h=0$ is within the closed surface $\mathbf{S}$, otherwise it is zero. If we consider the oscillation of skyrmions with an opposite handedness, namely, $\omega_{0}\rightarrow -\omega_{0}$, the Haldane Hamiltonian then becomes
\begin{equation}\label{Eq50}
 \mathcal{H}=\sum_{\langle jk\rangle}t_{1}\psi_{j}^{*}\psi_{k}-\sum_{\langle\langle jk\rangle\rangle}t_{2}\text{exp}(i2\pi\phi)\psi_{j}^{*}\psi_{k}+c.c.=\sum_{\langle jk\rangle}t_{1}\psi_{j}^{*}\psi_{k}+\sum_{\langle\langle jk\rangle\rangle}t_{2}\text{exp}[i(2\pi\phi+\pi)]\psi_{j}^{*}\psi_{k}+c.c..
\end{equation}
Effectively, the only change is an extra $\pi$ phase in the factor $2\pi\phi\rightarrow 2\pi\phi+\pi$. It is equivalent to implementing a time-reveral operation of Eq. \eqref{Eq47}: $\mathcal{H}(\mathbf{k})\rightarrow \mathcal{H}^{*}(-\mathbf{k})$, from which we obtain Eq. \eqref{Eq50}. Under this operation, the Berry curvature $\mathbf{\Omega}_{\eta}$ naturally changes its sign, and consequently, the chirality (Chern number) of the edge state in the band gap reverses.

\subsection{Higher-order topological phases}\label{section3.5}
In the previous sections, we have discussed the topological insulating phases in magnetic soliton system. All these phases, however, are first order by nature. In this section, we move on to the higher-order topological phase in magnetic soliton crystals, by presenting thorough calculations details in the breathing kagome \cite{Linpj2019}, honeycomb \cite{LiPRA2020}, and square \cite{LiPRB2020} lattices of magnetic vortex.

\subsubsection{Kagome lattice}\label{section3.5.1}
\begin{figure}[ptbh]
\begin{centering}
\includegraphics[width=0.9\textwidth]{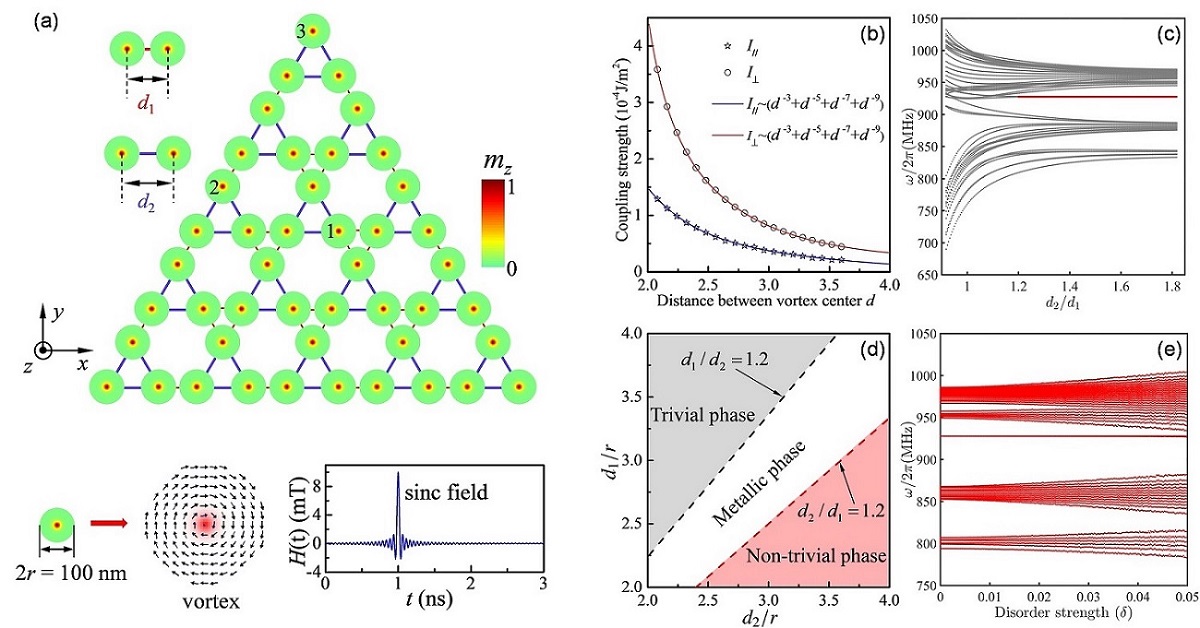}
\par\end{centering}
\caption{(a) Illustration of the triangle-shape breathing kagome lattice including 45 nanodisks of the vortex state, the radius and thickness of nanodisk are $r=50$ nm and $w=10$ nm, respectively. (b) Dependence of the coupling strength $I_{\parallel}$ and $I_{\perp}$ on the vortex-vortex distance $d$ (here $d$ is normalized by the disk radius $r$). Pentagrams and circles denote the micromagnetic simulation results and the solid curves represent the analytical fitting. (c) The eigenfrequencies of the system under the different ratios $d_{2}/d_{1}$ with the red segment labeling the corner state. (d) The phase diagram of the system. (e) The eigenfrequencies of the system under different disorder strengths. Source: The figures are taken from Ref. \cite{Linpj2019}}
\label{Figure36}
\end{figure}

We first consider a breathing kagome lattice of nanodisks with vortex states, as shown in Fig. \ref{Figure36}(a)  (the vortex topological charge $Q=1/2$), with $d_{1}$ and $d_{2}$ indicating the alternate distance between vortices. We start with the generalized Thiele's  equation derived in Section \ref{section3.4.2} to describe the collective dynamics of vortex lattice. Because there are two different bond distances ($d_{1}$ and $d_{2}$) in this model, the equation should be modified as:   
\begin{equation}\label{Eq51}
   \hat{\mathcal {D}}\psi_{j}=\omega_{K}\psi_{j}+\sum_{k\in\langle j\rangle, l}(\zeta_{l}\psi_{k}+\xi_{l} e^{i2\theta_{jk}}\psi^{*}_{k}),
\end{equation}
where $\hat{\mathcal {D}}=i\omega_{3}\frac{d^{3}}{dt^{3}}-\omega_{M}\frac{d^{2}}{dt^{2}}-i\frac{d}{dt}$, $\zeta_{l}=(I_{\parallel,l}-I_{\perp,l})/2|G| $, and $\xi_{l}=(I_{\parallel,l}+I_{\perp,l})/2|G|$, with $l=1$ (or $l=2$) representing the distance $d_{1}$ (or $d_{2}$) between the nearest neighbor vortices.

The coupling strengths $I_{\parallel}$ and $I_{\perp}$ strongly depend on the parameter $d$ ($d=d'/r$ with $d'$ the real distance between two vortices and $r$ being the radius of nanodisk) \cite{LeeJAP2011,SukhostavetsPRB2013,SinneckerJAP2014}. The analytical expression of $I_{\parallel}(d)$ and $I_{\perp}(d)$ are very important for calculating the spectra and the phase diagram. The eigenfrequencies of coupled two-vortex system can be expressed as $\omega=\omega_{0}\sqrt{(1\pm I_{\parallel}/K)(1\mp P_{1}P_{2}I_{\perp}/K)}$ \cite{LeeJAP2011}, where $P_{1}$ (or $P_{2}$) is either $+1$ or $-1$ depending on the vortex polarity. Therefore, once the frequencies of coupled modes for different combinations of vortex polarities ($P_{1}P_{2}=1$ or $P_{1}P_{2}=-1$) are determined from micromagnetic simulations, we can derive $I_{\parallel}$ and $I_{\perp}$ according to the dispersion relation. Moreover, it is found that coupling strengths $I_{\parallel}$ and $I_{\perp}$ are the functions of $d^{-3}$, $d^{-5}$, $d^{-7}$ and $d^{-9}$. Therefore, under the help of micromagnetic simulations for two vortices system with different combinations of vortex polarities, one can obtain the best fit of the numerical data \cite{Linpj2019}: $I_{\parallel}=\mu_{0}M_{s}^{2}r(-1.72064\times10^{-4}+4.13166\times10^{-2}/d^{3}-0.24639/d^{5}+1.21066/d^{7}-1.81836/d^{9}$) and $I_{\perp}=\mu_{0}M_{s}^{2}r(5.43158\times10^{-4}-4.34685\times10^{-2}/d^{3}+1.23778/d^{5}-6.48907/d^{7}+13.6422/d^{9}$), as shown in Fig. \ref{Figure36}(b), where the symbols and curves represent the simulation results and analytical formulas, respectively. In the calculations, the material parameters of Permalloy (Py: Ni$_{80}$Fe$_{20}$) \cite{VeltenAPL2017,YooAPL2012} were used. Therefore, we have $G=-3.0725\times10^{-13}$ J$\,$s$\,$rad$^{-1}$m$^{-2}$. Besides, the spring constant $K$, mass $M$, and non-Newtonian gyration $G_{3}$ can be obtained through the following relations \cite{IvanovJETPL2010,IvanovPRB1998}: $\omega_{0}= K/G$, $G_{3}\bar{\omega}^{2}=G$, $G_{3}(\Delta\omega+\omega_{0})=M$, $2\bar{\omega}=\omega_{1}+\omega_{2}$, and $\Delta\omega=|\omega_{2}-\omega_{1}|$, where $\omega_{0}$ is the frequency of the gyroscopic mode, $\omega_{1}$ and $\omega_{2}$ are the frequencies of the other two higher-order modes with opposite gyration handedness \cite{IvanovJETPL2010}. By analyzing the dynamics of a single vortex confined in the nanodisk \cite{Linpj2019}, we have: $K=1.8128\times10^{-3}$ J$\,$m$^{-2}$, $M=9.1224\times10^{-25}$ kg, and $G_{3}=-4.5571\times10^{-35}$J$\,$s$^{3}$rad$^{-3}$m$^{-2}$. With these parameters, by solving Eq. \eqref{Eq51} numerically, one can obtain the eigenfrequencies of the breathing kagome lattice for different values $d_{2}/d_{1}$, as shown in Fig. \ref{Figure36}(c), where $d_{1}$ is fixed to $2.2r$. By analyzing the spatial distribution of the eigenfunction for different modes, one can see that the second-order topological edge states (corner state) can exist only if $d_{2}/d_{1}>1.2$, this conclusion holds for different values of $d_{1}$. Furthermore, the complete phase diagram can be obtained by systematically changing $d_{1}$ and $d_{2}$, with results plotted in Fig. \ref{Figure36}(d). It can be seen that the boundary separating topologically non-trivial and metallic phases lies in $d_{2}/d_{1}=1.2$, while topologically trivial and metallic phases are separated by $d_{1}/d_{2}=1.2$. When $d_{2}/d_{1}>1.2$, the system is topologically non-trivial and can support second-order topological corner states.    

Topological corner states have the property of being immune from the bulk disorder. Figure \ref{Figure36}(e) plots the eigenfrequencies of the triangle-shape breathing kagome lattice of vortices under different strengths of disorder, with the geometric parameters $d_{1}=2.08r$ and $d_{2}=3.60r$ ($d_{2}/d_{1}=1.73>1.2$). The disorder is introduced by assuming the resonant frequency $\omega_{K}$ undergoes a random shift, i.e., $\omega_{0} \rightarrow \omega_{0}+\delta Z\omega_{0}$, where $\delta$ indicates the strength of the disorder and $Z$ is a uniformly distributed random number between $-1$ to $1$. It can be seen from Fig. \ref{Figure36}(e) that with the increasing of the disorder strength, the spectra for both edge and bulk states are significantly modified, while the corner states are quite robust.  
\begin{figure}[ptbh]
\begin{centering}
\includegraphics[width=0.9\textwidth]{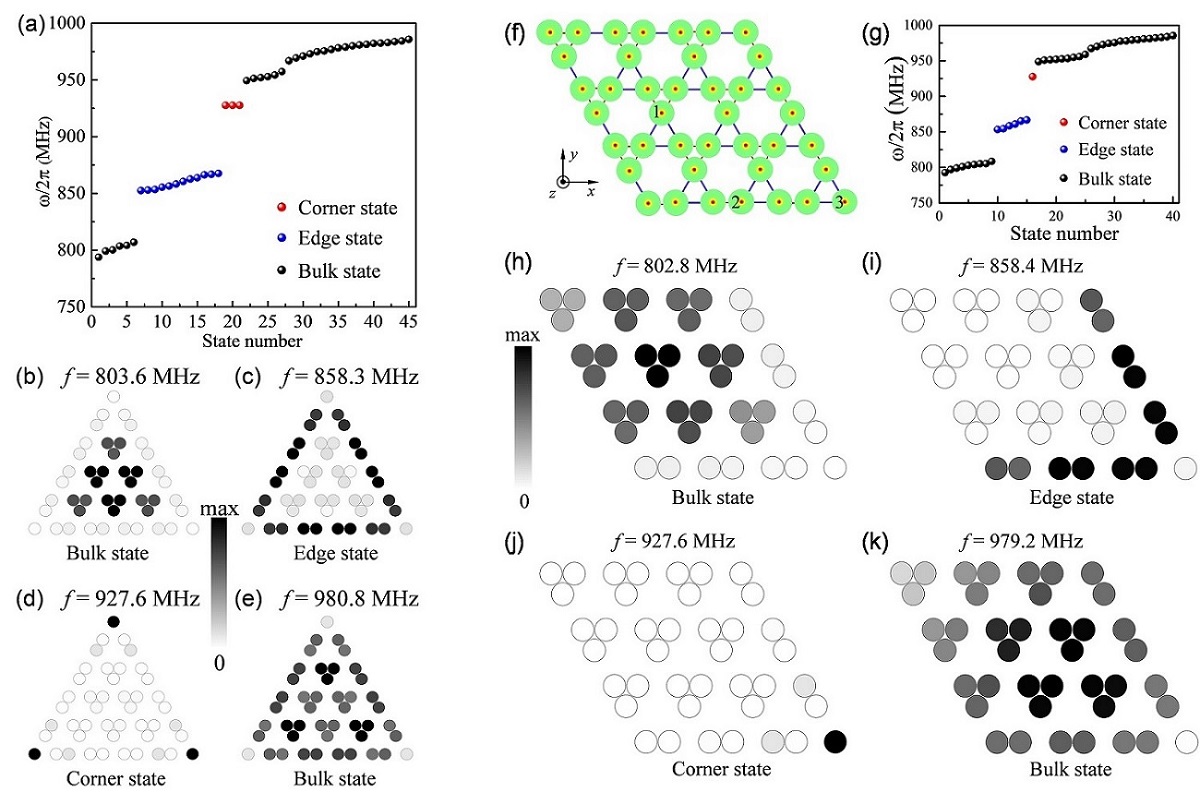}
\par\end{centering}
\caption{ (a) The eigenfrequencies of triangle-shape kagome vortex lattice with $d_{1}=2.08r$ and $d_{2}=3.60r$. The spatial distribution of vortex gyrations for the bulk [(b) and (e)], edge (c), and corner (d) states. The sketch for parallelogram-shaped breathing kagome lattice of vortices. (b) Numerically computed eigenfrequencies for parallelogram-shaped system. The spatial distribution of vortices oscillation for the bulk [(c) and (f)], edge (d), and corner (e) states. Source: The figures are taken from Ref. \cite{Linpj2019}}
\label{Figure37}
\end{figure}

The same geometric parameters as Fig. \ref{Figure36}(e) are chosen to visualize the different modes (including corner , edge, and bulk states). The eigenfrequencies and eigenmodes of the system are plotted in Figs. \ref{Figure37}(a) and \ref{Figure37}(b)-(e). It is found that there are three degenerate modes with the frequency equal to 927.6 MHz, represented by red balls. These modes are indeed second-order topological states (corner states) with oscillations being highly localized at the three corners; see Fig. \ref{Figure37}(d). The edge states are also identified, denoted by blue balls in Fig. \ref{Figure37}(a). The spatial distribution of edge oscillations are confined on three edges, as shown in Fig. \ref{Figure37}(c). However, these edge modes are Tamm-Shockley type \cite{TammPZS1932,ShockleyPR1939}, not chiral, which was confirmed by micromagnetic simulations \cite{Linpj2019}. Bulk modes are plotted in Figs. \ref{Figure37}(b) and \ref{Figure37}(e), where corners do not participate in the oscillations.

The other type of breathing kagome lattice of vortices (parallelogram-shape) also supports the corner states, with the sketch plotted in Fig. \ref{Figure37}(f). Here, the same parameters as those in the triangle-shape lattice are adopted. Figure \ref{Figure37}(g) shows the eigenfrequencies of system. Interestingly, it can be seen that there is only one corner state, represented by the red ball. Edge and bulk states are also observed, denoted by blue and black balls, respectively. The spatial distribution of vortices oscillation for different modes are shown in Figs. \ref{Figure37}(h)-\ref{Figure37}(k). From Fig. \ref{Figure37}(j), one can clearly see that the oscillations for corner state are confined to one acute angle and the vortex at the position of two obtuse angles hardly oscillates. The spatial distribution of vortex gyration for edge and bulk states are plotted in Figs. \ref{Figure37}(i), \ref{Figure37}(h), and \ref{Figure37}(k), respectively. Further, the robustness of the corner states are also confirmed \cite{Linpj2019}. 
\begin{figure}[ptbh]
\begin{centering}
\includegraphics[width=0.9\textwidth]{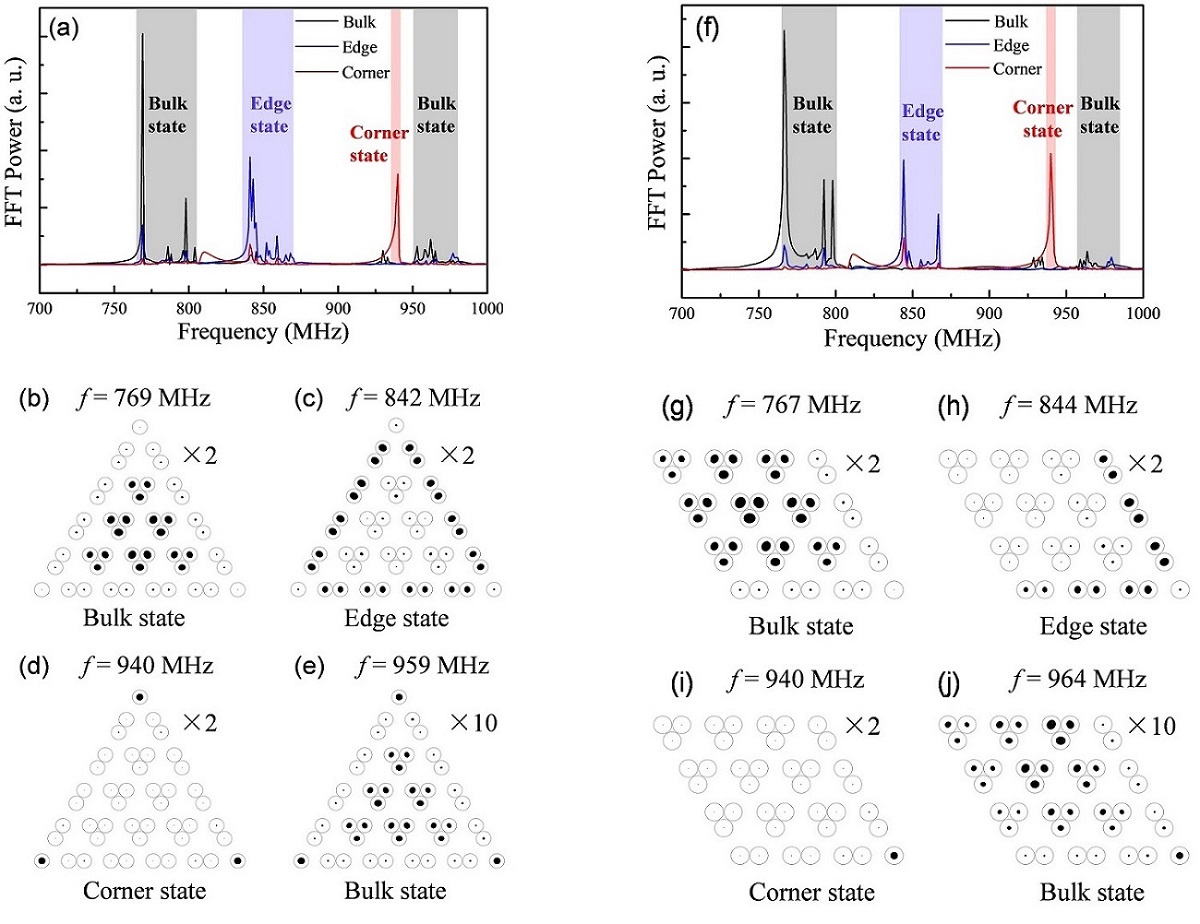}
\par\end{centering}
\caption{ Micromagnetic simulation of excitations in triangle-shape structure: (a) The temporal Fourier spectrum of the vortex oscillations at different positions. The spatial distribution of oscillation amplitude under the exciting field of various frequencies, 769 MHz (b), 842 MHz (c), 940 MHz (d), and 959 MHz (e). Micromagnetic simulation of excitations in parallelogram-shape structure: (f) The temporal Fourier spectrum of the vortex oscillations at different positions. The spatial distribution of oscillation amplitude under the exciting field with different frequencies, 767 MHz (b), 844 MHz (c), 940 MHz (d), and 964 MHz (e). The oscillation amplitudes of the vortices centers have been magnified in suitable multiples for better observation. Source: The figures are taken from Ref. \cite{Linpj2019}}
\label{Figure38}
\end{figure}

The higher-order topological properties can be interpreted in terms of the bulk topological index, i.e., the polarization \cite{King-SmithPRB1993,VanderbiltPRB1993}:
\begin{equation}\label{Eq52}
 P_{j}=\frac{1}{S}\int\!\!\!\int_{\rm{BZ}}A_{j}d^{2}k,
\end{equation}
where $S$ is the area of the first Brillouin zone, $A_{j}=-i\langle\psi|\partial k_{j}|\psi\rangle$ is Berry connection with $j=x,y$, and $\psi$ is the wave function for the lowest band. It is shown that $(P_{x},P_{y})=(0.499,0.288)$ for $d_{1}=2.08r$ and $d_{2}=3.60r$ and $(P_{x},P_{y})=(0.032,0.047)$ for $d_{1}=3r$ and $d_{2}=2.1r$. The former corresponds to the topological insulating phase while the latter is for the trivial phase. Theoretically, for breathing kagome lattice, the polarization $(P_{x}, P_{y})$ is identical to the Wannier center, which is restricted to two positions for insulating phases. If Wannier center coincides with (0, 0), the system is in trivial insulating phase and no topological edge state exists. Higher-order topological corner states emerge when the Wannier center lies at (1/2, 1/2$\sqrt{3}$) \cite{EzawaPRL2018,XueNM2019}.

Micromagnetic simulations can be used to verify the theoretical predictions of corner states. The triangle-shape and parallelogram-shape breathing kagome lattice of vortices are considered, as shown in Fig. \ref{Figure36}(a) and Fig. \ref{Figure37}(f), with the same geometric parameters as those in Fig. \ref{Figure37}(a) and Fig. \ref{Figure37}(g), respectively. Figure \ref{Figure38}(a) shows the temporal Fourier spectra of the vortex oscillations at different positions. One can immediately see that, near the frequency of 940 MHz, the spectrum for the corner has a very strong peak, which does not happen for the edge and bulk. It can be inferred that this is the corner-state band with oscillations localized only at three corners. Similarly, one can identify the frequency range that allows the bulk and edge states, as shown by shaded area with different colors in Fig. \ref{Figure38}(a). Four representative frequencies are chosen to visualize the spatial distribution of vortex oscillations for different modes: 940 MHz for the corner state, 842 MHz for the edge state, and both 769 MHz and 959 MHz for bulk states, and then stimulate their dynamics by a sinusoidal magnetic field $\textbf{h}(t)=h_0\sin(2\pi ft)\hat{x}$ with $h_{0}=0.1$ mT to the whole system for 100 ns. Figures \ref{Figure38}(b)-\ref{Figure38}(e) plot the spatial distribution of oscillation amplitude. One can clearly see the corner state in Fig. \ref{Figure38}(d). Spatial distribution of vortices motion for bulk and edge states are shown in Figs. \ref{Figure38}(b) and \ref{Figure38}(c), respectively. Figure \ref{Figure35}(e) plot the hybridized mode between the bulk and corner modes, since their frequencies are very close to each other, as shown in Figs. \ref{Figure37}(a) and \ref{Figure38}(a). The simulations of parallelogram-shaped lattice show similar results to triangle-shaped lattice. The spectra are shown in Fig. \ref{Figure38}(f). Shaded area with different colors denote different modes. The spatial distribution of oscillation amplitude is plotted in Figs. \ref{Figure38}(g)-\ref{Figure38}(j). Figure \ref{Figure38}(i) shows only one corner state at only one (bottom-right) acute angle. Spatial distribution of vortices gyration for bulk and edge states are shown in Figs. \ref{Figure38}(g) and \ref{Figure38}(h), respectively. Interestingly, the hybridization between bulk mode and corner mode occurs as well in parallelogram-shaped lattice, see Fig. \ref{Figure38}(j).

In recent years, nano-oscillators in magnetic systems have attracted great attention for potential applications. However, the working frequency of these oscillators is very sensitive to external disturbances. If the HOTI phase (corner state) is used, vortex-based nano-oscillators should have extraordinary stability against defects and disorder and should therefore have broader prospects for application as topological microwave sources. 

In condensed matter physics, besides kagome lattice, the topological properties in honeycomb lattice are also studied extensively. The rich topological phases (including first- and second-order) are confirmed in breathing honeycomb lattice of vortices, which will be introduced in the next section.

\subsubsection{Honeycomb lattice}\label{section3.5.2}
It is well known that the perfect graphene lattice has a gapless band structure with Dirac cones in momentum space \cite{NetoRMP2009}. When spatially periodic magnetic flux \cite{HaldanePRL1988} or spin-orbit coupling \cite{YaoPRB2007} are introduced, a gap will open at the Dirac point, leading to a FOTI. Interestingly, it have been shown that the gap opening and closing can be realized by tuning the intercellular and intracellular bond distances in photonic \cite{NohNP2018} and elastic \cite{FanPRL2019} honeycomb lattices, in which the HOTI appears. In this section, we show that the higher-order topological insulating phase do exist in a breathing honeycomb lattice of vortices. 
\begin{figure}[ptbh]
\begin{centering}
\includegraphics[width=0.95\textwidth]{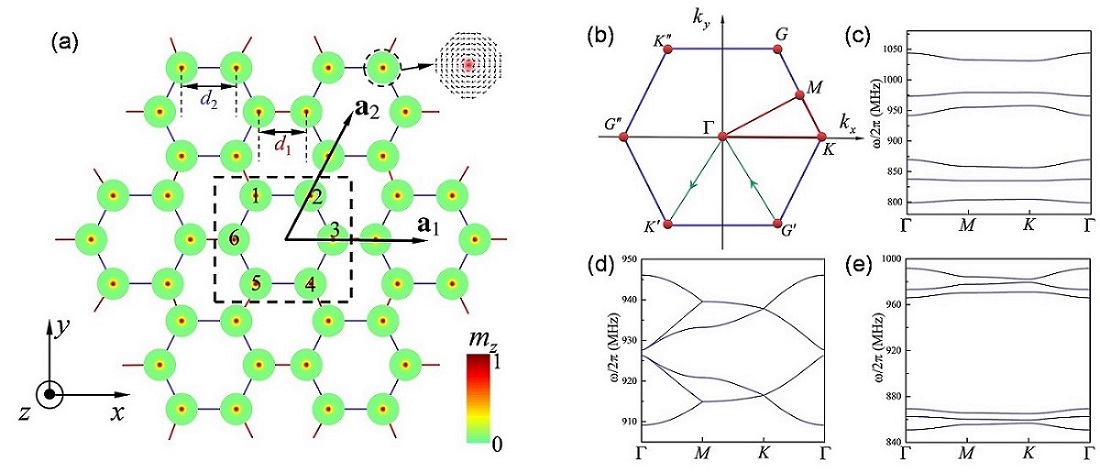}
\par\end{centering}
\caption{ Illustration of the breathing honeycomb lattice of magnetic vortices, with $d_{1}$ and $d_{2}$ denoting the alternating lengths of intercellular and intracellular bonds, respectively. The radius of each nanodisk is $r=50$ nm, and the thickness is $w=10$ nm. (b) The first Brillouin zone of the breathing honeycomb lattice with high-symmetry points. Band structures along the loop $\Gamma$-$M$-$K$-$\Gamma$ for different lattice parameters: $d_{1}=3.6r, d_{2}=2.08r$ (c), $d_{1}=d_{2}=3.6r$ (d), and $d_{1}=2.08r, d_{2}=3.6r$ (e). Source: The figures are taken from Ref. \cite{LiPRA2020}}
\label{Figure39}
\end{figure}

Figure \ref{Figure39}(a) shows a breathing honeycomb lattice of magnetic nanodisks with vortex states. The corresponding eigenvalue equation of the system can be obtained as follow (by using the similar method as mentioned above): 
\begin{equation}\label{Eq53}
%\begin{multline}
\hat{\mathcal {D}}\psi_{j}=(\omega_{K}-\frac{\xi^{2}_{1}+2\xi^{2}_{2}}{2\bar{\omega}_{K}})\psi_{j}+\zeta_{1}\sum_{k\in\langle j_{1}\rangle}\psi_{k}+\zeta_{2}\sum_{k\in\langle j_{2}\rangle}\psi_{k}\\
   -\frac{\xi_{1}\xi_{2}}{2\bar{\omega}_{K}}\sum_{s\in\langle\langle j_{1}\rangle\rangle}e^{i2\bar{\theta}_{js}}\psi_{s}-\frac{\xi^{2}_{2}}{2\bar{\omega}_{K}}\sum_{s\in\langle\langle j_{2}\rangle\rangle}e^{i2\bar{\theta}_{js}}\psi_{s},
%\end{multline} 
\end{equation}
with $\omega_{K}=\mathcal{K}/|\mathcal{G}|$ and $\bar{\omega}_{K}=\omega_{K}-\omega_{K}^{2}\omega_{M}$. For an infinite lattice, with the dashed black rectangle indicating the unit cell, as shown in Fig. \ref{Figure39}(a), $\textbf{a}_{1}=a\hat{x}$ and $\textbf{a}_{2}=\frac{1}{2}a\hat{x}+\frac{\sqrt{3}}{2}a\hat{y}$ are two basis vectors of the crystal, with $a=d_{1}+2d_{2}$. The band structure of system can be calculated by diagonalizing the Hamiltonian, 
\begin{equation}\label{Eq54}
 \mathcal {H}=\left(
 \begin{matrix}
   Q_{0} & \zeta_{2} & Q_{1}& Q_{2}& Q_{3}& \zeta_{2} \\
   \zeta_{2} & Q_{0} & \zeta_{2}& Q_{4}& \zeta_{1}\exp(i\textbf{k}\cdot\textbf{a}_{2})& Q_{5} \\
   Q_{1}^{*} & \zeta_{2} & Q_{0}& \zeta_{2}& Q_{6}& \zeta_{1}\exp(i\textbf{k}\cdot\textbf{a}_{1})\\
   Q_{2}^{*} & Q_{4}^{*} & \zeta_{2}& Q_{0}& \zeta_{2}& Q_{7}\\
   Q_{3}^{*} & \zeta_{1}\exp(-i\textbf{k}\cdot\textbf{a}_{2})& Q_{6}^{*} & \zeta_{2}& Q_{0}& \zeta_{2}\\
   \zeta_{2} & Q_{5}^{*} & \zeta_{1}\exp(-i\textbf{k}\cdot\textbf{a}_{1}) & Q_{7}^{*}& \zeta_{2}& Q_{0}
  \end{matrix}
  \right),
\end{equation}
where the elements can be expressed explicitly as. 
\begin{equation}\label{Eq55}
\begin{aligned}
Q_{0}&=\omega_{K}-\frac{\xi_{1}^{2}+2\xi_{2}^{2}}{2\bar{\omega}_{K}}, \\
Q_{1}&=-\frac{\xi_{1}\xi_{2}}{2\bar{\omega}_{K}}\exp(i\frac{2\pi}{3})\Big\{\exp[i\textbf{k}\cdot(\textbf{a}_{2}-\textbf{a}_{1})]+\exp(-i\textbf{k}\cdot\textbf{a}_{1})\Big\}-\frac{\xi^{2}_{2}}{2\bar{\omega}_{K}}\exp(i\frac{2\pi}{3}), \\
Q_{2}&=-\frac{\xi_{1}\xi_{2}}{2\bar{\omega}_{K}}\exp(-i\frac{2\pi}{3})\Big\{\exp[i\textbf{k}\cdot(\textbf{a}_{2}-\textbf{a}_{1})]+\exp(i\textbf{k}\cdot\textbf{a}_{2})\Big\}-\frac{\xi^{2}_{2}}{2\bar{\omega}_{K}}\exp(-i\frac{2\pi}{3}),\\
Q_{3}&=-\frac{\xi_{1}\xi_{2}}{2\bar{\omega}_{K}}\exp(i\frac{2\pi}{3})\Big\{\exp[i\textbf{k}\cdot(\textbf{a}_{2}-\textbf{a}_{1})]+\exp(i\textbf{k}\cdot\textbf{a}_{2})\Big\}-\frac{\xi^{2}_{2}}{2\bar{\omega}_{K}}\exp(i\frac{2\pi}{3}),\\
Q_{4}&=-\frac{\xi_{1}\xi_{2}}{2\bar{\omega}_{K}}\exp(-i\frac{2\pi}{3})[\exp(i\textbf{k}\cdot\textbf{a}_{2})+\exp(i\textbf{k}\cdot\textbf{a}_{1})]-\frac{\xi^{2}_{2}}{2\bar{\omega}_{K}}\exp(-i\frac{2\pi}{3}),\\
Q_{5}&=-\frac{\xi_{1}\xi_{2}}{2\bar{\omega}_{K}}\exp(i\frac{2\pi}{3})[\exp(i\textbf{k}\cdot\textbf{a}_{2})+\exp(i\textbf{k}\cdot\textbf{a}_{1})]-\frac{\xi^{2}_{2}}{2\bar{\omega}_{K}}\exp(i\frac{2\pi}{3}), \\
Q_{6}&=-\frac{\xi_{1}\xi_{2}}{2\bar{\omega}_{K}}\exp(i\frac{2\pi}{3})\Big\{\exp[i\textbf{k}\cdot(\textbf{a}_{1}-\textbf{a}_{2})]+\exp(i\textbf{k}\cdot\textbf{a}_{1})\Big\}-\frac{\xi^{2}_{2}}{2\bar{\omega}_{K}}\exp(i\frac{2\pi}{3}).
\end{aligned}
\end{equation}
The topological invariant Chern number is usually adopted to judge whether the system is in the FOTI phase \cite{WangPRB2017,AvronPRL1983}. However, to determine whether the system allows the HOTI phase, another different topological invariant should be considered. In addition to the bulk polarization, it has been shown that $\mathbb{Z}_{Q}$ Berry phase \cite{ZakPRL1989,KariyadoPRL2018,HatsugaiEPL2011,WakaoPRB2020,MizoguchiJPSJ2019,ArakiPRR2020,KudoPRL2019} is a powerful tool to characterize the HOTI.

In the presence of six-fold rotational ($C_{6}$) symmetry, the $\mathbb{Z}_{6}$ Berry phase is defined as follow:
\begin{equation}\label{Eq56}
{\theta}=\int_{L_{1}}\rm{Tr}[\mathbf{A}(\mathbf{k})]\cdot d\mathbf{k}\ \  (\rm{mod}\ 2\pi),
\end{equation}
where $\textbf{A}(\textbf{k})$ is the Berry connection:
\begin{equation}\label{Eq57}
 \mathbf{A}(\mathbf{k})=i\Psi^{\dag}(\mathbf{k})\frac{\partial}{\partial\mathbf{k}}\Psi(\mathbf{k}).
\end{equation}
Here, $\Psi(\textbf{k})=[\phi_{1}(\textbf{k})$,$\phi_{2}(\textbf{k})$,$\phi_{3}(\textbf{k})]$ is the 6 $\times$ 3 matrix composed of the eigenvectors of Eq. \eqref{Eq54} for the lowest three bands. $L_{1}$ is an integral path in momentum space $G^{\prime}\rightarrow \Gamma\rightarrow K^{\prime}$; see the green line segment in Fig. \ref{Figure39}(b). In addition, the six high-symmetry points $G$, $K$, $G^{\prime}$, $K^{\prime}$, $G^{\prime\prime}$, and $K^{\prime\prime}$ are equivalent, because of the $C_{6}$ symmetry. Therefore, there are other five equivalent integral paths ($L_{2}: K^{\prime}\rightarrow \Gamma\rightarrow G^{\prime\prime}$, $L_{3}: G^{\prime\prime}\rightarrow \Gamma\rightarrow K^{\prime\prime}$, $L_{4}: K^{\prime\prime}\rightarrow \Gamma\rightarrow G$, $L_{5}: G\rightarrow \Gamma\rightarrow K$, and $L_{6}: K\rightarrow \Gamma\rightarrow G^{\prime}$) leading to the identical $\theta$. It is also straightforward to see that the integral along the path $L_{1}+L_{2}+L_{3}+L_{4}+L_{5}+L_{6}$ vanishes. Thus, the $\mathbb{Z}_{6}$ Berry phase must be quantized as $\theta=\frac{2n\pi}{6}\ $ $(n=0,1,2,3,4,5)$. By simultaneously quantifying the Chern number $\mathcal{C}$ and the $\mathbb{Z}_{6}$ Berry phase $\theta$, the topological phases and their transition can be determined accurately.

Figures \ref{Figure39}(c)-\ref{Figure39}(e) show the bulk band structures under different lattice parameters. For $d_{1}=d_{2}=3.6r$ [see Fig. \ref{Figure39}(d)], the highest three bands and the lowest three bands merged separately, leaving a next-nearest hopping-induced gap centered at 927 MHz. In this case, the FOTI phase was anticipated \cite{KimPRL2017,LiPRB2018}. However, the six bands are separated from each other when considering the parameters $d_{1}\neq d_{2}$ [see Figs. \ref{Figure39}(c) and \ref{Figure39}(e)], indicating that the system is in the insulating state. These insulating phases and the phase transition point can be further distinguished by calculating Chern number and $\mathbb{Z}_{6}$ Berry phase.
\begin{figure}[ptbh]
\begin{centering}
\includegraphics[width=0.95\textwidth]{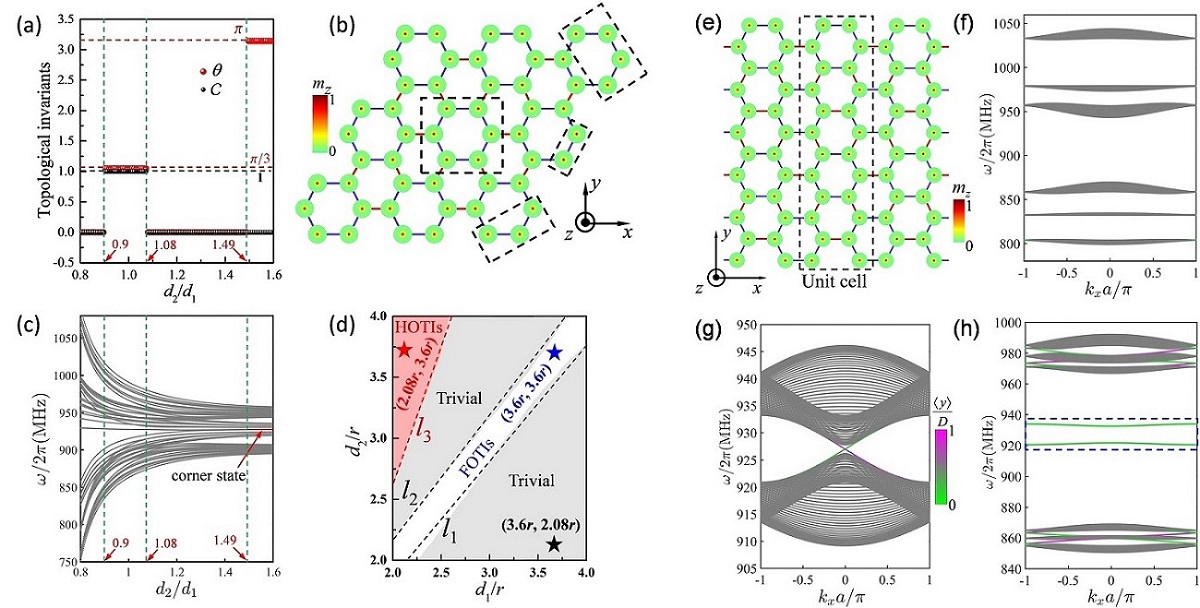}
\par\end{centering}
\caption{ (a) Dependence of the topological invariants Chern number and $\mathbb{Z}_{6}$ Berry phase on the ratio $d_{2}/d_{1}$ when $d_{1}$ is fixed at $2.5r$. (b) Schematic plot of the parallelogram-shaped vortex lattice with armchair edges. (c) Eigenfrequencies of collective vortex gyration under different ratios $d_{2}/d_{1}$ with the red segment denoting the corner state phase. (d) Phase diagram of the system. (e) Nanoribbon with armchair edges. Band dispersions for different parameters (f) $d_{1}=3.6r, d_{2}=2.08r$; (g) $d_{1}=d_{2}=3.6r$; (h) $d_{1}=2.08r, d_{2}=3.6r$. The dashed blue frame in (h) indicates the band of non-chiral edge states. $D$ is the width of the nanoribbon. Source: The figures are taken from Ref. \cite{LiPRA2020}}
\label{Figure40}
\end{figure}

Figure \ref{Figure40}(a) shows the dependence of the Chern number ($\mathcal{C}$) and the $\mathbb{Z}_{6}$ Berry phase ($\theta$) on the parameter $d_{2}/d_{1}$. Here the material parameters of Py (Ni$_{80}$Fe$_{20}$) \cite{VeltenAPL2017,YooAPL2012} are used and $d_{1}$ is fixed to $2.5r$. In addition, the eigenfrequencies for a parallelogram-shaped [see Fig. \ref{Figure40}(b)] structure are also shown in Fig. \ref{Figure40}(c). One can see that the system is in the trivial phase when $d_{2}/d_{1}<0.9$ and $1.08<d_{2}/d_{1}<1.49$, in the FOTI phase when $0.9<d_{2}/d_{1}<1.08$, and in the HOTI phase when $d_{2}/d_{1}>1.49$. The complete phase diagram of system can be obtained by systematically changing $d_{1}$ and $d_{2}$, with the results plotted in Fig. \ref{Figure40}(d). The boundary for the phase transition between trivial and FOTI phases depends only weakly on the choice of the absolute values of $d_{1}$ and $d_{2}$ but is (almost) solely determined by their ratio, as indicated by dashed black lines ($l_{1}: d_{2}/d_{1}=0.94$ and $l_{2}: d_{2}/d_{1}=1.05$) in the figure. While the boundary for the phase transition between trivial and HOTI phases is a linear function $l_{3}: d_{2}=2.24d_{1}-1.88$. From Eq. \eqref{Eq53}, we can see that the topological charge of the vortex has no influence on higher-order topology for the reason that the sign of topological charge just determines the direction (clockwise or anti-clockwise) of gyration. However, it indeed can affect the chiral edge state (first-order topology). Namely the chirality of edge state will be reversed if the topological charge changes.

The existence of symmetry-protected states on boundaries is the hallmark of a topological insulating phase. Figures. \ref{Figure40}(f)-\ref{Figure40}(h) show the energy spectrum of the ribbon configuration with armchair edges [see Fig. \ref{Figure40}(e)] for different choices of $d_{1}$ and $d_{2}$. For $d_{1}=3.6r$ and $d_{2}=2.08r$, the system is in the trivial phase without any topological edge mode [see Fig. \ref{Figure40}(f)]. For $d_{1}=d_{2}=3.6r$, the lattice considered is identical to a magnetic texture version of graphene. In contrast to the gapless band structure for perfect graphene nanoribbons, the imaginary second-nearest hopping term opens a gap at the Dirac point and supports a topologically protected first-order chiral edge state \cite{KimPRL2017,LiPRB2018}. For $d_{1}=2.08r$ and $d_{2}=3.6r$, one can clearly see two distinct edge bands, in addition to bulk ones, as shown in Fig. \ref{Figure40}(h). These localized modes are actually not topological because they maintain the bidirectional propagation nature, which is justified by the fact that the wave group-velocity $d\omega/dk_{x}$ can be either positive or negative at different $k_{x}$ points. However, the higher-order topological corner states will emerge around these edge bands when the system is decreased to be finite in both dimensions.

A parallelogram-shaped vortex lattice is considered to visualize the second-order corner states, where $d_{1}=2.08r$ and $d_{2}=3.6r$. 
\begin{figure}[ptbh]
\begin{centering}
\includegraphics[width=0.9\textwidth]{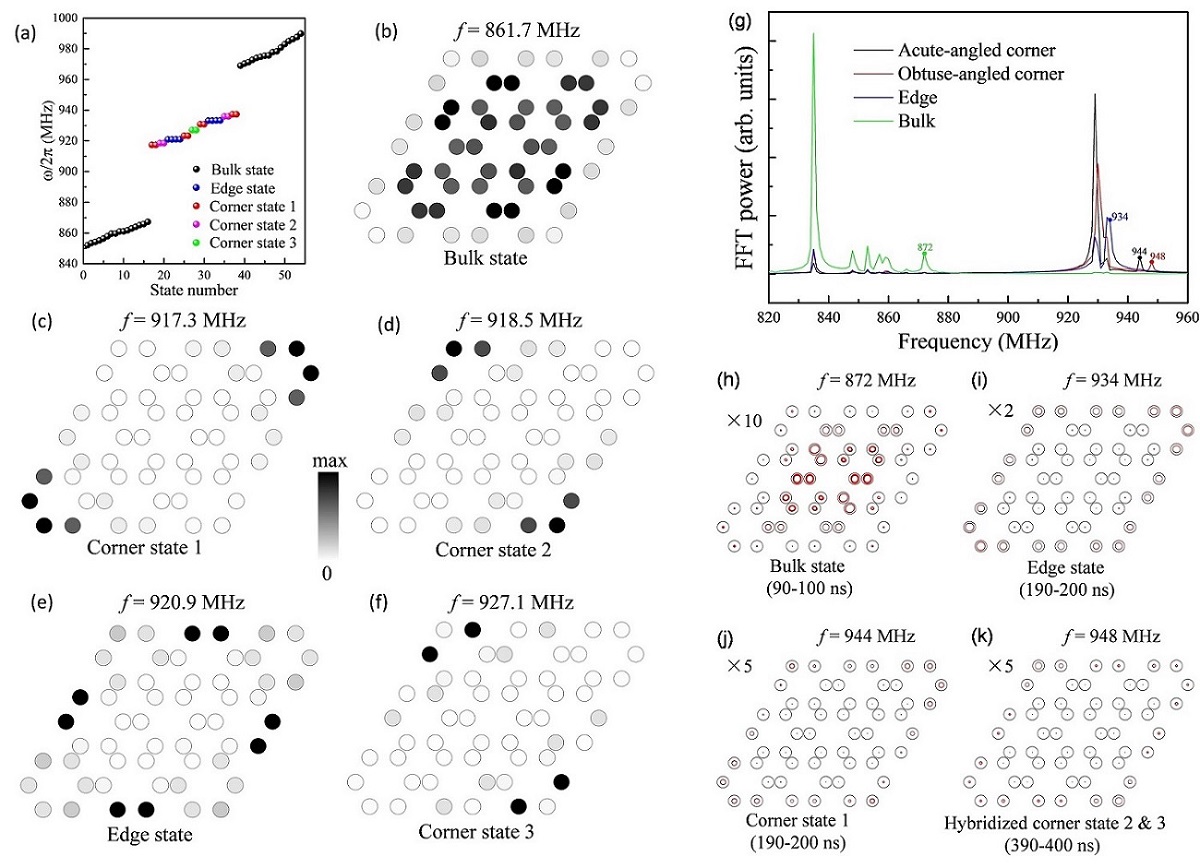}
\par\end{centering}
\caption{ (a) Eigenfrequencies of the finite system with parameters $d_{1}=2.08r$ and $d_{2}=3.6r$ for the parallelogram-shaped structure. The spatial distribution of vortex gyrations for the bulk (b), corner (c, d and f), and edge (e) states with five representative frequencies. (g) The spectra of the vortex oscillations at different positions. The gyration path for all vortices under excitation fields with different frequencies, 872 MHz (h), 934 MHz (i), 944 MHz (j), and 948 MHz (k). Source: The figures are taken from Ref. \cite{LiPRA2020}}
\label{Figure41}
\end{figure}
From the spectrum [see Fig. \ref{Figure41}(a)], one can clearly see that there exist a few degenerate modes in the band gap. The spatial distribution of vortex gyrations are plotted for each mode in Figs. \ref{Figure41}(b)-\ref{Figure41}(f) to distinguish these states. Three types of corner states are confirmed, all of which have oscillations highly localized at obtuse-angled or acute-angled corners [see Figs. \ref{Figure41}(c), \ref{Figure41}(d), and \ref{Figure41}(f)]. The bulk and edge modes are also identified, as shown in Figs. \ref{Figure41}(b) and \ref{Figure41}(e), respectively. Further, it have been confirmed \cite{LiPRA2020} when the moderate defects and disorder are introduced into the system, corner state 3 at the obtuse-angled corner is well confined around 927 MHz, which means that this corner state is suitably immune from external frustrations. This feature is due to the topological protection from the generalized chiral symmetry \cite{LiPRA2020}. However, the frequencies of other corner modes have obvious shifts, revealing that these crystalline-symmetry-induced modes are sensitive to disorder. The origin of the edge state is attributed to the so-called Tamm-Shockley mechanism \cite{TammPZS1932,ShockleyPR1939}.   

To verify theoretical predictions, one can implement full micromagnetic simulations. Here, the parallelogram-shaped breathing honeycomb lattice of magnetic vortices with an armchair edge is considered, as shown in Fig. \ref{Figure40}(b). Figure \ref{Figure41}(g) shows the temporal Fourier spectra of the vortex oscillations at different positions. It can be seen that around the frequency of 944 MHz (948 MHz), the spectra for acute-angled corner (obtuse-angled corner) have an obvious peak, which does not happen for the spectra for edge and bulk bands. Therefore, these two peaks denote two different corner states that are located at acute-angled or obtuse-angled corners. Similarly, the frequency range for bulk and edge states also can be identified. Further, to visualize the spatial distribution of the vortex oscillations for different modes, four representative frequencies are chosen: 872 MHz for the bulk state, 934 MHz for the edge state, 944 MHz for the acute-angled corner state, and 948 MHz for the obtuse-angled corner state, respectively. We then stimulate their dynamics by applying a sinusoidal field to the whole system. The 10 ns gyration paths of all vortices are plotted in Figs. \ref{Figure41}(h)-\ref{Figure41}(k) when the excitation field drives a steady-state vortex dynamics. The spatial distribution of vortices motion for the bulk and edge states are shown in Fig. \ref{Figure41}(h) and Fig. \ref{Figure41}(i), respectively. We observe type I corner state with vortex oscillation localized at the acute-angled corner in Fig. \ref{Figure41}(j). Interestingly, one can note a strong hybridization between the type II and type III corner states, as shown in Fig. \ref{Figure41}(k), which is because their frequencies are very close to each other and their wavefunctions have a large overlap [see Figs. \ref{Figure41}(d) and \ref{Figure41}(f)].

Corner states are topologically protected and are deeply related to the symmetry of Hamiltonian \eqref{Eq54}. Below, we prove that the emergence of topological zero modes is protected by the generalized chiral symmetry. First of all, because $(\xi_{1}^{2}+2\xi_{2}^{2})/2\omega_{K}\ll\omega_{0}$, the diagonal element of $\mathcal {H}$ can be regarded as a constant, i.e., $Q_{0}=\omega_{0}$, which is the ``zero-energy" of the original Hamiltonian. $Q_{1,2,3,4,5,6}$ are the next-nearest hopping terms. At first glance, the system does not possess any chiral symmetry to protect the ``zero-energy" modes because the breathing honeycomb lattice is not a bipartite lattice. Here, we generalize the chiral symmetry for a unit cell containing six sites by defining
\begin{equation}\label{Eq58}
\begin{aligned}
 \Gamma_{6}^{-1}\mathcal {H}_{1}\Gamma_{6}&=\mathcal {H}_{2},\\
 \Gamma_{6}^{-1}\mathcal {H}_{2}\Gamma_{6}&=\mathcal {H}_{3},\\
 \Gamma_{6}^{-1}\mathcal {H}_{3}\Gamma_{6}&=\mathcal {H}_{4},\\
 \Gamma_{6}^{-1}\mathcal {H}_{4}\Gamma_{6}&=\mathcal {H}_{5},\\
 \Gamma_{6}^{-1}\mathcal {H}_{5}\Gamma_{6}&=\mathcal {H}_{6},\\
 \mathcal {H}_{1}+\mathcal {H}_{2}+\mathcal {H}_{3}+\mathcal {H}_{4}&+\mathcal {H}_{5}+\mathcal {H}_{6}=0,
 \end{aligned}
\end{equation}
where the chiral operator $\Gamma_{6}$ is a diagonal matrix, and $\mathcal{H}_{1}=\mathcal{H}-Q_{0}\text{I}$. Here, to prove the system has generalized chiral symmetry, we divide the system into six subgroups with the components of matrix Hamiltonian being nonzero only between different subgroups, such a property is essential for chiral symmetry and indicates no interaction within sublattices. Upon combining the last equation with the previous five in Eqs. \eqref{Eq58}, we have $\Gamma_{6}^{-1}\mathcal {H}_{6}\Gamma_{6} =\mathcal {H}_{1}$, implying that $[\mathcal {H}_{1},\Gamma_{6}^{6}]=0$; thus, $\Gamma_{6}^{6}=\text{I}$, which is completely analogous to the SSH model \cite{SuPRL1979}. Hamiltonians $\mathcal {H}_{1,2,3,4,5,6}$ each have the same set of eigenvalues $\lambda_{1,2,3,4,5,6}$. The eigenvalues of $\Gamma_{6}$ are $1,\ \exp(2\pi i/6),\ \exp(4\pi i/6),\ \exp(\pi i),\ \exp(8\pi i/6)$, and $\exp(10\pi i/6)$. Therefore, we can write
\begin{equation}\label{Eq59}
 \Gamma_{6}=\left(
 \begin{matrix}
   1 & 0 & 0& 0& 0& 0 \\
   0 & e^{\frac{2\pi i}{6}} & 0& 0& 0& 0 \\
   0 & 0 & e^{\frac{4\pi i}{6}}& 0& 0& 0\\
   0 & 0 & 0& e^{\pi i}& 0& 0\\
   0 & 0 & 0 & 0& e^{\frac{8\pi i}{6}}& 0\\
   0 & 0 & 0 & 0& 0& e^{\frac{10\pi i}{6}}
  \end{matrix}
  \right),
\end{equation}
in the same bases as that for expressing Hamiltonian \eqref{Eq54}. By taking the trace of the sixth line from Eqs. \eqref{Eq58}, we can obtain $\sum_{i=1}^{6}\text{Tr}(\mathcal {H}_{i})=6\text{Tr}(\mathcal {H}_{1})=0$, which indicates that the sum of the six eigenvalues vanishes $\sum_{i=1}^{6}\lambda_{i}=0$. Given an eigenstate $\phi_{j}$ that has support in only sublattice $j$, it will satisfy $\mathcal {H}_{1}\phi_{j}=\lambda\phi_{j}$ and $\Gamma_{6}\phi_{j}=\exp[2\pi i(j-1)/6]\phi_{j}$ with $j=1,2,3,4,5,6$. From these formulas and Eqs. \eqref{Eq58}, we obtain $\sum_{i=1}^{6}\mathcal {H}_{i}\phi_{j}=\sum_{i=1}^{6}\Gamma_{6}^{-(i-1)}\mathcal {H}_{1}\Gamma_{6}^{i-1}\phi_{j}=6\lambda\phi_{j}=0$, indicating $\lambda=0$ for any mode that has support in only one sublattice, i.e., zero-energy corner state. 

\begin{figure}[ptbh]
\begin{centering}
\includegraphics[width=0.90\textwidth]{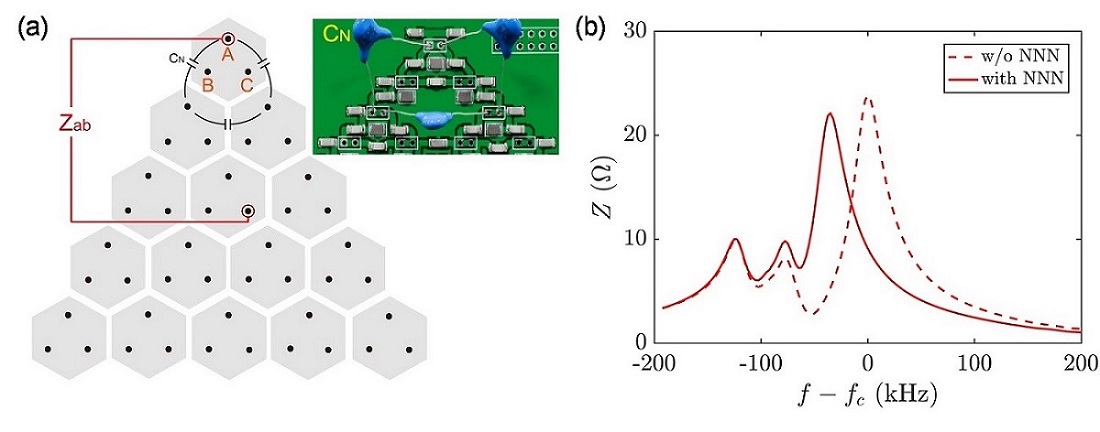}
\par\end{centering}
\caption{ (a) Locally connecting three capacitors $C_{N}$ within sublattices $A$ in the top corner. The inset shows the experiment setup. (b) The experimental results for configuration considered in (a). The dashed curve is the measurement without $C_{N}$. Source: The figures are taken from Ref. \cite{YangPRR2020}}
\label{Figure42}
\end{figure}
The corner states are protected by the generalized chiral symmetry. To prove this point experimentally, one can observe whether the frequencies of corner states are robust when the generalized chiral symmetry is broken by introducing the NNN hopping terms in the specific lattice. However, it is rather difficult to introduce NNN hopping by a designed manner in magnetic and condensed matter systems. Very recently, it is shown that the emerging topolectrical circuits can solve this problem for the reason that the coupling between any two lattice can be easily realized by adding extra circuit elements (such as capacitor and inductor). Based on the breathing kagome topolectrical circuit, Yang \emph{et al.} \cite{YangPRR2020} observed the symmetry-protected zero modes (corner states). They proved that the frequency of corner states suffers from a obvious shift when the NNN hopping is introduced by connecting the capacitor $C_{N}$ within the $A$ sites in the corner. The illustration and experiment setup are shown in Fig. \ref{Figure42} (a). Figure \ref{Figure42} (b) plots the experimental measurements of impedance $Z$ (between corner and bulk) with and without $C_{N}$. Furthermore, it is also comfirmed that the frequency of corner states does not change when the NNN hopping is located in the edge or bulk. These experimental results therefore substantiate the conclusion that corner states are indeed protected by the generalized chiral symmetry.       

\subsubsection{Square lattice}\label{section3.5.3}
In previous sections, we have discussed the HOTI phase in the breathing kagome and honeycomb lattice of vortices. On the one hand, the square lattice is also widely studied in different systems \cite{XiePRB2018,ImhofNP2018,XueNC2020}. On the other hand,  it is well known that topological states of many Hamiltonians, which support topological states in honeycomb lattice or other lattices, disappear immediately when the lattice deform into a square lattice. Thus the study of topological states in square lattice is important and may be non-trivial in this aspect. Moreover, previous studies are focused on corner states only with a single frequency. Since the generalized Thiele's equation contains higher-order terms, the topologically stable multimode corner phases may exist in magnetic soliton lattice. In this subsection, we show that the multimode HOTI phase indeed emerges in a breathing square lattice of vortices. Besides, it is demonstrated that the HOTI phase based on square lattice is convenient for the application of display.   
\begin{figure}[ptbh]
\begin{centering}
\includegraphics[width=0.90\textwidth]{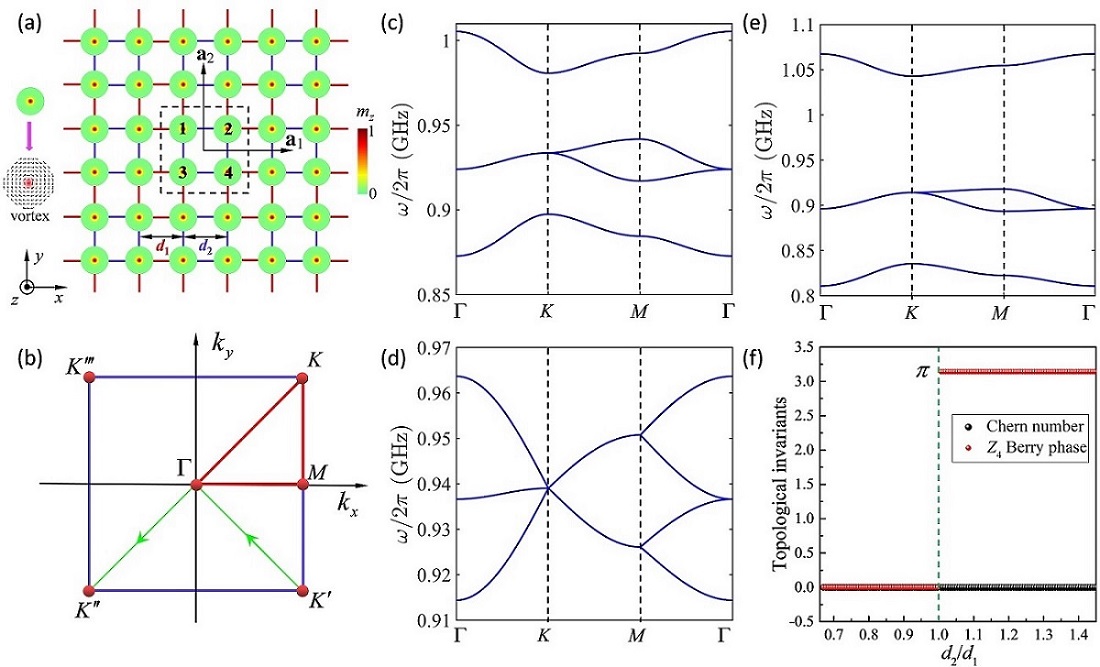}
\par\end{centering}
\caption{ (a) Illustration of the breathing square lattice of magnetic vortices, with $\mathbf{a}_1$ and $\mathbf{a}_2$ denoting the basis vectors.(b) The first Brillouin zone, with the high-symmetry points, the green line segment is the integral path for calculating $\mathbb{Z}_{4}$ Berry phase. The (lowest-four) band structures along the path $\Gamma$-$K$-$M$-$\Gamma$ for different geometric parameters: $d_{1}=3.6r, d_{2}=2.4r$ (c), $d_{1}=d_{2}=3.6r$ (d), and $d_{1}=2.08r, d_{2}=3.6r$ (e). (f) Dependence of Chern number and $\mathbb{Z}_{4}$ Berry phase on the ratios $d_{2}/d_{1}$ with $d_{1}$ being fixed to $3r$. Source: The figures are taken from Ref. \cite{LiPRB2020}}
\label{Figure43}
\end{figure}

The breathing square lattice of magnetic nanodisks with vortex states is shown in Fig. \ref{Figure43}(a). Similarly, the generalized Thiele's equation is adopted to describe the collective dynamics of the vortex lattice and the eigenvalue equation of the system can be obtained: 
\begin{equation}\label{Eq60}
  \begin{aligned}
 \hat{\mathcal {D}}\psi_{j}=(\omega_{K}-\frac{\xi^{2}_{1}+\xi^{2}_{2}}{\bar{\omega}_{K}})\psi_{j}+\zeta_{1}\sum_{k\in\langle j_{1}\rangle}\psi_{k}+\zeta_{2}\sum_{k\in\langle j_{2}\rangle}\psi_{k}
   -\frac{\xi_{1}\xi_{2}}{2\bar{\omega}_{K}}\sum_{s\in\langle\langle j_{1}\rangle\rangle}e^{i2\bar{\theta}_{js}}\psi_{s}-\frac{\xi^{2}_{2}}{2\bar{\omega}_{K}}\sum_{s\in\langle\langle j_{2}\rangle\rangle}e^{i2\bar{\theta}_{js}}\psi_{s}-\frac{\xi^{2}_{1}}{2\bar{\omega}_{K}}\sum_{s\in\langle\langle j_{3}\rangle\rangle}e^{i2\bar{\theta}_{js}}\psi_{s},
  \end{aligned}
\end{equation}

For an infinite lattice, the dashed black rectangle indicates the unit cell, as shown in Fig. \ref{Figure43}(a). $\textbf{a}_{1}=a\hat{x}$ and $\textbf{a}_{2}=a\hat{y}$ are two basis vectors, with $a=d_{1}+d_{2}$. The matrix form of the Hamiltonian in momentum space can be obtained by considering a plane wave expansion of $\psi_{j}=\phi_{j}\exp(i\omega t)\exp\big[i(n\mathbf{k}\cdot\textbf{a}_{1}+m\textbf{k}\cdot\textbf{a}_{2})\big]$, where $\textbf{k}$ is the wave vector, $n$ and $m$ are two integers:

\begin{equation}\label{Eq61}
 \mathcal {H}=\left(
 \begin{matrix}
   Q_{0} & \zeta_{2}+\zeta_{1}\exp(-i\textbf{k}\cdot\textbf{a}_{1}) & \zeta_{2}+\zeta_{1}\exp(i\textbf{k}\cdot\textbf{a}_{2}) & Q_{1} \\
   \zeta_{2}+\zeta_{1}\exp(i\textbf{k}\cdot\textbf{a}_{1}) & Q_{0} & Q_{2} & \zeta_{2}+\zeta_{1}\exp(i\textbf{k}\cdot\textbf{a}_{2}) \\
   \zeta_{2}+\zeta_{1}\exp(-i\textbf{k}\cdot\textbf{a}_{2}) & Q_{2}^{*} & Q_{0}& \zeta_{2}+\zeta_{1}\exp(-i\textbf{k}\cdot\textbf{a}_{1})\\
   Q_{1}^{*} & \zeta_{2}+\zeta_{1}\exp(-i\textbf{k}\cdot\textbf{a}_{2})& \zeta_{2}+\zeta_{1}\exp(i\textbf{k}\cdot\textbf{a}_{1})& Q_{0}
  \end{matrix}
  \right),
\end{equation}
with elements expressed as
\begin{equation}\label{Eq62}
\begin{aligned}
Q_{0}&=\omega_{K}-\frac{\xi_{1}^{2}+\xi_{2}^{2}}{\bar{\omega}_{K}}-\frac{\xi_{1}\xi_{2}}{\bar{\omega}_{K}}[\cos(\textbf{k}\cdot\textbf{a}_{1})+\cos(\textbf{k}\cdot\textbf{a}_{2})], \\
Q_{1}&=\frac{\xi_{1}\xi_{2}}{\bar{\omega}_{K}}[\exp(i\textbf{k}\cdot\textbf{a}_{2})+\exp(-i\textbf{k}\cdot\textbf{a}_{1})]+\frac{\xi^{2}_{1}}{\bar{\omega}_{K}}\exp[i\textbf{k}\cdot(\textbf{a}_{2}-\textbf{a}_{1})]+\frac{\xi^{2}_{2}}{\bar{\omega}_{K}}, \\
Q_{2}&=\frac{\xi_{1}\xi_{2}}{\bar{\omega}_{K}}[\exp(i\textbf{k}\cdot\textbf{a}_{2})+\exp(i\textbf{k}\cdot\textbf{a}_{1})]+\frac{\xi^{2}_{1}}{\bar{\omega}_{K}}\exp[i\textbf{k}\cdot(\textbf{a}_{2}+\textbf{a}_{1})]+\frac{\xi^{2}_{2}}{\bar{\omega}_{K}}.
\end{aligned}
\end{equation}

The bulk band structures with various geometric parameters ($d_{1}$ and $d_{2}$) are shown in Figs. \ref{Figure43}(c)-\ref{Figure43}(e). For $d_{1}=d_{2}=3.6r$ [see Fig. \ref{Figure43}(d)], all bands merge together, leading to a gapless band. However, when $d_{1}\neq d_{2}$, two gaps open and they locate between 1st and 2nd bands, 3rd and 4th bands, respectively. Interestingly, the 2nd and 3rd bands are always merged no matter what values $d_{1}$ and $d_{2}$ take. The topological invariants Chern number and $\mathbb{Z}_{4}$ Berry phase can be used to further distinguish whether these insulating phases are topologically protected.

Figure \ref{Figure43}(f) plots the dependence of the Chern number $\mathcal{C}$ and the $\mathbb{Z}_{4}$ Berry phase $\mathcal{\theta}$ on the ratio $d_{2}/d_{1}$ with $d_{1}$ fixed to $3r$. One can clearly see that the $\mathbb{Z}_{4}$ Berry phase is quantized to 0 when $d_{2}/d_{1}<1$ and to $\pi$ otherwise, showing that $d_{2}/d_{1}=1$ is the phase transition point separating the trivial and topological phases. Furthermore, the Chern number vanishes for all ratios $d_{2}/d_{1}$, indicating that the system has no first-order TI phase.  A simple way to understand the zero Chern number is that all elements of the lattice Hamiltonian are real numbers apart from the phase factor $\text{exp}(\pm i\mathbf{k}\cdot\textbf{a}_{1,2})$, which naturally leads to a vanishing Chern number. Therefore, we conclude that the system is in the HOTI phase when $d_{2}/d_{1}>1$, and in the trivial phase when $d_{2}/d_{1}<1$. This conclusion holds independent of the $d_{1}$ value, since the 2D breathing square lattice can be viewed as two copies of SSH chains along the horizontal and vertical directions, respectively.
\begin{figure}[ptbh]
\begin{centering}
\includegraphics[width=0.95\textwidth]{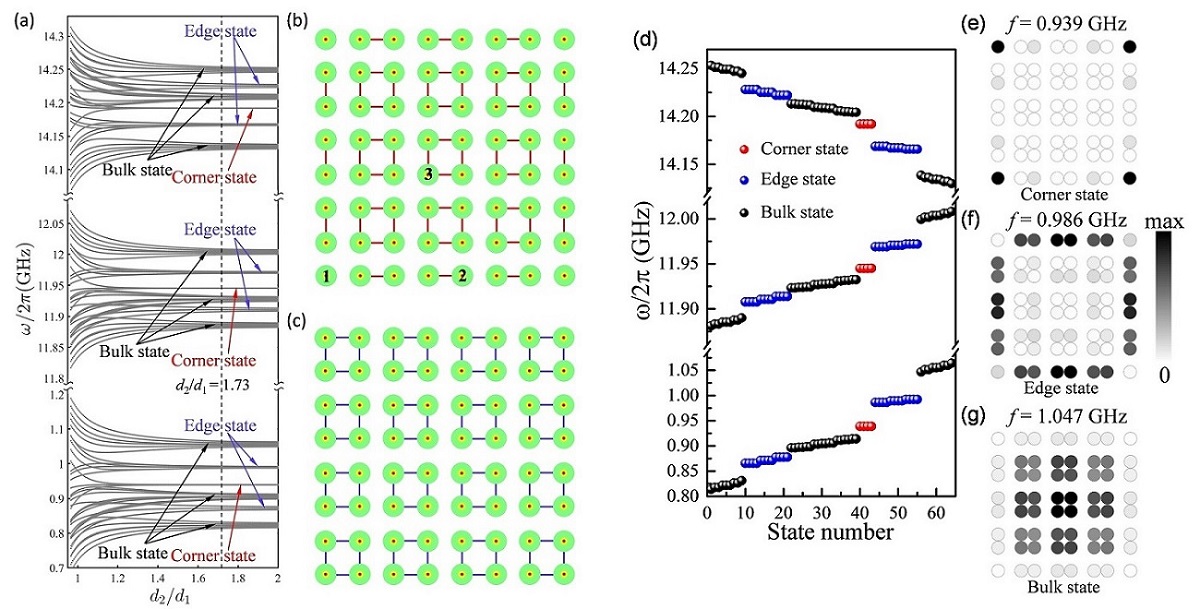}
\par\end{centering}
\caption{(a) Eigenfrequencies of collective vortex gyrations under different ratios $d_{2}/d_{1}$ in a finite lattice of the size $(3d_{1}+4d_{2})\times(3d_{1}+4d_{2})$, with $d_{1}$ fixed to $2.08r$. The schematic plot of the vortex lattice in (a) for two limit cases, $d_{2}\rightarrow\infty$ (b) and $d_{1}\rightarrow\infty$ (c). (d) Eigenfrequencies of square-shape vortex lattice with $d_{1}=2.08r$ and $d_{2}=3.6r$. The spatial distribution of vortex gyrations for the corner (e), edge (f), and bulk (g) states with different frequencies. Source: The figures are taken from Ref. \cite{LiPRB2020}}
\label{Figure44}
\end{figure}

The dynamics of finite vortex lattice can be used to directly confirm the existence of corner states. The eigenfrequencies as a function of $d_{2}/d_{1}$ for a finite square-shaped lattice [see Figs. \ref{Figure44}(b) and \ref{Figure44}(c)] are ploted in Fig. \ref{Figure44}(a). The bulk, edge, and corner states are marked by black, blue, and red arrows in Fig. \ref{Figure44}(a), respectively. The intuitive understanding why these corner states only appear in the special parameter region ($d_{2}/d_{1}>1$) is as follow: on the one hand, the configuration shown in Fig. \ref{Figure44}(b) is in the HOTI phase. In such a case, one can clearly identify four isolated vortices at corners. Thus the localized corner states will appear; on the other hand, in the limit $d_{1}\rightarrow\infty$ [see Fig. \ref{Figure44}(c)], there are no uncoupled vortices, thus no corner states. The system is therefore in the trivial phase.

The square-shaped vortex lattice with $d_{1}=2.08r$ and $d_{2}=3.6r$ is considered to visualize the second-order corner states. Figures \ref{Figure44}(d)-\ref{Figure44}(g) show the computed eigenfrequencies and eigenmodes. It is found that there exist three corner states with different frequencies (0.939 GHz, 11.945 GHz, and 14.192 GHz), represented by red balls in Fig. \ref{Figure44}(d). The spatial distribution of the corner state shows that its oscillation is highly localized at four corners [see Fig. \ref{Figure44}(e)]. The spatial distribution of the edge and bulk states are also shown in Figs. \ref{Figure44}(f) and \ref{Figure44}(g), respectively. However, these edge states are Tamm-Shockley type \cite{TammPZS1932,ShockleyPR1939} and are not topologically protected because of the vanishing Chern number [see Fig. \ref{Figure43}(f)].
\begin{figure}[ptbh]
\begin{centering}
\includegraphics[width=0.98\textwidth]{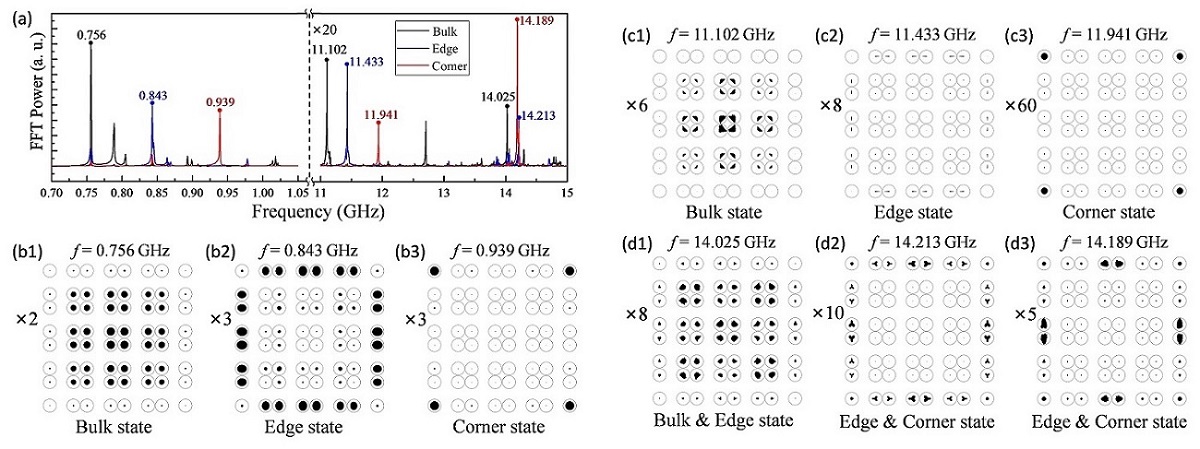}
\par\end{centering}
\caption{ (a) The temporal Fourier spectra of the vortex oscillations at different positions. (b1-d3) The spatial distribution of oscillation amplitude under the exciting field with different frequencies indicated in (a). The oscillation amplitudes of the vortices centers have been magnified in suitable multiples for better observation. Source: The figures are taken from Ref. \cite{LiPRB2020}}
\label{Figure45}
\end{figure}

Micromagnetic simulation results are plotted in Fig. \ref{Figure45} for a comparison. The spectra of the vortex oscillations at different positions are shown in Fig. \ref{Figure45}(a). One can clearly see that near the eigenfrequencies of a single vortex gyration (0.939 GHz and 11.941 GHz), the spectrum for the corner has two very strong peaks, which do not exist for edge and bulk bands. Similarly, the frequency range supporting the bulk and edge states also can be identified. Interestingly, for the 14.189 GHz peak, although the spectrum in the corner has a strong peak, the oscillation amplitude at the edge is sizable as well, which indicates a strong coupling between edge and corner oscillations. Similar mode hybridization occurs at 14.025 GHz and 14.213 GHz, too. The spatial distribution of oscillation amplitude for different frequencies are ploted in Figs. \ref{Figure45}(b1)-(d3), from which one can distinguish the bulk states, edge states, and corner states. The hybridized modes are observed as well: bulk \& edge state [Fig. \ref{Figure45}(d1)] and edge \& corner state [Figs. \ref{Figure45}(d2) and \ref{Figure45}(d3)]. The mode hybridization results from the fact that the frequencies of these different states are so close, see Fig. \ref{Figure45}(a).

\begin{figure}[ptbh]
\begin{centering}
\includegraphics[width=0.65\textwidth]{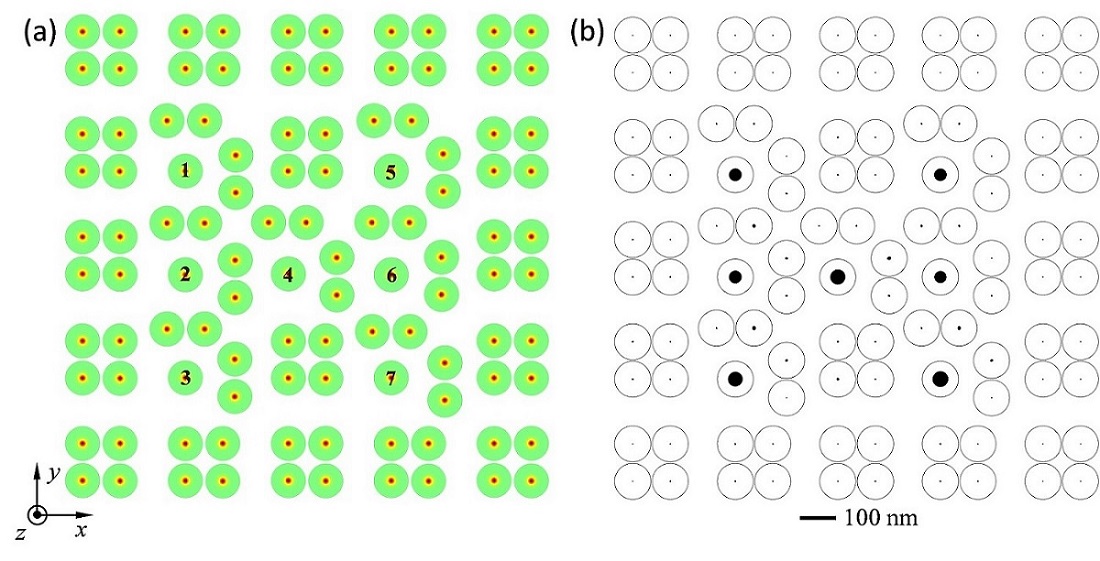}
\par\end{centering}
\caption{(a) The schematic plot of the vortex lattice for ``H'' display. (b) Micromagnetic simulation of vortex gyrations with frequency $f=0.939$ GHz. The oscillation amplitudes of the vortex core have been magnified by 3 times. Source: The figures are taken from Ref. \cite{LiPRB2020}}
\label{Figure46}
\end{figure}

The emerging HOTI in vortex lattice can be used to design topological devices. Figure \ref{Figure46} show a display device based on vortex lattice. The desired display “H” in the HOTI phase is surrounded by another vortex lattice in the trivial phase, as shown in Fig. \ref{Figure46}(a). The display points are marked by arabic numbers $1-7$. The collective dynamics of the whole system is stimulated by applying a sinusoidal magnetic field with the frequency $f=0.939$ GHz. Figure. \ref{Figure46}(b) plots the spatial distribution of the oscillation amplitudes, from which one can clearly see that only the vortices at the desired display points have sizable oscillations, while the other vortices do not participate in the display. We point out that other display shapes can be realized by a similar method, too.     

\section{Conclusion and outlook}\label{section4}

We have reviewed the recent progress on topological insulator and semimetal phases in magnon and soliton based crystals. These studies not only deepen our understanding on topological physics and its manifestation in classical magnetism, but also hold promise for future robust spintronic devices. 

The topological insulating phases of magnons appear in gapped system and can support topologically protected spin wave modes confined in the boundaries or corners. Dirac and Weyl magnons emerge in gapless band structures and magnon arc states are topologically protected in the surface. The generation and detection of topological edge spin wave can be achieved with the same techniques used in conventional magnonic devices. For example, topological edge spin wave can be excited by the antenna microwave magnetic field and can be detected by the Brillouin light-scattering spectroscopy \cite{KruglyakJPD2010,SergaJPD2010}. Remarkably, Bonetti \emph{et al.} \cite{BonettiNC2015} report that the real-space spin wave movie can be created by using a high-sensitivity time-resolved magnetic X-ray microscopy, which can be conveniently adopted to detect the topological edge spin wave. For application, the spin-wave edge states can be used to design various topological magnonic devices, including spin-wave diode, spin-wave beam splitters, and spin-wave interferometers, etc.
\begin{figure}[ptbh]
\begin{centering}
\includegraphics[width=0.85\textwidth]{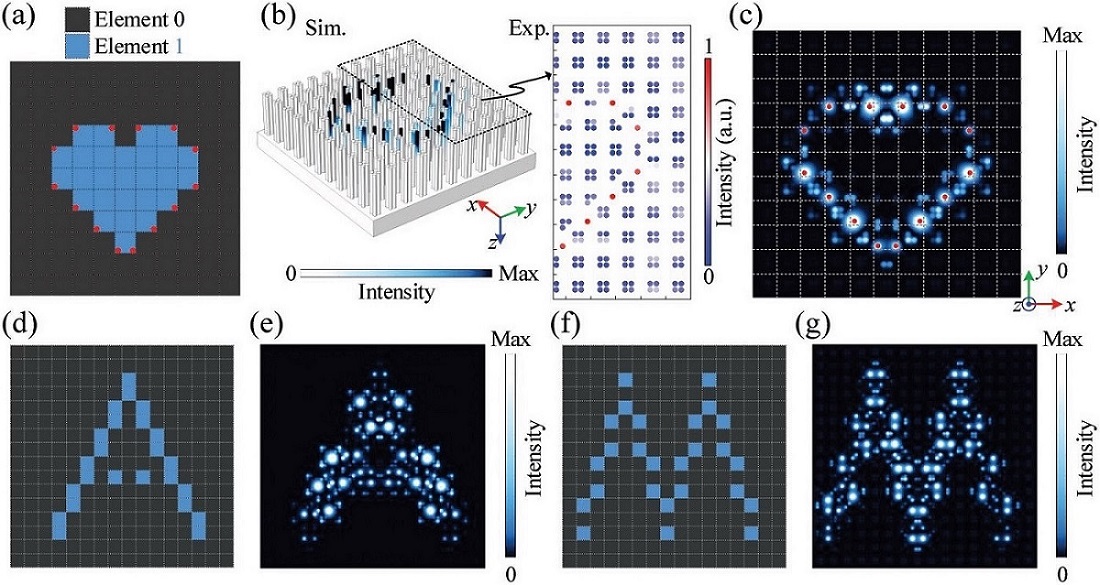}
\par\end{centering}
\caption{(a) Programmable subwavelength acoustic second-order TI imaging device composed of a heart-like nontrivial region (elements 1) surrounded by trivial region (elements 0). (b) Simulated (left panel) and experimentally measured (right panel) distributions of the sound intensity fields for the heart-like image. (c) Sound intensity distributions in the xy-plane at a distance of 0.1 mm above the holey concentric square crystal surface. The red dots represent the corners that should be excited. (d) Coding arrays and (e) sound intensity distributions on the cross section of the acoustic image for letter “A”. (f) (g) Same as (d) and (e), but for imaging letter “M”. Source: The figures are taken from Ref. \cite{ZhangAM2019}}
\label{Figure47}
\end{figure}

The predicted second-order topological insulating phase based on magnetic soliton lattice can facilitate the design of different spintronic devices: (i) For the application aspect, it is still a challenging issue to realize stable display function in natural and artificial materials. The main obstacle lies in the difficulty for precisely controlling both the position and the frequency of the local oscillations. Topological insulators provide a new route for that purpose. The chiral topological edge states can perfectly localize the energy at the boundary of the system. If the boundary of the system is set to a specific shape, the display function can be realized under the external excitation of chiral edge modes. However, the proposal suffers from some disadvantages. On the one hand, to form a clear picture, a very large system needs to be conceived. On the other hand, the target picture must be continuous and it is difficult to display discrete pictures. The emerging HOTI can well solve these problems. Zhang \emph{et al}. \cite{ZhangAM2019} designed the programmable imaging device based on the acoustic second-order topological insulators, as shown in Fig. \ref{Figure47}. The imaging device consists of two subwavelength digital elements of "0" and "1", which correspond to trivial and second-order topological nontrivial state, respectively. Figure \ref{Figure47}(a) illustrates a heart-like acoustic profile. When the corner states [labeled by red dots in Fig. \ref{Figure47}(a)] are excited at the specific frequency, the simulated and experimental data verify that most energy is confined at the corners, which realized stable topological acoustic imaging device, as shown in Figs. \ref{Figure47}(b) and \ref{Figure47}(c). Similarly, the acoustic imaging of characters can be realized, see Figs. \ref{Figure47}(d)-\ref{Figure47}(g). For magnetic system, the imaging device based on HOTI in magnetic vortices lattice is discussed as well [see Fig. \ref{Figure46}]. Remarkably, these imaging device are topologically protected and can immune from external disturbances, which makes magnetic HOTIs have potential applications for designing topological spintronic imaging elements. (ii) Moreover, in recent years, nano-oscillators in magnetic systems have attracted great attention for potential applications \cite{KakaN2005,AwadNP2017,DemidovNM2010,ZengNS2013}. In particular, oscillators based on magnetic-soliton structures \cite{PribiagNP2007,HrkacJPD2015,ZhangNJP2015,HamadehPRL2014,DussauxNC2010,MartinezPRB2011} can be used as good microwave sources owing to their outstanding characteristics of small size, easy manipulation, and high tunability. However, the working frequency of these oscillators is very sensitive to external disturbances. If the HOTI phase is used, vortex-based nano oscillators should have extraordinary stability against defects and disorder and should therefore have broader prospects for application as topological microwave sources. (iii) The multiband nature of the corner modes (with a spectrum ranging from less than 1 GHz to dozens of gigahertz) is very useful for designing broadband topological devices.

\begin{figure}[ptbh]
\begin{centering}
\includegraphics[width=0.85\textwidth]{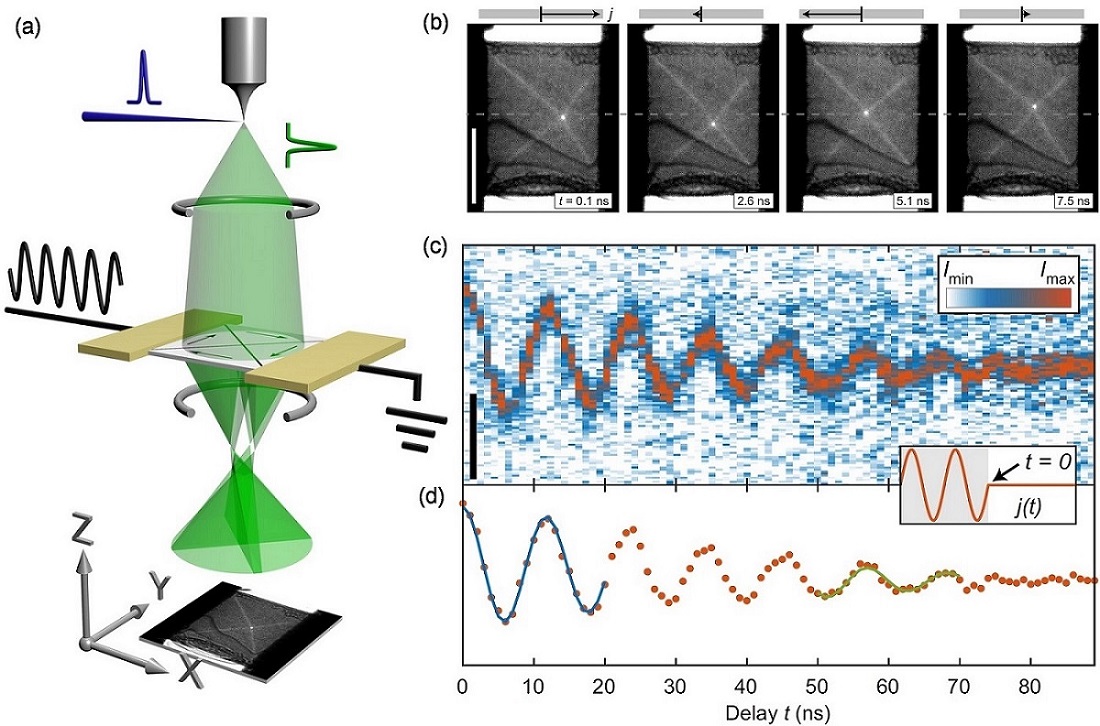}
\par\end{centering}
\caption{(a) Schematic of time-resolved Lorentz microscopy combined with synchronous radiofrequency sample excitation. An ultrashort electron pulse (green) is generated via a linear photoemission process from an optical laser pulse (blue) inside the electron gun. A magnetic sample (light gray) is stroboscopically illuminated with a nearparallel electron beam and imaged onto a transmission electron microscope camera. Dynamics in the sample are excited in situ with radiofrequency currents (black) phaselocked to multiples of the laser repetition rate, creating the appearance of a static image. (b) Four time-resolved Lorentz micrographs acquired in time steps of 2.5 ns at a defocus of 520 $\mu$m (scale bar: 1 $\mu$m). (c) Profiles of the vortex core along the x-axis, illustrating the damped sinusoidal motion towards the equilibrium position (scale bar: 200 nm). (d) Red: Tracked x-position of the vortex core as function of t. Blue/green: First/last result of the moving-window fit to the vortex trajectory. Source: The figures are taken from Ref. \cite{MollerCP2020}}
\label{Figure48}
\end{figure}

From an experimental point of view, we note that both fabricating the metamaterials of magnetic solitons and detecting the highly spatially localized corner modes are already within the reach of current technology. On the one hand, by using electron-beam lithography \cite{HanSR2013,BehnckePRB2015,SunPRL2013} or X-ray illumination \cite{GuangNC2020}, the artificial magnetic soliton with different lattices can be created. On the other hand, by tracking the nanometer-scale vortex orbits using the ultrafast Lorentz microscopy technique in a time-resolved manner \cite{MollerCP2020}, one can directly observe the second-order topological corner states of magnetic soliton lattice. Figure \ref{Figure48} shows the schematic representation of time-resolved Lorentz microscopy and the related experimental results for tracking vortex core. 

\begin{figure}[ptbh]
\begin{centering}
\includegraphics[width=0.95\textwidth]{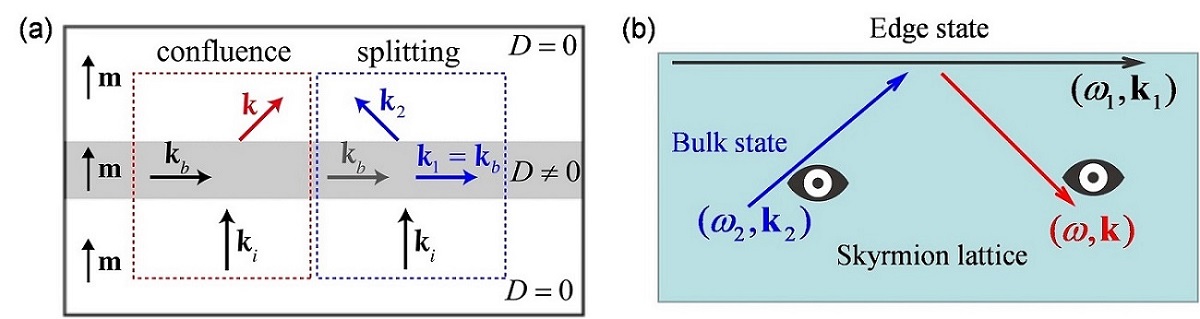}
\par\end{centering}
\caption{(a) Schematic picture of nonlinear three-magnon processes in the DM interaction nanostrip. In the dashed red square, it shows the three-magnon confluence of $\mathbf{k}_{i}$ and $\mathbf{k}_{b}$ into $\mathbf{k}$. In the dashed blue square,it plots the stimulated three-magnon splitting of $\mathbf{k}_{i}$ into two modes $\mathbf{k}_{i}=\mathbf{k}_{b}$ and $\mathbf{k}_{2}$, assisted by a localized magnon $\mathbf{k}_{b}$. (b) Illustration for detecting the edge state of skyrmion lattice. Source: The figures are taken from Ref. \cite{WangPRAP2018}.}
\label{Figure49}
\end{figure}

Moreover, the detection of soliton lattice edge states can be realized by using the interaction between edge and bulk waves which is similar to the nonlinear three-magnon process \cite{WangPRAP2018}. Figure \ref{Figure49}(a) shows the schematic picture of  nonlinear three-magnon processes in the DM interaction nanostrip, where a propagating spin wave ($\omega_{i},\mathbf{k}_{i}$) interact with the localized spin wave ($\omega_{b},\mathbf{k}_{b}$) bounded in the nanostrip. In general, two kinds of three-magnon processes (confluence and splitting) can occur, as illustrated in Fig. \ref{Figure49} (a). By comparing $\mathbf{k}_{i}$ and $\mathbf{k}$ (or $\mathbf{k}_{2}$), one can obtain the information of bounded state spin wave. Similarly, if we send both a bulk mode ($\omega_{2}$,$\mathbf{k}_{2}$) and an edge mode ($\omega_{1}$,$\mathbf{k}_{1}$) simultaneously in the soliton lattice, by probing the reflected bulk mode ($\omega$,$\mathbf{k}$) as shown in Fig. \ref{Figure49}(b), we can identify the information of edge state based on the energy-momentum conservation law $\omega=\omega_{2}+\omega_{1}$ and $(\mathbf{k}_{2}+\mathbf{k}_{1}-\mathbf{k})\cdot \hat{x}=0$, and $\omega=\omega_{2}-\omega_{1}$ and $(\mathbf{k}_{2}-\mathbf{k}_{1}-\mathbf{k})\cdot \hat{x}=0$ for three-wave confluences and for three-wave splittings, respectively.  

Finally, we comment that the study on topological phase and phase transitions in magnon and soliton metamaterials is quite active but is still in the initial stage, and many open questions are yet to be answered and more new phenomena is to be disvovered: 

(1) In the original calculations of magnetic soliton HOTIs, identical nanodisks are assumed. However, if the translational symmetry is broken, for instance by introducing Kekulé distortions into the disk sizes, one may realize topological devices supporting robust Majorana-like zero modes localized in the device’s geometric center \cite{GaoPRL2019}. 

(2) The third-order TIs have been realized in other systems \cite{XuePRL2019,XueNC2020,NiNC2020,WeinerSA2020,BaoPRB2019}, while there are no counterparts reported in magnetic system.

(3) The twisted bilayer graphene structure has attracted a lot of attention over the past few years for the exotic physical properties \cite{CaoN2018_1,CaoN2018_2}. We envision that the topological property of twisted bilayer of honeycomb lattice based on magnetic solitons (or spins) is also an appealing research topic. 

(4) The practical applications of topological insulating phases (especially for HOTI) in magnetic system are still lacking, since the experimental detection of these topological phases and the device design are challenging.

(5) The LLG equation describing the dynamics of magnetic moment is intrinsically nonlinear, however, the analytical theory about the topological magnons is based on the linear approximation. When the oscillation amplitude of magnetic moment or soliton is large enough, the nonlinear effect should be considered. The influence of nonlinearity on the  topology phase and phase transition is an interesting issue for future study.  

(6) In order to observe the Weyl points in magnonic system, the excitation sources with very high frequency (hundreds of gigahertz) are demanded, which, is not compatible with the mature microwave antenna technology. Magnetic soliton as another important excitation in magnetic system, holds much lower frequency (by one order of magnitude) than magnons. Therefore, realization Weyl semimetals in magnetic soliton crystals is an appealing research topic, too. 

\addcontentsline{toc}{section}{Declaration of competing interest}
\section*{Declaration of competing interest}
The authors declare no competing financial interests that could have appeared to influence the work reported in this paper.
\addcontentsline{toc}{section}{Acknowledgments}
\section*{Acknowledgments}
P. Y. would like to thank Beining Zhang, Zhenyu Wang, Huanhuan Yang, Chen Wang, Lingling Song, Xiaofan Wang, Wenrui Yang, Weiwei Bao, Tianlin Yu, Yuanyuan Jiang,  Huaiyang Yuan, Hang Li, Xiansi Wang, Ke Xia, and Xiangrong Wang for stimulating discussions and collaborations. This work was supported by the National Natural Science Foundation of China (NSFC) (Grants No. 12074057, No. 11604041, and No. 11704060). Z.-X. Li acknowledges financial support from the China Postdoctoral Science Foundation (Grant No. 2019M663461) and the NSFC (Grant No. 11904048).

\end{document}